\documentclass[traditabstract,doublecolumn]{aa}

\usepackage{graphicx}
\usepackage[intlimits,sumlimits]{amsmath}
\usepackage{deluxetable}
\usepackage{times,epsfig} 
\usepackage{amssymb}
\usepackage{mathrsfs}  

\usepackage{natbib}
\bibpunct{(}{)}{;}{a}{}{,} 
\usepackage{caption}

\usepackage{amsmath}
\usepackage[english]{babel}
\usepackage{longtable}
\usepackage{url}
\usepackage[pagebackref=true]{hyperref}
\usepackage{hyperref}
\usepackage{amssymb}
\usepackage{color}

\newcommand{\nosne}{34}
\newcommand{\nosnenir}{26}

\newcommand{\beqa}{\begin{eqnarray}} 
\newcommand{\eeqa}{\end{eqnarray}}

\newcommand{\bsub}{\begin{subequations}}
\newcommand{\esub}{\end{subequations}}
\newcommand{\beal}{\begin{align}}
\newcommand{\ealn}{\end{align}}

\authorrunning{Taddia et al.}
\titlerunning{CSP-I SE~SN light-curve analysis}
\begin{document}
\title{The Carnegie Supernova Project~I: \\ analysis of stripped-envelope supernova light curves\thanks{Based on observations collected at  Las Campanas Observatory.}}

\author{
F. Taddia\inst{1} 
\and M.~D. Stritzinger\inst{2}
\and M. Bersten\inst{3,4,5}
\and E. Baron\inst{6}
\and C. Burns\inst{7}
\and C. Contreras\inst{8,2}
\and S. Holmbo\inst{2}
\and E.~Y. Hsiao\inst{9}
\and N. Morrell\inst{8}
\and M.~M. Phillips\inst{8}
\and J. Sollerman\inst{1}
\and N.~B. Suntzeff\inst{10}
}

\institute{
Department of Astronomy, The Oskar Klein Center, Stockholm University, AlbaNova, 10691 Stockholm, Sweden
\and
Department of Physics and Astronomy, Aarhus University, Ny Munkegade 120, DK-8000 Aarhus C, Denmark 
\and 
 Facultad de Ciencias Astron\'{o}micas y Geof\'{i}sicas, Universidad
Nacional de La Plata, Paseo del Bosque S/N, B1900FWA La Plata, Argentina
\and
Instituto de Astrof\'isica de La Plata (IALP),
  CONICET, Argentina
  \and 
Kavli Institute for the Physics and Mathematics of
the Universe, Todai Institutes for Advanced Study, University of
Tokyo, 5-1-5 Kashiwanoha, Kashiwa, Chiba 277-8583, Japan
\and
Homer L. Dodge Department of Physics and Astronomy, University of Oklahoma, 440 W. Brooks, Rm 100, Norman, OK 73019-2061, USA
\and
Observatories of the Carnegie Institution for Science, 813 Santa Barbara St., Pasadena, CA 91101, USA
\and
Las Campanas Observatory, Carnegie Observatories, Casilla 601, La Serena, Chile.
\and
Department of Physics, Florida State University, 77 Chieftain Way, Tallahassee, FL, 32306, USA
\and
George P. and Cynthia Woods Mitchell Institute for Fundamental Physics and Astronomy, Department of Physics and Astronomy, Texas A\&M University, College Station, TX 77843, USA 
}

\date{Received XXX ; Accepted XXX}

\abstract{Stripped-envelope (SE)  supernovae (SNe) include H-poor (Type
  IIb), H-free (Type Ib) and He-free (Type Ic) events thought to be associated with the deaths of massive stars. 
  The exact nature of their progenitors is a matter of debate with several lines of evidence pointing towards intermediate
  mass ($M_{\rm init}<20~M_{\odot}$) stars in binary systems, while in other cases they may be 
  linked to single massive Wolf-Rayet stars. 
  Here we present the analysis of the light curves of 34 SE~SNe published by the \textit{Carnegie Supernova Project} (CSP-I), which are unparalleled in terms of photometric accuracy and  wavelength range.
Light-curve parameters are estimated through the fits of an analytical function and trends are searched for among the resulting fit parameters. Detailed inspection of the dataset suggests a tentative correlation between the peak absolute $B$-band  magnitude and $\Delta m_{15}(B)$, while the post maximum light curves reveals a correlation between the late-time linear slope and $\Delta m_{15}$. 
Making use of the full set of optical \textit{and} near-IR photometry, combined with robust host-galaxy extinction corrections, comprehensive bolometric light curves are constructed  and compared to both analytic and hydrodynamical models. This analysis finds consistent results among the two different modeling techniques and from the hydrodynamical models we obtained ejecta masses of $1.1-6.2$~$M_{\odot}$, $^{56}$Ni masses of $0.03-0.35$~$M_{\odot}$, and explosion energies (excluding two SNe~Ic-BL)  of $0.25-3.0\times10^{51}$~erg. 
Our analysis indicates that adopting $\kappa = 0.07$ cm$^{2}$~g$^{-1}$ as the mean opacity serves to be a suitable assumption when comparing Arnett-model results to those obtained from hydrodynamical calculations. 
We also find that adopting \ion{He}{i} and \ion{O}{i} line velocities to infer the expansion velocity in He-rich and He-poor SNe, respectively,  provides ejecta masses relatively  similar to those obtained by using the \ion{Fe}{ii} line velocities, although the use of \ion{Fe}{ii} as a diagnostic does imply higher explosion energies.
The inferred range of ejecta masses are compatible with  intermediate mass ($M_{ZAMS} \leq 20$ $M_{\sun}$) progenitor stars in  binary systems for the majority of SE~SNe. Furthermore, our hydrodynamical modeling of the bolometric light curves suggest  a significant fraction of the sample may have experienced significant mixing of $^{56}$Ni, particularly in the case of SNe~Ic. }
\keywords{supernovae: general, supernovae: individual: 
SN~2004ew, SN~2004ex, SN~2004fe, SN~2004ff, SN~2004gq, SN~2004gt, SN~2004gv, SN~2005Q, SN~2005aw, SN~2005bf, SN~2005bj, SN~2005em, SN~2006T, SN~2006ba, SN~2006bf, SN~2006ep, SN~2006fo, SN~2006ir, SN~2006lc, SN~2007C, SN~2007Y, SN~2007ag, SN~2007hn, SN~2007kj, SN~2007rz, SN~2008aq, SN~2008gc, SN~2008hh, SN~2009K, SN~2009Z, SN~2009bb, SN~2009ca, SN~2009dp, SN~2009dt
} 

\maketitle

\section{Introduction}


Stripped-envelope (SE) core-collapse supernovae (SNe) are associated
with the deaths of massive stars that have experienced significant
mass loss over their evolutionary lifetimes. The severity of the mass loss drives to first order the contemporary spectroscopic classification sequence of Type
 IIb$\rightarrow$Ib$\rightarrow$Ic \citep[e.g.,][]{filippenko97,galyam16, prentice17, shivvers17}. The progenitors of SN~IIb are thought to  retain a residual amount ($\sim 0.01~M_{\sun}$) of hydrogen prior to exploding, and as an outcome they exhibit hydrogen features in pre-maximum spectra. However, soon after maximum ($t_{\rm max}$) their spectra typically evolve to resemble normal SNe~Ib \citep[e.g., SN~1993J;][]{filippenko93}, exhibiting conspicuous helium features and only traces (if any signatures at all) of hydrogen. Rounding out the spectroscopic sequence are SNe~Ic, which lack hydrogen and helium spectral features, and in some instances show exceedingly broad-lined (BL) spectral features. Some SNe~Ic-BL have been discovered to emerge from long-duration gamma-ray bursts \citep[e.g., SN~1998bw;][]{galama98}.
 
A large number of single-object studies of SE~SNe exists, especially of
events that occurred in nearby galaxies. Examples of SNe~IIb that were comprehensively studied in single-object papers are: 
SN~1993J \citep{filippenko93,filippenko94},
SN~2008ax \citep{pastorello08,chornok11,taubenberger11,folatelli15},
SN~2011dh \citep{bersten12,ergon14,ergon15},   
SN~2011hs \citep{bufano14}, 
SN~2010as \citep{folatelli14b}, and 
PTF12os \citep{fremling16}.
 Among the studies of SNe~Ib, we find:  
SN~1996N \citep{sollerman98}, 
SN~1999dn \citep{cano14},
SN~2005bf \citep{anupama05,tominaga05,folatelli06},
SN~2007Y \citep{stritzinger09}, 
SN~2008D \citep{soderberg08,mazzali08,modjaz09,malesani09,bersten13},
SN~2009jf \citep{valenti11}, 
SN~2013ge \citep{drout16}, amd 
iPTF13bvn \citep{cao13,fremling14,bersten14,fremling16}.
Finally, SNe~Ic and Ic-BL examined in dedicated papers are: 
SN~1994I \citep{filippenko95},
SN~1997ef \citep{iwamoto00,mazzali00}
SN~1998bw \citep{galama98,patat01}, 
SN~2002ap \citep{foley03}
SN~2003jd \citep{valenti08a}
SN~2004aw \citep{taubenberger06},
SN~2006aj \citep{mirabal06,modjaz06,pian06,sollerman06,campana06,mazzali06},
SN~2007gr \citep{valenti08b,hunter09}, 
SN~2009bb \citep{pignata11},
SN~2010bh \citep{cano11,bufano12}, 
PTF10vgv \citep{corsi12}, 
SN~2011bm \citep{valenti12}, 
PTF11mnb \citep{taddia16b}, and
iPTF15dtg \citep{taddia16a}.

The light curves of SE~SNe are mainly powered by thermalized energy originating from the radioactive decay chain $^{56}$Ni $\rightarrow$ $^{56}$Co $\rightarrow$ $^{56}$Fe.
Given the amount of $^{56}$Ni synthesized in SE~SNe, their relatively low ejecta masses, and the compact radii of their progenitors, they almost always display bell-shaped light curves peaking a few weeks after explosion. 
For a handful of cases the SE~SNe were discovered 
within hours to days after explosion. 
 In some of 
these cases an initial peak has been documented, followed by a rapid drop in luminosity. 
 This early emission is believed to be driven by the shock-wave   breaking out through the progenitors surface or through an extended envelope surrounding the progenitor \citep[e.g.,][]{arnett76,ensman92,woosley94,bersten12,piro13,nakar14,piro15}. The early luminosity is mainly dependent on the progenitor radius. 
Evidence of this phenomenon 
was first documented in the peculiar Type~II SN~1987A
\citep[e.g.,][]{catchpole87}, the Type~IIb SN~1993J
\citep[e.g.,][]{vandriel93}, the Type~Ib/c SN~1999ex
\citep{stritzinger02} and the Type~Ib SN~2008D
\citep[e.g.,][]{mazzali08,soderberg08}. 
Recently, with the advent of both amateur and professional transient surveys, a handful of additional SE~SNe  have been discovered in the midst of its initial peak/adiabatic-cooling phase,  including for example: SN~2009K \citep{stritzinger17a}, SN~2011hs \citep{bufano14}, SN~2011dh \citep{arcavi11}, PTF11mnb \citep{taddia16b}, and iPTF15dtg \citep{taddia16a}. 

In recent years, several studies have presented  
expanded SE~SN samples. \citet{richardson06} presented the analysis of a sample of $V$-band light curves for 27 SE~SNe. 
\citet{drout11} published the first multi-band ($V$ and $R$ bands) sample of SNe~Ib/c, studying 25 SNe; more recently, \citet{bianco14}, \citet{modjaz14}, and \citet{liu16} have published optical and near-infrared light curves and visual-wavelength spectroscopy of $>$~60 SE~SNe followed by the Center for Astrophysics (CfA) SN group.  
\citet{taddia15} studied the $ugriz$ light curves of a sample of 20 SNe~Ib/c obtained by the
Sloan-Digital-Sky-Survey II (SDSS-II) SN survey. Additionally,
\citet{cano13}, \citet{lyman16}, and \citet{prentice16} have used
large SE~SN samples (61, 38 and 85 SNe, respectively) based on collections of optical data from the
literature to constrain explosion and progenitor properties.
From these studies, SE~SNe are found to be characterized by relatively small average ejecta masses ($M_{ej}$) ranging between 1--5~$M_{\odot}$, average explosion energies ($E_K$) of a few 10$^{51}$ erg, and average $^{56}$Ni masses of $\approx$0.1--0.3~$M_{\odot}$. Hydrodynamical modeling of several specific SE~SN indicate similar values for the explosion properties. For example, SN~2011dh, modelled by \citet{bersten12} and \citet{ergon14}, was characterized by $M_{ej}~=~1.8-2.5~M_{\odot}$, energy 0.6--1.0$\times10^{51}~$erg, and $^{56}$Ni mass of 0.05--0.10~$M_{\odot}$.  Furthermore, light-curve and spectral modeling reveals that in several cases the $^{56}$Ni is mixed into the outer SN ejecta \citep[e.g.,][]{bersten12,cano14,taddia15}.  
As compared to SNe~IIb, Ib and Ic, 
SNe~Ic-BL are generally characterized by higher $E_K$ and larger $^{56}$Ni masses \citep[see, e.g.,][]{cano13,taddia15,lyman16}. 

The fact that SE~SNe generally have small ejecta masses suggests a large fraction of them do not arise from very massive stars ($>$~25--30~$M_{\odot}$), whose mass-loss rates would not be high enough to strip most of the outer layers and leave these low ejecta masses. 
Therefore, it is more likely that they arise from binary systems, where the SN progenitor is an intermediate-mass star ($M_{ZAMS}~\lesssim$~20~$M_{\odot}$) that experiences significant
mass loss to its companion over its evolutionary lifetime \citep[see, e.g.,][and references therein]{yoon15}.
In the case of SN~1993J the companion was even identified in  images a decade after its explosion
\citep{maund04,fox14}. Furthermore, a possible detection of the companion of SN~2011dh's progenitor star was suggested by  \citet{folatelli14a}. iPTF13bvn was the first SN~Ib whose progenitor (a relatively low-mass star) was detected \citep{cao13,fremling14}, as recently confirmed by its disappereance in HST post-explosion observations \citep{folatelli16,eldridge16}.

The analysis of late-phase nebular spectra of SE~SNe also indicates
relatively low-mass progenitors, particularly in the case of SNe~IIb
\citep{jerkstrand15}. Specifically, mass constraints of SE~SN
progenitors obtained from oxygen-abundance determinations by
modeling late-phase spectroscopy point towards progenitors
characterized by $M_{ZAMS}~\approx12-13$ $M_{\sun}$   \citep[see, e.g.,][]{jerkstrand15}. 
This is corroborated by the lack of detections of bright Wolf-Rayet (WR) stars in pre-explosion images of nearby SE~SNe \citep{eldridge13}, as well as by the relatively high rate of SE~SNe \citep{smith11_rates,shivvers16}. 
 However,
a few SE~SNe with large ejecta masses (corresponding to broad light
curves) have been suggested, such as SN~2005bf
\citep[e.g.,]{folatelli06}, SN~2011bm \citep{valenti12},  iPTF15dtg
\citep{taddia16a}, PTF11mnb \citep{taddia16b}, and SN~2012aa \citep{roy16}. 
These objects could have possibly arisen from massive  ($M_{ZAMS}~>~30~M_{\odot}$) single
stars. 

Studies of the environments of SE~SNe suggested a difference in metallicity between SNe~Ib and Ic, with the latter being richer in metals \citep{anderson10,modjaz11}. This suggests an important role for line driven winds in the stripping of the SE~SN progenitors, as naturally expected for single massive stars. However, other works did not confirm this difference \citep{leloudas11,sanders12}.

Between 2004 and 2009 the {\it Carnegie Supernova Project} (CSP-I; \citealt{hamuy06}) conducted follow-up observations of over two hundred SNe using mainly facilities at Las Campanas Observatory (LCO). A chief aim of the CSP-I was to construct a SE~SN sample obtained on a homogeneous, stable and well-understood photometric system.
By the completion of the CSP-I follow-up program,  optical broad-band
observations of \nosne\ spectroscopically classified SE~SNe were obtained, with a subset of \nosnenir\ objects having at least some near-infrared (NIR) imaging.  Definitive photometry of the sample is presented by \citet[][]{stritzinger17a}, while additional companion papers by \citet{stritzinger17b} and \citet{holmbo17}
study the color/reddening properties and visual-wavelength spectroscopy, respectively. 
In this paper we present the analysis of the light-curve properties and construct comprehensive bolometric light curves, which are used to estimate key explosion parameters via semi-analytical and hydrodynamical modeling.
 
We note that up to now much of the photometry found in the literature of  SE SN suffer   a number of issues related   to data quality and/or poor photometric calibration. To list just a few of the issues plaguing the quality of the literature-based sample include: data obtained from sites with poor seeing and often of low signal-to-noise, a general lack of (or even no) understanding of the  photometric systems used to obtained the data,  the improper use of color terms to calibrate SN photometry, and incomplete attempts to correct for host-galaxy reddening. An overall goal of the CSP-I is to obtain photometry of a variety of  SNe types on a stable, homogeneous,  and well-understood photometric system. Fortunately, the stability of the observing conditions offered by LCO and its facilities, combined with our dedication to leave no stone unturned in our efforts to understand the CSP-I photometry system \citep[see][]{krisciunas17}, enabled the computation of photometry with an accuracy and wavelength coverage unparalleled in other samples. 
Combining the photometry of the CSP-I SE SN sample with the robust host-galaxy reddening corrections  (see \citealp{stritzinger17b}), we  construct comprehensive UVOIR (UltraViolet-Optical-near-InfraRed) bolometric light curves, which are modeled using both semi-analytical and hydrodynamical modeling. The  consistency of the inferred explosion parameters between the two methods is also investigated.
 
The organization of this paper is as follows. Section \ref{sec:sample} provides a brief summary of the CSP-I SE~SN sample, including  pertinent details regarding each SN. 
Section \ref{sec:lc} contains the detailed analysis of the light-curve shape properties.
This is followed by Section \ref{sec:abspeak} which examines the absolute magnitudes. 
Subsequently, in Section \ref{sec:boloprop} spectral energy distributions (SEDs) are used to construct UVOIR light curves, from which progenitor and explosion parameters are estimated in Section \ref{sec:model}. 
Finally, a discussion on our results is presented in Sections \ref{sec:discussion} and conclusion are given in Section \ref{sec:conclusion}.

\section{The CSP-I stripped-envelope~supernova sample}
\label{sec:sample}

Table~\ref{listsn} contains the list of the \nosne\ SE~SNe followed by the CSP-I \citep{stritzinger17a}.
Twenty-nine of the objects have $ugriBV$-band light curves, 5 objects lack $u$-band photometry  (i.e., SN~2004ew, SN~2006bf, SN~2007ag, SN~2007rz, and SN~2009dp), and 26 objects have $YJH$-band photometry. 

The sample consists of 10 SNe~IIb, 11 SNe~Ib, and 13 SNe~Ic, with the classification of all of the objects based on visual-wavelength spectra obtained by the CSP-I \citep{holmbo17}.
Among the SN~Ib sub-sample is the peculiar SN~2005bf, which is characterized by a prominent second peak, which has never been seen before in these objects. Given its uniqueness, it is omitted from our light-curve analysis. However, a detailed study of it based on CSP-I light curves and spectroscopy has been presented by \citet{folatelli06}, in addition to earlier papers by \citet{anupama05} and \citet{tominaga05}. 
In addition, among the SN~Ic sub-sample both SN~2009bb \citep{pignata11} and SN~2009ca are broad-lined objects. Beside SN~2005bf and 2009bb, the CSP-I data of SN~2007Y were published and analyzed in a single object paper by \citet{stritzinger09}. A number of SNe in our sample were observed by other groups or included in literature sample analysis. Specifically, \citet{drout11} observed and analyzed the $V$ and $R$ band light curves of SNe~2004fe, 2004ff,  2004gq, 2004gt, 2004gv, and 2007C. In \citet{taddia15} we published SDSS $ugriz$ light curves of SNe~2006lc and 2006fo. \citet{lyman16} collected literature data and analyzed the bolometric light curve of SNe~2004fe, 2004ff, 2004gq, 2005bf, 2006T, 2006ep, 2007C, 2007Y, and 2009bb. \citet{bianco14} published the CfA optical and NIR light curves of a number of the SNe which are also in common with our sample, namely SNe~2004fe, 2004gq, 2005bf, 2006T, 2006ep, 2006fo, 2006lc, 2007Y, and 2009bb. \citet{prentice16} analyzed a collection of light curves from the literature including  SNe  2004fe, 2004gq, 2005bf, 2006T, 2006ep, 2006fo, 2006lc, 2007Y, and 2009bb. In total, we provide new additional data and a robust analysis of the light curves of eight-teen SNe already presented in the literature, of which three are originally from previous CSP-I papers.

Basic information for each SN and its host galaxy were compiled using
the NASA/IPAC Extragalactic Database (NED) and the Asiago Supernova Catalogue  \citep{barbon99}, and compiled into Table~\ref{listsn}.
This includes SN designation, coordinates and  spectral type,  host-galaxy designation and coordinates, Galactic visual extinction, redshift and distance. 
Values are also provided for semi-major and semi-minor axes,
morphological type and position angle (PA) of the host galaxy, as well
as the de-projected SN distance from the host-galaxy center. 

Milky Way extinction values ($A_X^{MW}$, where $X$ corresponds to a
given passband) are obtained from
NED\footnote{\href{http://ned.ipac.caltech.edu}{http://ned.ipac.caltech.edu}}'s
listings of the \citet{schlafly11} recalibration of the
\citet{schlegel98} dust maps.  

 Host-galaxy reddening values are estimated through the comparison of observed optical and NIR  colors to intrinsic color-curve templates  constructed from sub-samples of minimally-reddened CSP-I SE~SNe \citep{stritzinger17b}. Nine minimally-reddened events were selected among those with no or little \ion{Na}{I}~D absorption, with the observed bluest $B-V$ color at 10 days past peak, located far from their host-galaxy centers and in galaxies which are not strongly tilted. 
For seven highly-reddened objects, we directly determined the reddening parameter $R_V^{\rm host}$ and the $A_V^{\rm host}$ extinction by fitting their measured color excesses with a \citet[][hereafter F99]{fitzpatrick99} reddening law. These values are taken from \citet[][last two columns of their Table 3]{stritzinger17b}.
For objects suffering lower amounts of extinctions we adopted the average $R_V^{\rm host}$ value listed in \citet[][last column of their Table 4]{stritzinger17b}.

As explained by \citet{stritzinger17b}, the $R_V^{\rm host}$ values  used differ for each of the SE~SN subtypes. 
Specifically, for SNe~Ib suffering low reddening we adopt the $R_V^{\rm host}$ value obtained from the most reddened member of this subtype (i.e., SN~2007C). 
This approach is also followed for the other SE~SN subtypes.  
 As demonstrated in \citet{stritzinger17b}, nine objects are identified to be minimally-reddened and they are used to construct intrinsic color-curve templates.  
 When the photometry of an object could not be used to estimate the reddening via comparison of observed and intrinsic color due to poor follow-up,  we 
instead turn to estimates obtained from the equivalent width of the 
\ion{Na}{i}~D  ($EW_{\ion{Na}{i}~D}$) feature. 
Combining the $EW_{\ion{Na}{i}~D}$ measurements (\citealp[][their Table 1]{stritzinger17b}) 
and the relation between this quantity and $A_V^{\rm host}$ as derived in \citet{stritzinger17b} (i.e., $A_V^{\rm host}{\rm[mag]} = 0.78 \cdot EW_{\ion{Na}{i}~D}$ [\AA]), we obtain an estimate of the host extinction. We note that \citet{phillips13} showed that estimating extinction (even in our galaxy) via  $EW_{\ion{Na}{i}~D}$ implies large uncertainty.  
 The adopted values of $R_V^{\rm host}$  and  $A_V^{\rm host}$ for each SE~SN are summarized in Table~\ref{listsn}.

The listed redshifts and direct distance estimates are from the NED 
and NED-D catalogs. Direct distance measurements are adopted (mainly obtained through the Tully-Fisher method) when available. If not, NED-based luminosity distances are adopted assuming cosmological parameters $\Omega_{m} = 0.27$, $\Omega_{\Lambda} = 0.73$ \citep{komatsu08} and $H_0$ =  $73.8\pm2.4$~km~s$^{-1}$~Mpc$^{-1}$ \citep{riess11},
 and corrections for peculiar velocity based on Virgo, Great Attractor (GA) and Shapley flow models \citep{mould00}. 

NED also provides values for the major (2a) and minor (2b) galaxy diameters, while the morphological t-type and PA of each galaxy are adopted from the Asiago Supernova Catalogue \citep{barbon99}.
Following \citet{hakobyan09} and \citet{hakobyan12},  de-projected and diameter-normalized SN distances from the host-galaxy center ($d_{\rm SN}$) were computed and are listed in the last column of Table~\ref{listsn}. In this table we also report the values of the galaxy diameters, position angles and coordinates as well as the SN coordinates that we used to compute the de-projected and diameter-normalized distances. 
These parameters allowed us to confirm that each of the minimally-reddened SE~SNe was located far from its host's center \citep[see][]{stritzinger17b}.  

In the following, all the light curves are corrected for time dilation 
and K corrected. 
Given the redshifts of the SNe,
the time dilation corrections are $<$~3\%, with the exception of SNe~2008gc and 2009ca whose time corrections are $\approx$5\%\ and $\approx$10\%, respectively. 
The K corrections were computed following the method described by \citet{hsiao07}. 
The visual-wavelength K corrections were calculated using the Nugent SN~Ibc spectral template\footnote{Available at: \href{https://c3.lbl.gov/nugent/nugent_templates.html}{https://c3.lbl.gov/nugent/nugent\_templates.html}}. 
As the Nugent templates extend out to +70d and spectral features evolve slowly at late time, we use the last spectrum for anything beyond this epoch. 
At NIR wavelengths, K corrections were computed using the NIR spectroscopic time series of SN~2011dh  \citep{ergon14}. 
In the vast majority of objects, the K corrections are on the order of $<$~0.05~mag in the $V$ band, with the median of all the K correction in the $V$ band being 0.03~mag. 
\citet{oates12} found similar $V$-band K correction values  form the Type~IIb SN~2009mg located at z$=$0.0076, which is about half of the median redshift range of the CSP-I sample.

\section{Light-curve shape properties}
\label{sec:lc}

\subsection{Light-curve fits}
\label{sec:lc_fit}

The broad wavelength coverage afforded by the CSP-I SE~SN sample enables the light-curve shapes to be studied in nine photometric passbands extending from $u$ to $H$ band. 
To facilitate comparison of the various filtered light curves among the entire sample, each filtered light curve was fit with an analytic function. The adopted function works well with decently time-sampled SN follow-up, providing a continuous description of the data and a set of parameters describing the shape of the light curve that are useful for comparison. 

The shape of SE~SN light curves can be represented in terms of three components consisting of: (i) an initial exponential rise, (ii) a Gaussian-like peak, and (iii) a late linear decay.  
The functional form of the analytic function is expressed as

\begin{equation}
m(t)=\frac{y_0+m(t-t_0)+g_0 {\rm exp}[-(t-t_0)^2 / 2\sigma_0^2]}{1-{\rm exp}[(\tau-t) / \theta]}.
\label{eq:1}
\end{equation}

\noindent Here $y_0$ is the intercept of the linear decline, characterized by slope $m$. The final term in the numerator corresponds to the Gaussian-peak, normalized to phase ($t_0$), amplitude ($g_0$) and width ($\sigma_0$). The denominator corresponds to the exponential rise, where $\theta$ is a characteristic time, and $\tau$ is a separate phase zero-point.
This function was originally introduced by \citet[][]{vacca96} to study the light-curve properties of thermonuclear supernovae (see additional applications to SN~Ia studies in papers by  \citealp{contardo00} and \citealp{stritzinger05}). 

\begin{figure}
 \centering
\includegraphics[width=16cm]{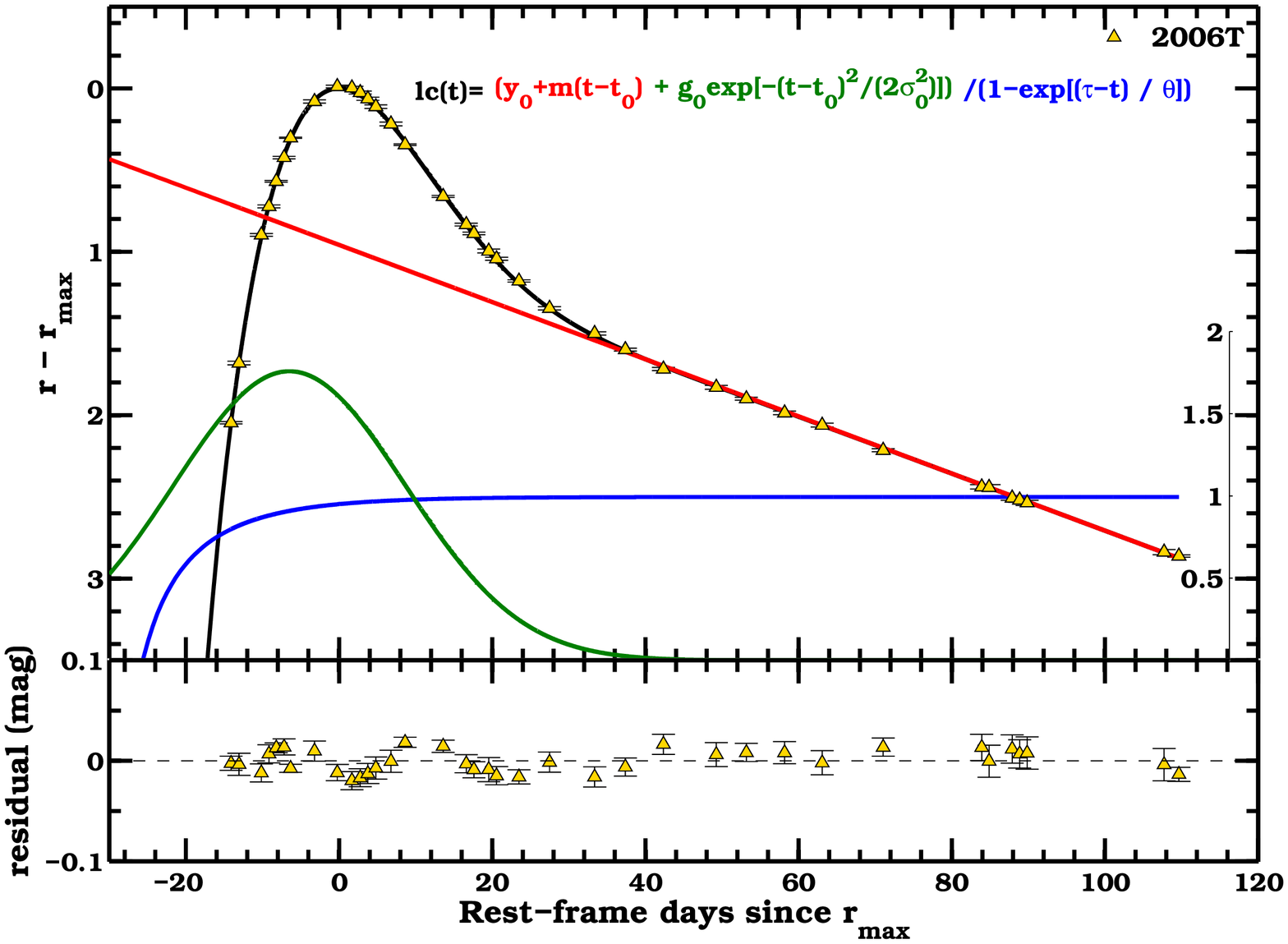}  
  \caption{\label{example06Tlc} 
  Best fit of Eq.~\ref{eq:1} (black solid line) to the $r$-band light curve (triangles) of SN~2006T normalized to its peak brightness. The three components of the analytic fit are shown: blue for the exponential rise, green for the Gaussian peak, and red for the linear decay. The corresponding terms in Eq.~\ref{eq:1} are color coded accordingly. The residuals between the fit and the photometry are plotted in the bottom panel, and in this case they never exceed 0.03~mag.}
 \end{figure}

Plotted in Fig.~\ref{example06Tlc} is the best fit of Eq.~\ref{eq:1} to the 
$r$-band light curve of SN~2006T.  
The fit clearly provides a smooth representation of the light curve, and this is particularly the case when the rise-to and subsequent fall-from peak brightness is well sampled. 
Some of the objects in the CSP-I sample were observed slightly prior to t$_{\rm max}$. 
In these cases 
  the denominator of Eq.~\ref{eq:1} was set to unity in order to ensure convergence of the fit.   
In addition, for SNe observed only around peak and for less than seven epochs, the functional fits to the light curve was limited to a single Gaussian.
Shown in Fig.~\ref{residual} are the best fits of Eq.~\ref{eq:1} to the optical and NIR light curves of the CSP-I SE~SN sample. 
Overall, regardless of filter, the light curves are characterized by a single Gaussian-shape peak, followed a few weeks past t$_{\rm max}$ by a linear declining phase. 
In Sect.~\ref{sec:lc} we only consider those SE~SNe whose light-curve data begin before or just at maximum brightness at least in one band (26 events).

\subsection{Light-curve peak epochs}
\label{sec:lc_peak_epochs}

\begin{figure}
\centering
\includegraphics[width=12cm]{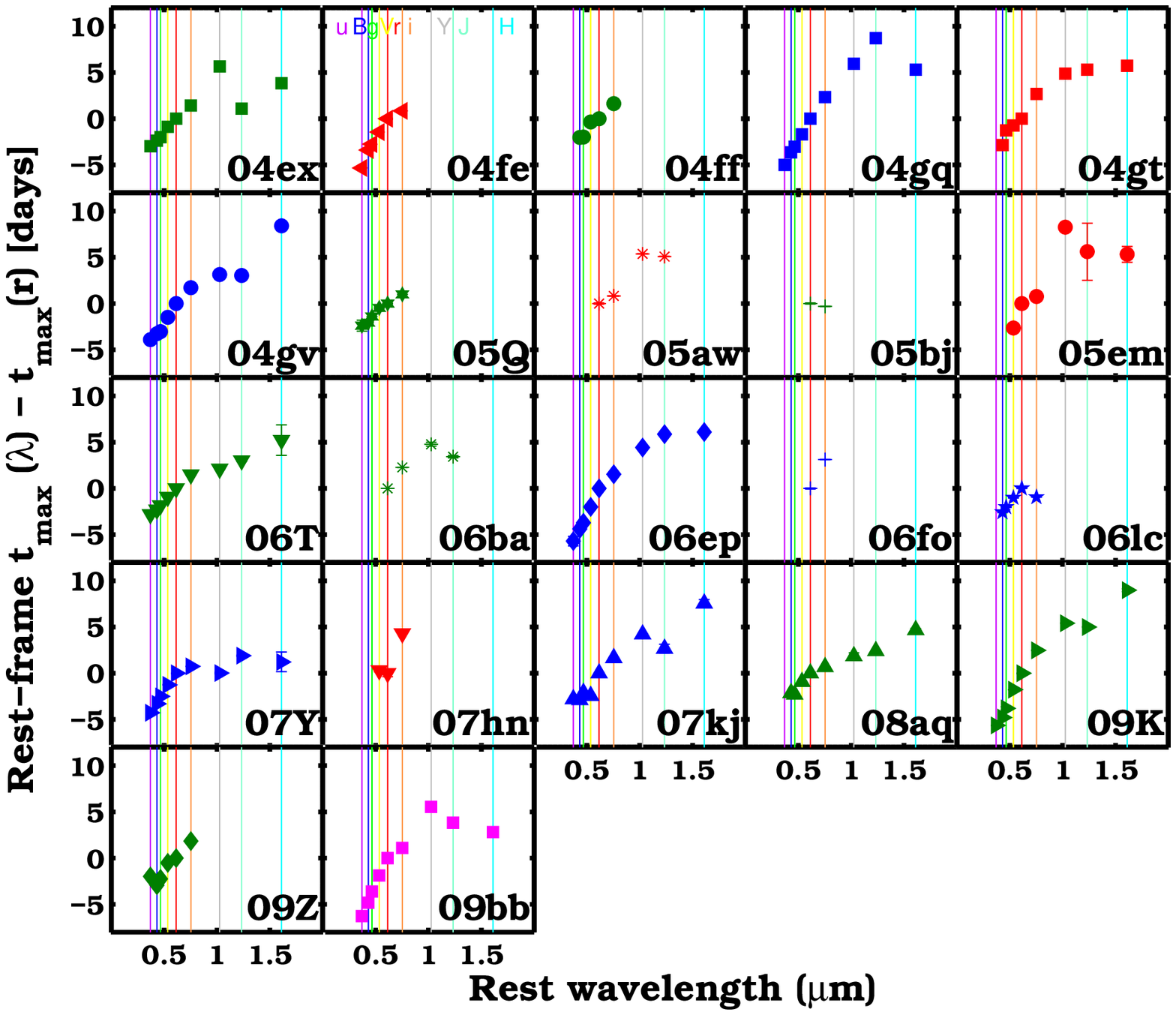}\\
\includegraphics[width=12cm]{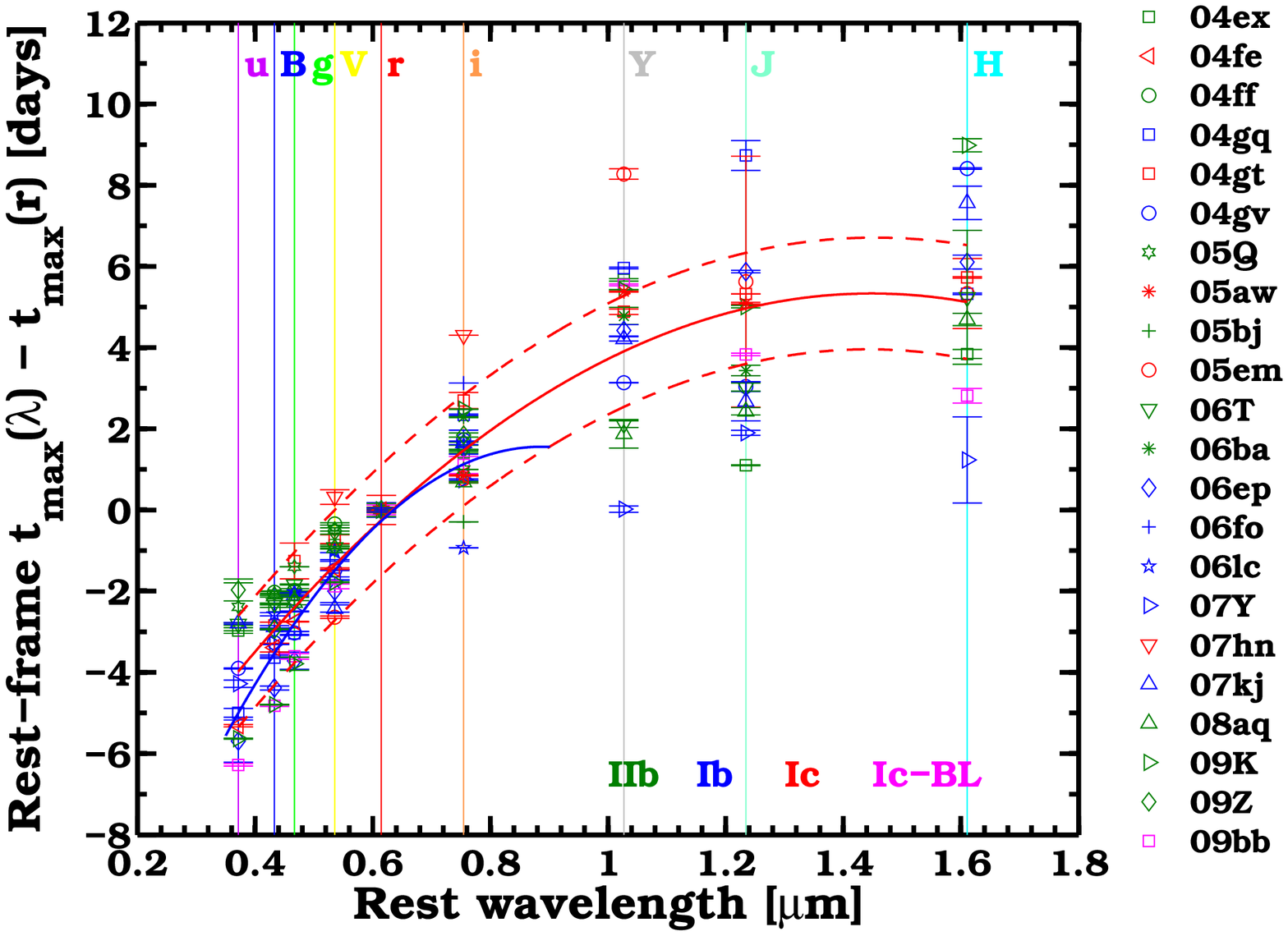}
\caption{\label{maxvslambda} \textit{(Top panel)} Epoch of maximum light (relative to t$_{\rm max}$ in the $r$ band) as a function of wavelength, where the effective wavelengths of the CSP-I passbands are indicated with solid vertical lines. 
Included here are 22 objects whose light curves cover the $r$-band maximum. The SE~SN subtype of each object is indicated by the color of its name with green, blue, red, and magenta corresponding to Type~IIb, Type~Ib, Type~Ic and Type~Ic-BL, respectively.
 Bluer optical bands peak prior to redder optical bands, while in the NIR,  t$_{\rm max}$ is nearly coeval amongst the $Y$, $J$ and $H$ passbands. 
 \textit{(Bottom panel)} Same as in the top panel, but here  all the SNe are plotted together. The red solid line corresponds to a low-order polynomial fit, with the associated fit uncertainty of $\approx$1.4 days indicated by dashed red lines. 
  The functional form of the polynomial fit is: $t_{\rm max}(\lambda)-t_{\rm max}(r)=-$8.0285$\lambda^2+$23.234$\lambda-$11.476, with time in days and $\lambda$ in $\mu$m.
The solid blue line corresponds to the polynomial fit obtained from the SDSS-II SE~SN sample  \citep{taddia15}.}
\end{figure}

An important parameter computed from the light-curve fitting described  in 
Sect.~\ref{sec:lc_fit} is $t_{\rm max}$. 
Estimates of $t_{\rm max}$ for the filtered light curve of each SN with pre-maximum follow-up observations are reported in Table~\ref{tmax_kcorr}. 
Plotted in Fig.~\ref{maxvslambda} is $t_{\rm max}$ for each observed passband [normalized to 
$t(r)_{\rm max}$: $t_{\rm max}-t(r)_{\rm max}$] vs. wavelength, where the effective wavelength of each CSP-I passband is indicated with a solid vertical line.
 In the top panel, the data are plotted individually for each SN, while in the bottom panel  all of the data are combined into one figure. 

Each SN reaches $t_{\rm max}$ first in the $u$ band, and subsequently peaks in red optical passbands sequentially with increasing wavelength  from the $B$ to $i$ bands. 
Close inspection of Fig.~\ref{maxvslambda} indicates that the NIR passbands peak after the optical passbands, but the  $J$- and/or $H$-band light curves often reach $t_{\rm max}$ simultaneously or even prior to $t(Y)_{\rm max}$. This holds independent of the SE~SN subtype. The dispersion around the general trend of later peak epochs at longer wavelength is larger in the NIR. 

The behavior of the blue passbands peaking prior to the red passbands confirms a trend noted in the 
 SDSS-II SE~SN sample \citep{taddia15}, and is a reflection of the rapid cooling of the SN ejecta around maximum (see Sect.~\ref{sec:T}). 
Compared to the SDSS-II SE~SN sample, the CSP-I SE~SN sample 
extends the observed wavelength coverage out through 1.8 microns. 
This is highlighted by the solid red line in the bottom panel of Fig.~\ref{maxvslambda}, corresponding to a low-order polynomial fit to the data, as compared to the solid blue line which is a similar fit to the SDSS-II SN survey's SE~SN sample.
The function is steep at optical wavelengths and turns over to being nearly  flat at  NIR wavelengths. In the caption of Fig.~\ref{maxvslambda} we report the expression of this best polynomial fit. The extended fit allows for the prediction (with $\approx\pm1.4$~d uncertainty) of $t_{\rm max}$  for SE~SNe with light curves observed prior to maximum in the red passbands, but lack pre-maximum observations in the blue passbands.
In Table~\ref{tmax_kcorr} a star indicates any peak epochs  derived using this method. 

\subsection{Light-curve decline-rate parameter $\Delta m_{15}$}
\label{sec:dm15}

A common light-curve decline-rate parameter to characterize the light curves of thermonuclear Type~Ia supernovae (SNe~Ia) is $\Delta m_{15}$ \citep{phillips93}. By definition $\Delta m_{15}$ is the difference in the brightness of a SN between peak and 15 days later. 
In the case of SNe~Ia, the luminosity-decline rate is known to correlate with luminosity in the sense that smaller $\Delta m_{15}$  values correspond to more luminous objects \citep{phillips93}.

\begin{figure}
\centering
\includegraphics[width=12cm]{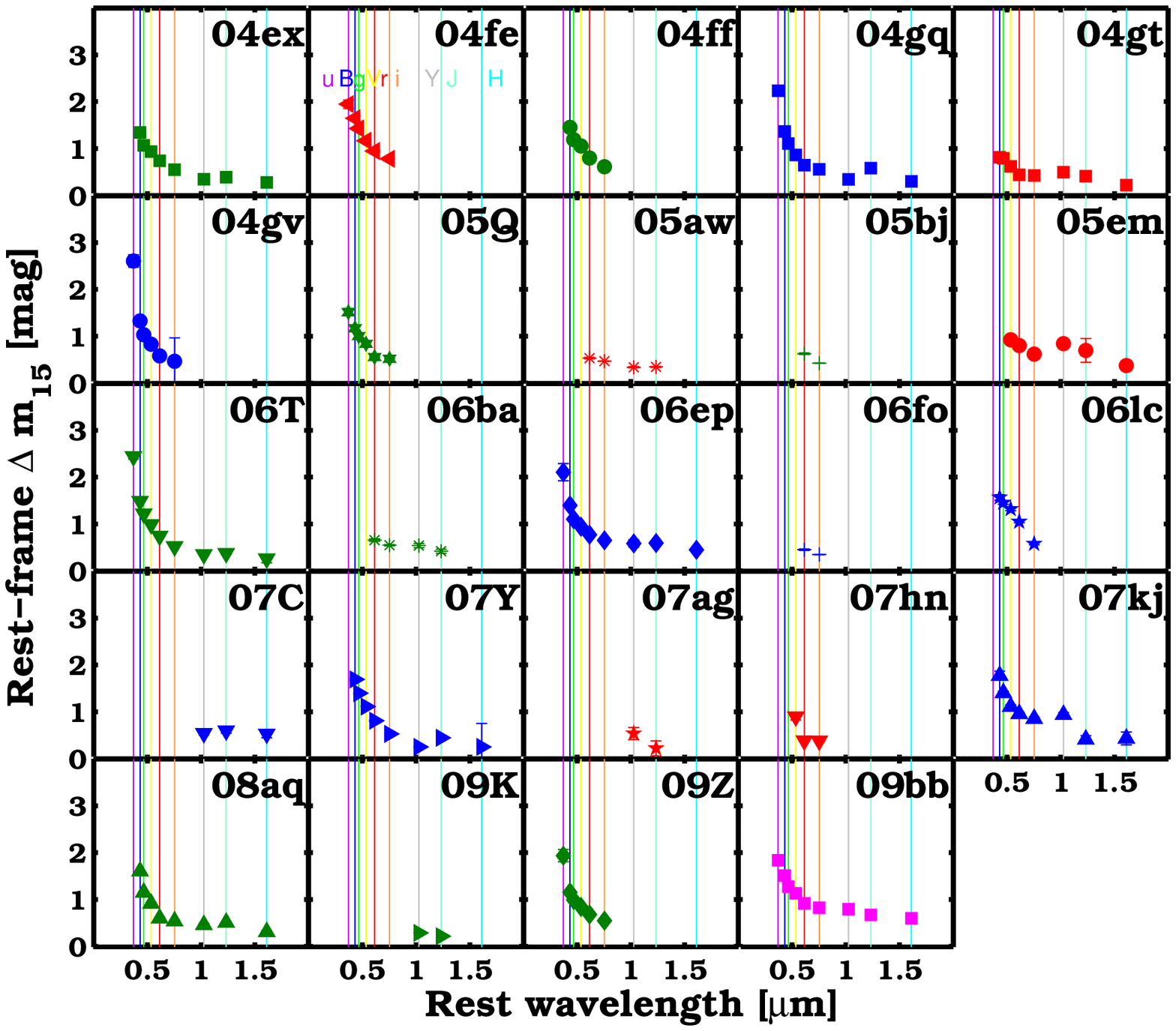}\\
\includegraphics[width=12cm]{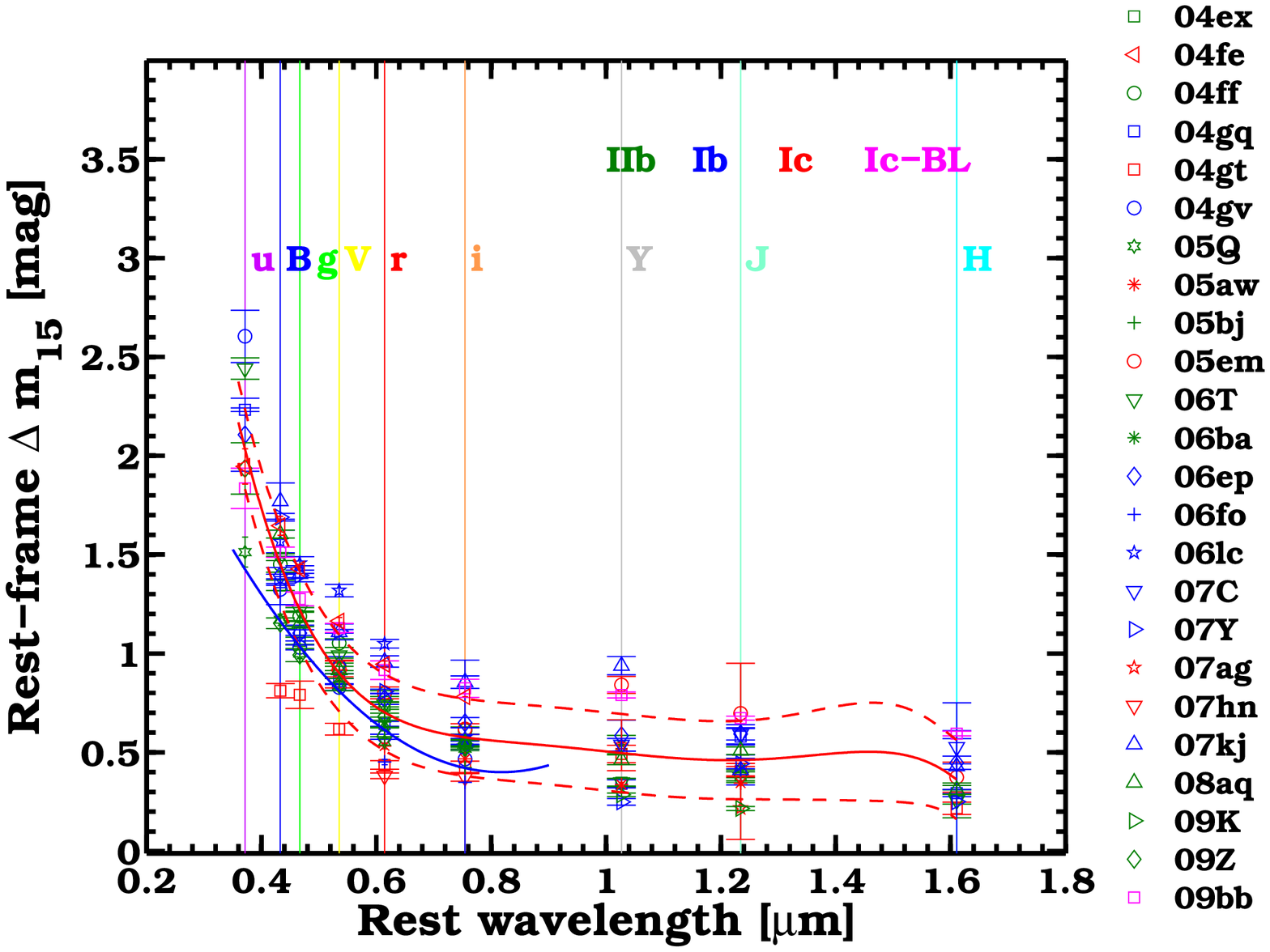}
\caption{\label{dm15vslambda} \textit{(Top panel)} The light-curve decline-rate parameter, $\Delta m_{15}$, plotted as a function of wavelength where the effective wavelengths of the CSP passbands are indicated by vertical lines. 
At optical wavelengths the blue bands exhibit larger  $\Delta m_{15}$ values than the red bands, while in the NIR $\Delta m_{15}$ is similar among the different bands. \textit{(Bottom panel)} Same as in the top panel but with all SNe plotted in one panel, along with a low-order polynomial  fit (solid red line) and  its associated 1$\sigma$ uncertainty ($\approx$0.2~mag; dashed red line).  The functional form of the polynomial fit is given by: $\Delta m_{15}(\lambda)=$ 
$-$11.88$\lambda^5+$63.74$\lambda^4-$134.17$\lambda^3+$138.81$\lambda^2-$71.00$\lambda+$15.00. Here $\lambda$ is in units of  $\mu$m and $\Delta m_{15}$ in units of magnitude.
Shown in blue is a polynomial fit obtained from the same analysis of the SDSS-II SE~SN sample \citep{taddia15}.}
\end{figure}

The light-curve parameter $\Delta m_{15}$ is readily computed from the light-curve fits presented in Sect.~\ref{sec:lc_fit}, and the resulting values are listed in Table~\ref{dm15_kcorr} and plotted in Fig.~\ref{dm15vslambda}. 
Plotted individually in the top panel of Fig.~\ref{dm15vslambda} is $\Delta m_{15}$ vs. wavelength for 24 SE~SNe (those with observed maxima in the light curves and with at least 15 days of observations after peak), where the effective wavelengths of the  CSP-I passbands are indicated with vertical lines.  
Clearly passbands with bluer effective wavelengths
exhibit higher $\Delta m_{15}$ values, implying faster declining light curves. 
This trend holds irrespective of SE~SN subtype, with an average $\Delta m_{15}(u) \approx2.0$~mag and an average $\Delta m_{15}(H) \approx 0.4$~mag. 
Plotted in the bottom panel of Fig.~\ref{dm15vslambda} are all of the SNe along with a low order polynomial fit to the data (solid red line, reported in the caption).

Examination of the distributions of $\Delta m_{15}$ values yields no significant differences between the different SE~SN subtypes, which is in agreement with previous studies by \citet{drout11} and \citet{taddia15}.

\subsection{Light curves beyond a month past maximum}
\label{sec:latelc}

\begin{figure*}
 \centering
\includegraphics[width=18cm]{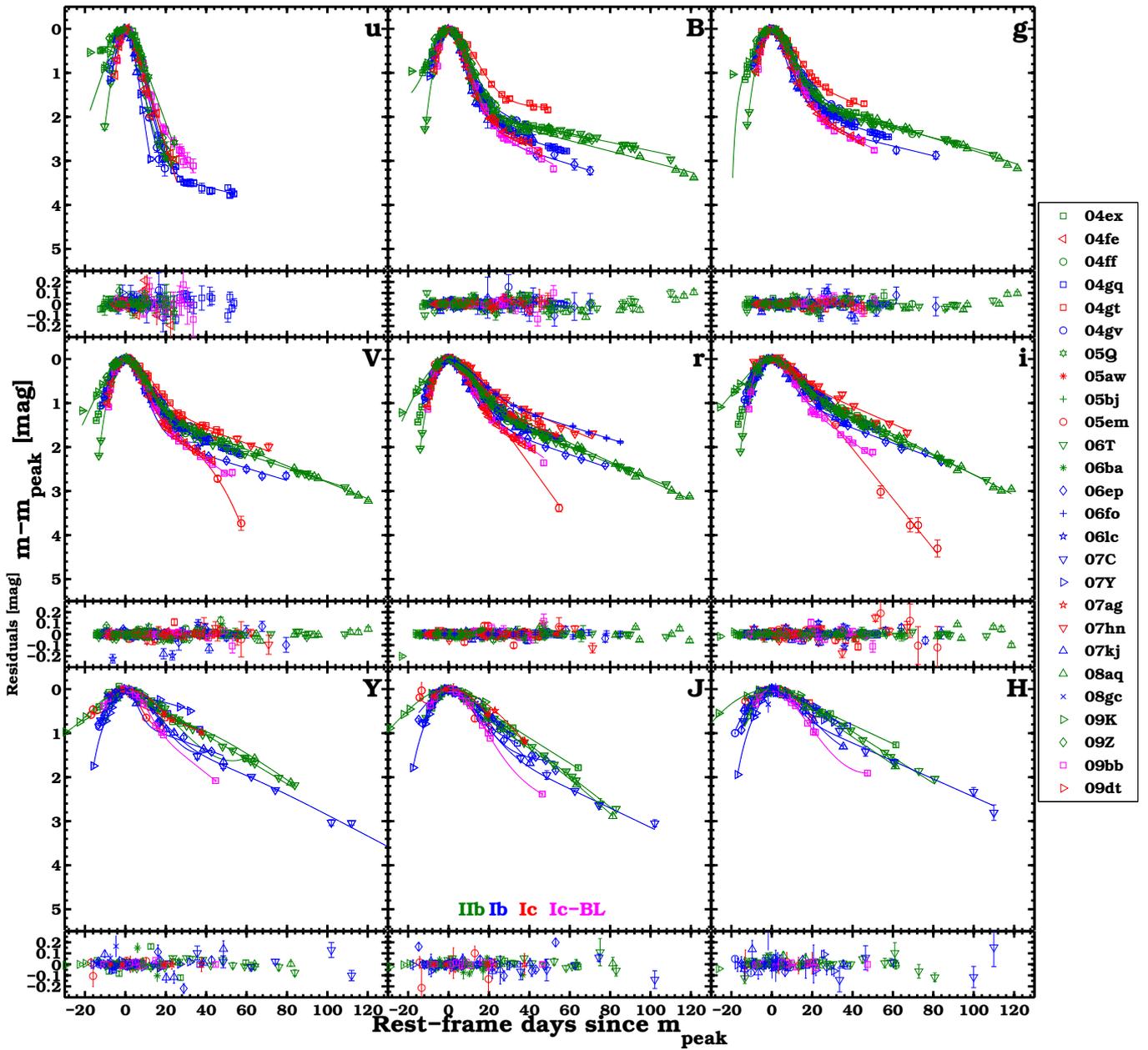}    
  \caption{\label{residual}  $u$- to $H$-band light curves of 26 SE~SNe with data obtained prior to t$_{\rm max}$ in at least one band. 
  Each filtered light curve is normalized to peak brightness and aligned to t$_{\rm max}$  estimated from the best fit of Eq.~\ref{eq:1} (colored solid lines) to the observed photometry. 
  Shown below each light-curve panel are the residuals of the light-curve fits. 
  Objects are color-coded based on their subtype: SNe~IIb are green, SNe~Ib are blue, SNe~Ic are red, and SNe~Ic-BL are magenta.} 
 \end{figure*}

Beginning around $\approx$3 weeks past maximum, the light-curve evolution of SE~SNe begins to show significant diversity (see Fig.~\ref{residual}). 
This motivated us to consider the alternative light-curve parameter $\Delta m_{40}$. $\Delta m_{40}$ measurements from the $r$-band light curves are found to show standard deviation of 0.31 and 0.50~mag for the SN~Ib and SN~Ic sub-samples, and 0.07~mag for the SN~IIb sub-sample. The fact that the light curves of SNe~IIb are more uniform than those of the other SE~SN subtypes was recently noted by \citet{lyman16}, and this applies to all of the optical band light curves. The average  $\Delta m_{40}$ values are similar among the three different subtypes (1.5--1.7~mag in $r$ band).

\begin{figure}
\centering
$\begin{array}{c}
\includegraphics[width=14cm]{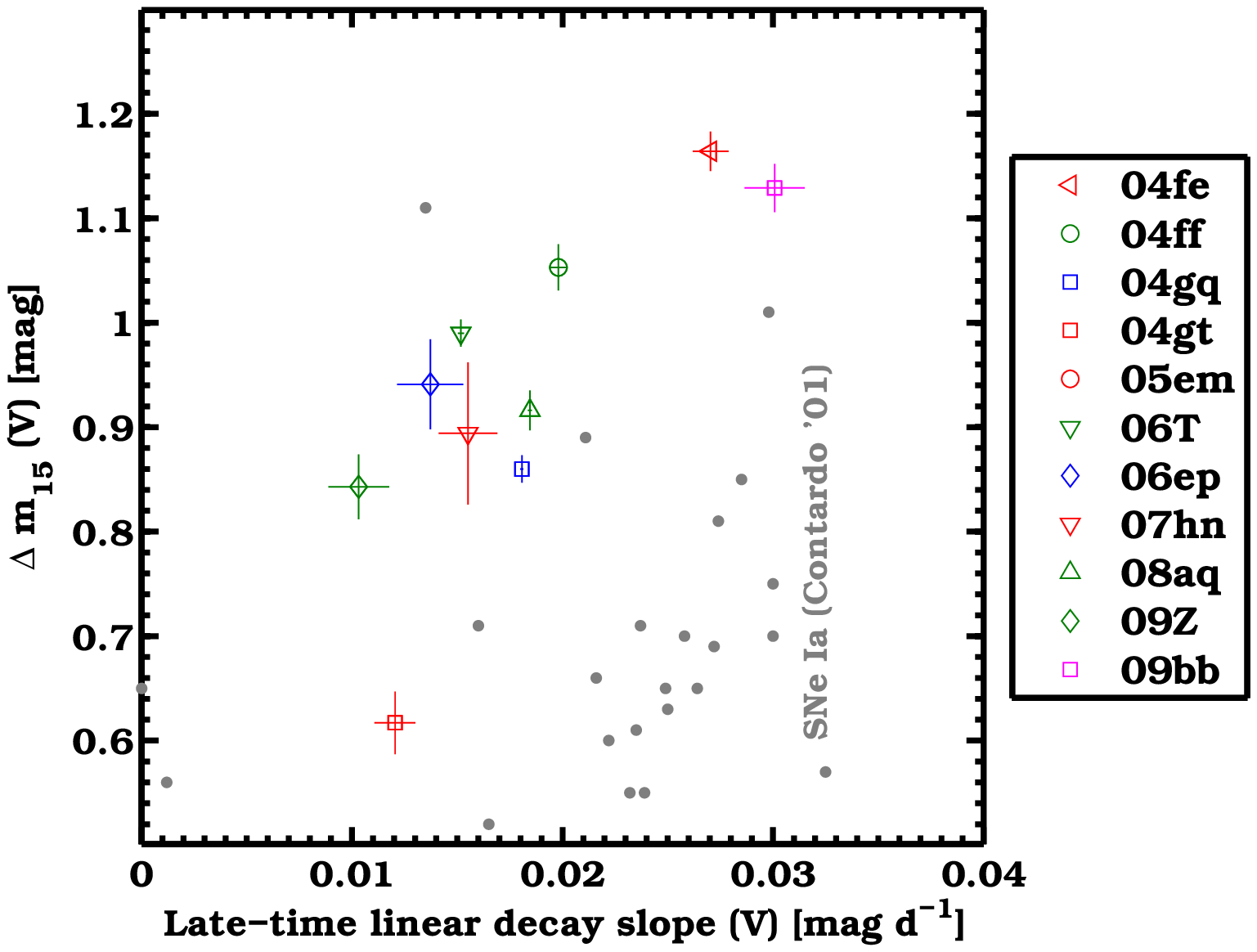}\\
\includegraphics[width=14cm]{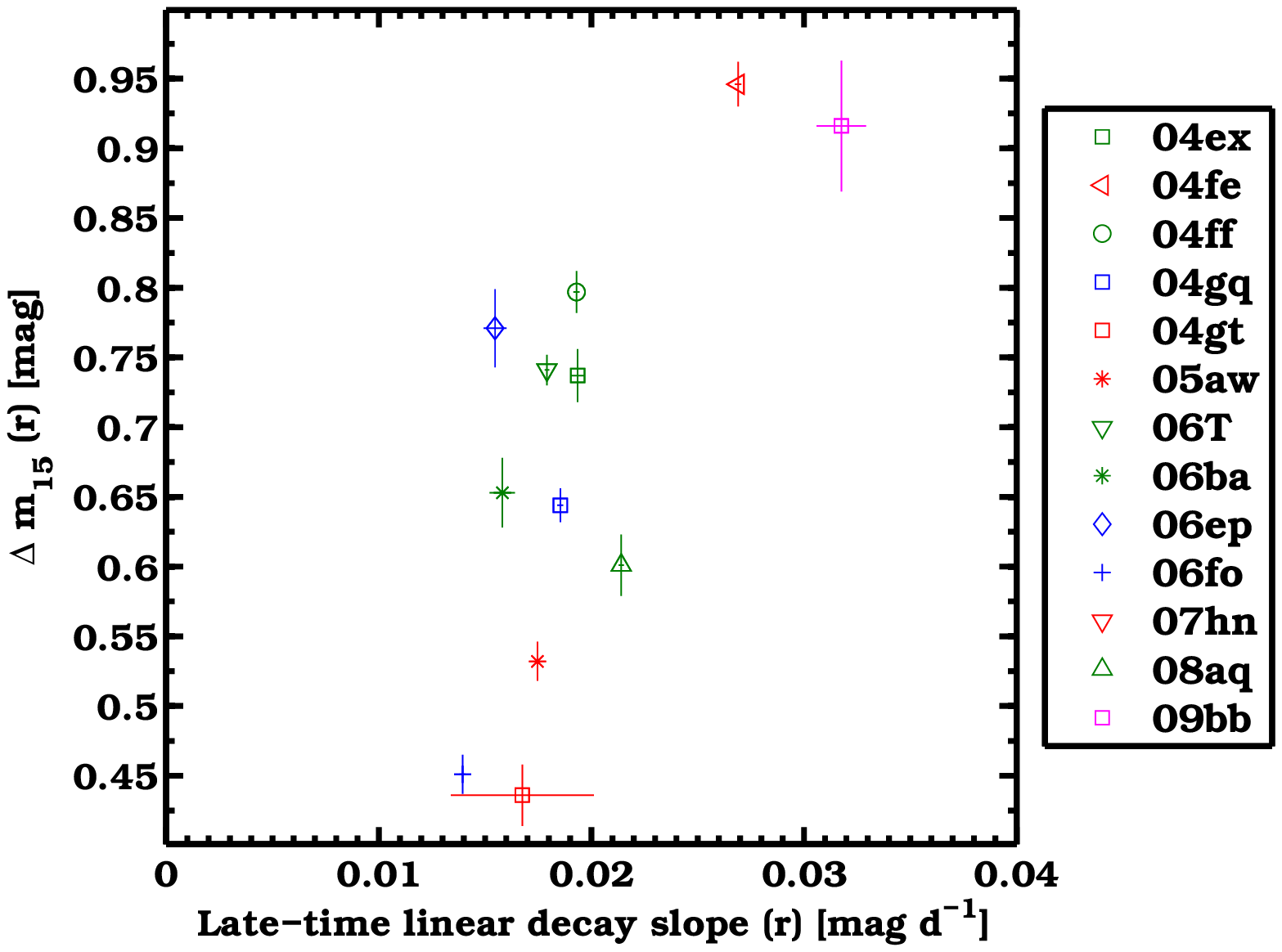}
\end{array}$
\caption{\label{slopevsdm15} Late-time linear decay slope in $V$ and $r$ band versus $\Delta m_{15}$ for the CSP-I SE~SNe with both their  peak luminosity covered and their last observation being $> +$40d days post maximum. 
Faster declining light curves (higher $\Delta m_{15}$) tend to decline faster at late phases.  Objects with both large uncertainties on the slope and on $\Delta m_{15}$ are excluded from the figure.  SN~IIb, Ib, Ic, and Ic-BL are represented in green, blue, red and magenta, respectively.
SN~2005em is not included and falls far from the correlation due to its large late time slope. With gray points we represent the results for the SNe~Ia fit by \citet{contardo01}, which do not show any clear trend.}
\end{figure}

Further inspection of the $r$-band light-curve fits in Fig.~\ref{residual}  indicates that the majority of objects with observations up to at least $+$40d follow a similar linear decline rate of $\approx$2~mag per one hundred days at late epochs. The post maximum linear decline phase marks the time when energy deposition is dominated by the $^{56}$Co $\rightarrow$ $^{56}$Fe decay chain. 
In principle, steeper slopes in the light curves correspond to events with higher gamma-rays escape fractions due to higher explosion energy to ejecta mass ratios and/or to higher degrees of $^{56}$Ni mixing (defined as the fraction of the total ejecta mass enclosed in the maximum radius reached by 
radioactive $^{56}$Ni). 
SNe~IIb, Ib and Ic exhibit rather uniform slopes quantified by 0.016--0.021~mag~d$^{-1}$, 0.014--0.018~mag~d$^{-1}$ and 0.017--0.027~mag~d$^{-1}$, respectively (see Fig.~\ref{slopevsdm15}).
These values are also consistent with the slopes measured in the other optical light curves. Comparing the late phase decline rates of our sample to that of the  $^{56}$Co to  $^{56}$Fe decay chain show differences of $\approx$50\%, suggesting   
a significant  fraction of  gamma rays are not deposited into the SN ejecta. We will return to this  issue in Sect.~\ref{sec:discussion}. In comparison, the $r$-band decline rate of normal SNe~Ia is slower with a value of $\approx$ 0.014 mag~d$^{-1}$ \citep[e.g.,][]{stritzinger02,lair06,leloudas09}.

As evident from Fig.~\ref{residual}, the light curves of Type~Ic SN~2005em evolve very rapidly over its post maximum decline phase.  Indeed this object appears similar to a sub-class of fast evolving Type~Ic objects that includes the well-studied SN~1994I \citep[see, e.g.,][]{clocchiatti11}.

 In Fig.~\ref{slopevsdm15} we plot the late-time linear decay slope (parameter $m$ in Eq.~\ref{eq:1})
for those SNe with observations extending out to $+$40d  past $V$- and $r$-band maximum versus $\Delta m_{15}$ in the same bands. 
This figure suggests a trend in the sense that objects characterized by faster decline rates in the two weeks after peak are also characterized by steeper slopes at later phases. This trend is also present in the $i$ band (albeit less striking), whereas it is less clear if it is present in bluer bands and in the NIR bands, where we have less late-time data. 
A possible interpretation of this trend is provided in Sect.~\ref{sec:boloprop}.  
Finally, we note for comparison as indicated in the the top panel of Fig.~\ref{slopevsdm15}, SNe~Ia do not follow the same behavior in the  $V$ band.

\subsection{Light-curve templates}
\label{sec:templates}

\begin{figure}
 \centering
\includegraphics[width=18cm]{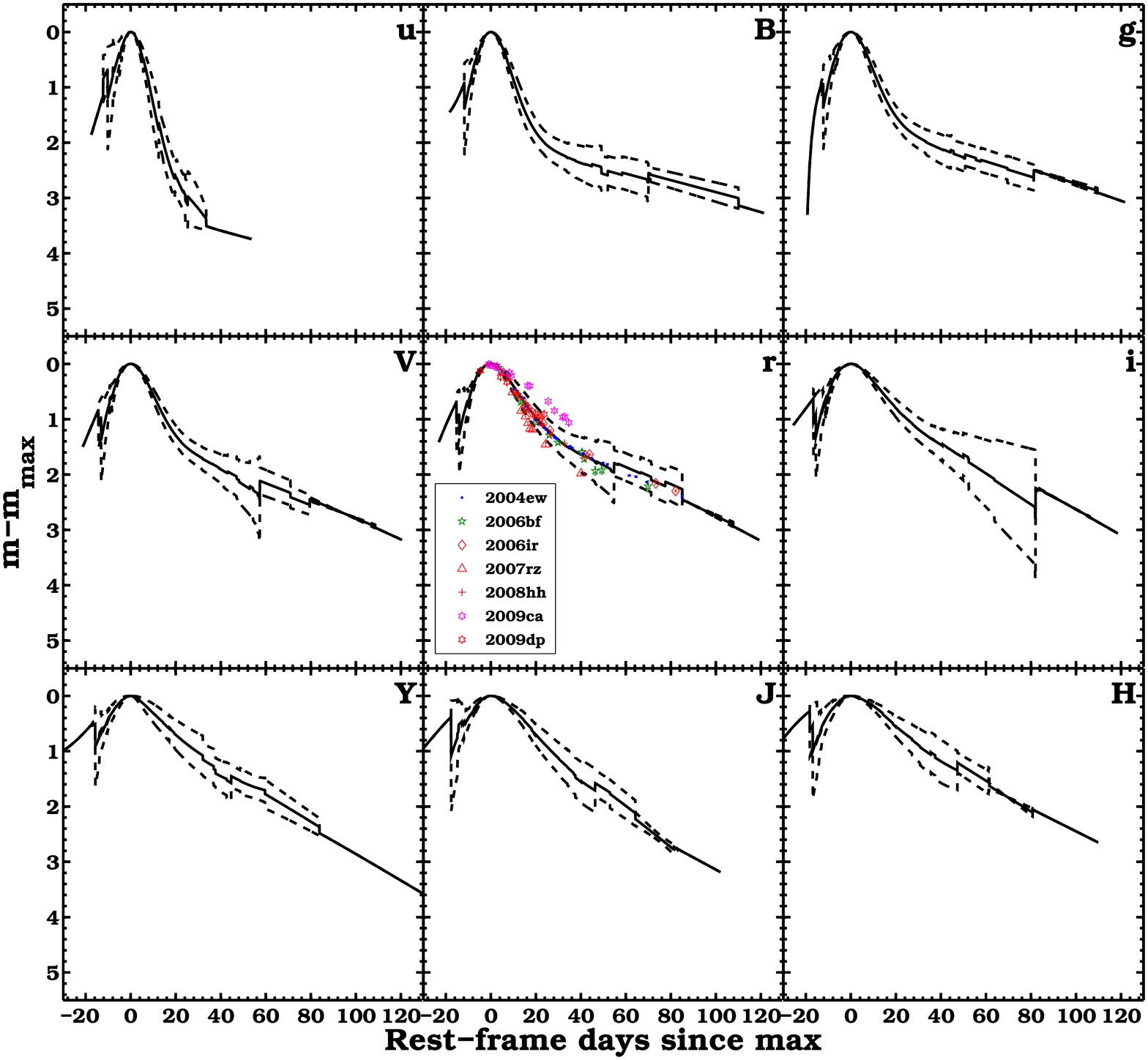} 
  \caption{\label{template} SE~SN light-curve templates. The templates were constructed by averaging the light-curve fits plotted in Fig.~\ref{residual}. The templates are represented by solid lines, their uncertainties by dashed lines. The $r$-band template light curve was used to estimate $t(r)_{\rm max}$ for seven objects having follow-up observations beginning past peak (see central panel). Depending on their subtype, i.e., IIb, Ib, Ic, and Ic-BL, these objects are represented in green, blue, red and magenta, respectively.}
 \end{figure}

Armed with the light-curve fits presented in Section~\ref{sec:lc_fit}, we proceed to construct template light curves covering the assortment of passbands used to observe SNe by the CSP-I. 
The resulting template light curves are plotted in Fig.~\ref{template}.
 Templates were constructed by taking the average of the fits to the observed light curves, while the associated uncertainty is defined by the standard deviation of these fits. 

The fit to the light curves that we used to build the templates are normalized to peak luminosity, so the templates show small dispersion around peak. After $\approx+$20d the uncertainties of the templates become 
more significant given the  large variety of decline rates which characterize the light curves 
(see Sect.~\ref{sec:dm15}) and the small sample size.
Clearly the templates are broader in the red bands compared to the blue bands around maximum brightness.

With the $r$-band template light curve in hand, $t(r)_{\rm max}$ is estimated for seven objects whose maximum was not entirely observed in the optical and/or NIR passbands. 
Estimates of $t(r)_{\rm max}$ were obtained by fitting the $r$-band template (in the range between $-5$d and $+30$d) to the observed light curves, and the best estimates of $t(r)_{\rm max}$ are indicated in Table~\ref{tmax_kcorr} with a double star. We allow fits to the template from $-$5d, as in some cases (e.g., SN~2009dp) the light curve around peak was poorly observed and the first detection may actually have occurred before peak.

 The best fits are shown in the central panel of Fig.~\ref{template}. 
An uncertainty of 1.5 days is adopted for all $t_{\rm max}$ values inferred from the template light-curve fits. 
For these seven objects, $t_{\rm max}$ in the other bands were then determined  using the relation shown in Fig.~\ref{maxvslambda}. 

The $r$-band template is used to establish $t_{\rm max}$ in the bands without maximum coverage as r-band maximum occurs relatively late compared to the other optical passbands. Furthermore, for several objects their $r$-band light curves exhibit smaller scatter compared to the NIR light curves. 

The light curve templates are electronically\footnote{The template light curves can be downloaded in electronic format from the Padadena-based CSP-I webpage: \href{http://csp.obs.carnegiescience.edu/}{http://csp.obs.carnegiescience.edu/data/} and point out to any potential users the templates are on the CSP-I photometric system.}
 available and can be used to constrain the phase and magnitude of peak for SE SNe  observed after peak, as demonstrated in our analysis for several objects, and can also be used to aid in photometric classification of SE SNe.

\section{Absolute magnitude light curves}
\label{sec:abspeak}
 
\begin{figure*}
 \centering
\includegraphics[width=18cm]{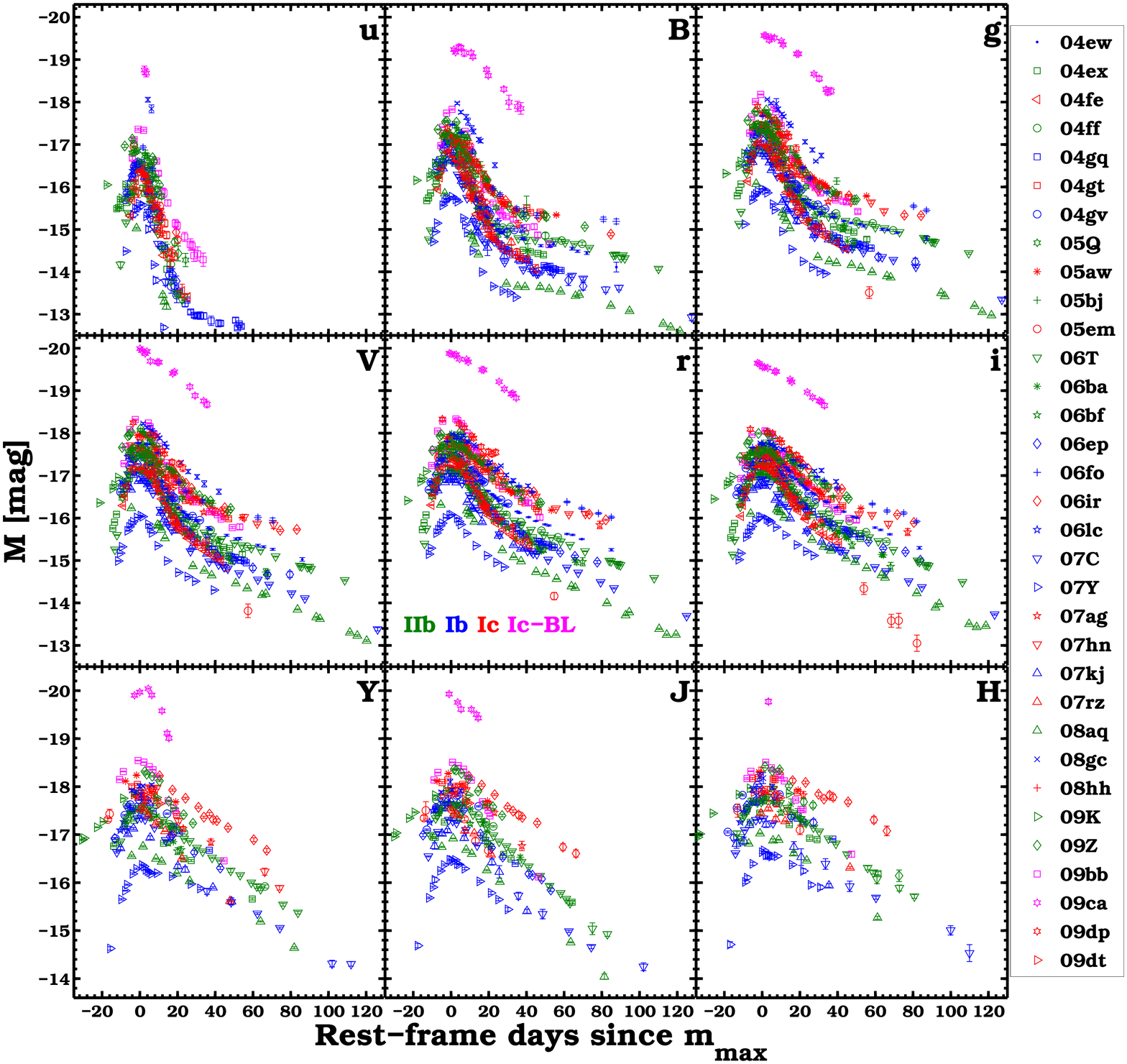}   
  \caption{\label{abspeak}Absolute magnitude $uBgVriYJH$-band  light curves of the full CSP-I SE~SN sample.  
 SN~IIb, Ib, Ic, and Ic-BL are represented in green, blue, red and magenta, respectively.}
 \end{figure*}

Absolute magnitudes are computed from all apparent magnitudes corrected for reddening 
(see Sect.~\ref{sec:sample} and Table~\ref{listsn}) and adopting the distances 
to their hosts
given in Table~\ref{listsn} to set the absolute flux scale. 
The resulting absolute magnitude light curves are plotted in Fig.~\ref{abspeak}, and the peak absolute magnitude for each filtered light curve is reported in Table~\ref{tab:abspeak_kcorr}.
The majority of objects (16 objects out of 22 in $r$ band)
reach peak absolute magnitudes ranging between $-17$~mag to $-18$~mag. 
The Type~Ic-BL SN~2009ca is a significant outlier, reaching a maximum brightness $M_{r} \approx -20$~mag, while the other Type~Ic-BL in the sample, SN~2009bb, only lies at the bright end of the normal luminosity distribution of the sample. 

\begin{figure}
 \centering
\includegraphics[width=18cm]{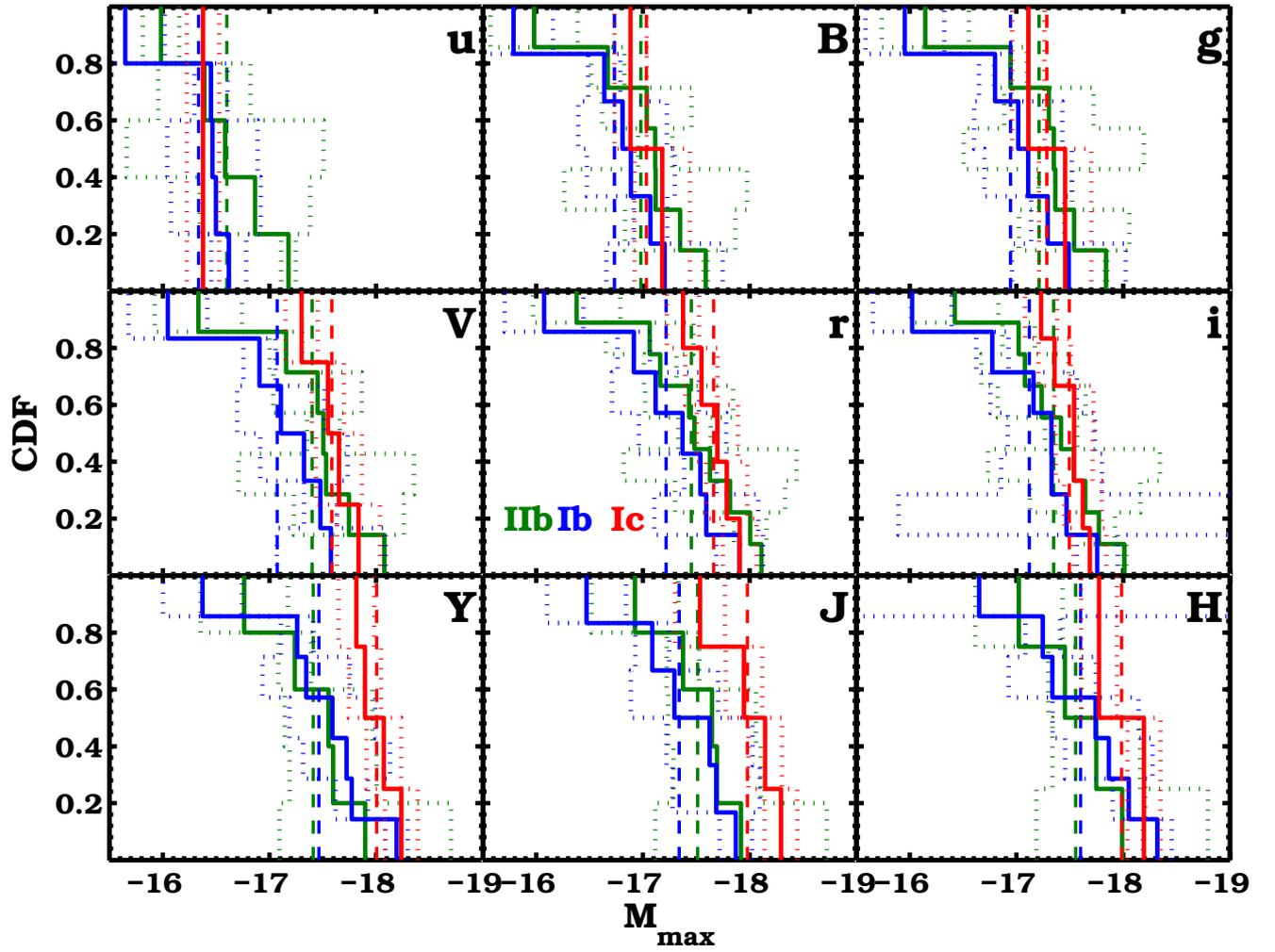}   
  \caption{\label{abspeakcdf}Absolute peak magnitude distributions for the $uBgVriYJH$-bands for the CSP-I SE~SN sample.  
 SNe~IIb, Ib, and Ic are represented in green, blue, and red, respectively. The two SNe~Ic-BL in our sample have been omitted. The average peak magnitudes of each subtype are marked by vertical dashed lines. The uncertainty of each peak value is marked by a dotted line.}
 \end{figure}

To display the
distribution of luminosities amongst SN~IIb, SN~Ib, and normal SN~Ic subtypes, shown in  
Fig.~\ref{abspeakcdf} are the cumulative distribution functions (CDFs) of the peak absolute magnitudes for each of the CSP-I passbands. 
The CDFs for the SNe~Ib and SNe~Ic are consistent with those  obtained from the SDSS-II SN~Ib/c sample \citep{taddia15}.
Inspection of the CDFs reveals no significant difference amongst the different subtypes. 
Indeed, a Kolmogorov-Smirnov (K-S) test reveals p-values $>$~0.05 in all but the $Y$ band, where the comparison between SNe~Ib and SNe~Ic indicates  p$=$0.02 with SNe~Ic being more luminous on average. The average peak absolute magnitudes for each subtype are reported in Table~\ref{tab:abspeak_kcorr} and indicated by dashed vertical lines in Fig.~\ref{abspeakcdf}. 

In Table~\ref{tab:abspeak_kcorr}, next to the absolute-magnitude peak averages, we also report the associated dispersions for each band and each SE SN subtype. These are obtained from the standard deviations of the peak magnitudes.  For instance, the dispersion of the $r$-band peak magnitudes are 0.54/0.60/0.21~mag for SNe~IIb, Ib and Ic, respectively. We investigated if these dispersions are mainly intrinsic or if they are mostly associated with uncertainties in the adopted distance and extinction. 
First, for each object we computed the uncertainty in the peak absolute magnitude, which is reported next to each peak magnitude in Table~\ref{tab:abspeak_kcorr}. These uncertainties are also reported as dotted lines in the cumulative distribution plots in Fig.~\ref{abspeakcdf}, next to each peak magnitude value. The uncertainty of the peak of the absolute magnitudes were obtained by summing in  quadrature the uncertainties associated with (i) the inferred peak apparent magnitude  (see Table~\ref{tab:apppeak_kcorr}), (ii) the extinction, and (iii) the distance (see Table~\ref{listsn}). We found, for example, that the $r$-band peak magnitudes extend from $-$17.38 to $-$17.91~mag for SNe~Ic (so there are 0.53~mag between the faintest and the brightest object of the SN Ic sample), but when we consider the uncertainty in their peak magnitudes, their confidence intervals do not completely overlap only in the region between $-$17.53 and $-$17.59~mag. This implies that accounting for the uncertainty of the extinction and on the distance  might reduce the observed difference among SN Ic peak to a very tiny intrinsic difference. However, for SNe~Ib and IIb the range where their peak $r$-band magnitude confidence intervals do not completely overlap is rather wide, ranging between $-$17.69 and $-$16.44~mag, and between $-$18.10 and $-$16.69~mag, respectively.  Therefore, the dispersion in their peak luminosities is not only driven by the uncertainties on the distance and on the extinction, but reflects an intrinsic difference. 

\begin{figure}
\centering
\includegraphics[width=18cm]{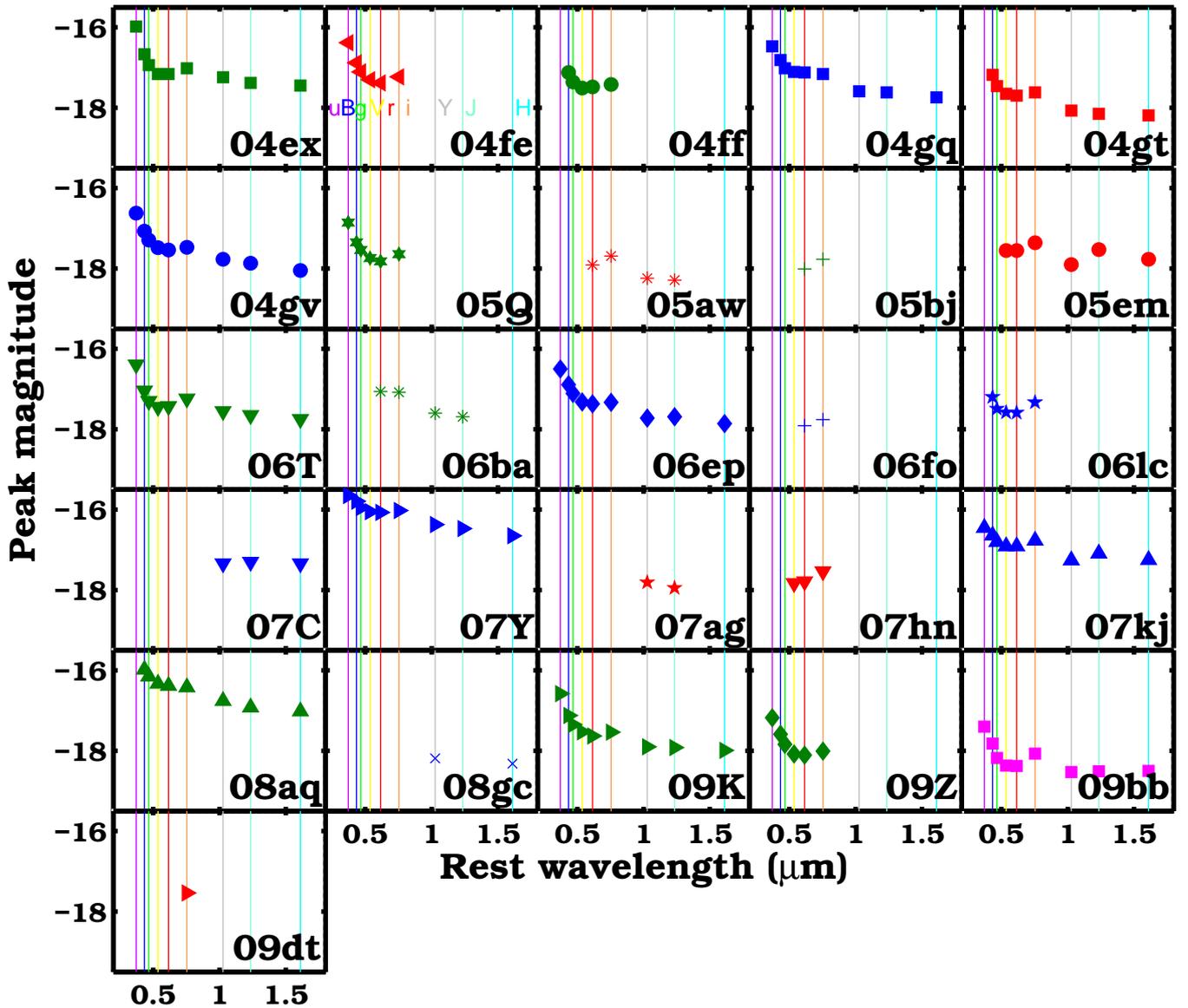}
\caption{\label{absmaxvslambda}  Peak absolute magnitudes plotted as a function of wavelength, where the effective wavelengths of the CSP-I passbands are indicated with vertical lines. At optical wavelengths, SE~SNe generally show higher values in the blue passbands  compared to the red passbands.  Among the NIR passbands, SE~SNe exhibit similar peak absolute magnitudes and the values at NIR wavelengths are lower than those in the optical. SNe~IIb, Ib, Ic, and Ic-BL are represented in green, blue, red and magenta, respectively.}
\end{figure}

We now turn to the absolute peak magnitudes as a function of wavelength as plotted in Fig.~\ref{absmaxvslambda}.
Strikingly, within the visual-wavelength region the peak luminosities are found to be dependent on the wavelength in the sense that red passbands tend to exhibit lower peak absolute magnitudes than the blue bands. 
Moving out to the NIR wavelengths  the peak magnitudes continue to follow a trend of reaching lower values, though these values appear to be insensitive to the exact wavelength interval contained between $\approx$1.1 to 1.8~$\mu$m. 
Figure~\ref{absmaxvslambda} suggests that
 the flux (in erg~s$^{-1}$~\AA$^{-1}$) at the effective wavelength of each passband and at the time of maximum in each specific band is
higher at shorter effective wavelength. We note that since the peak magnitudes are measured at different epochs it is not a spectral energy distribution of the SN  shown in the figure.

\begin{figure}
\centering
\includegraphics[width=18cm]{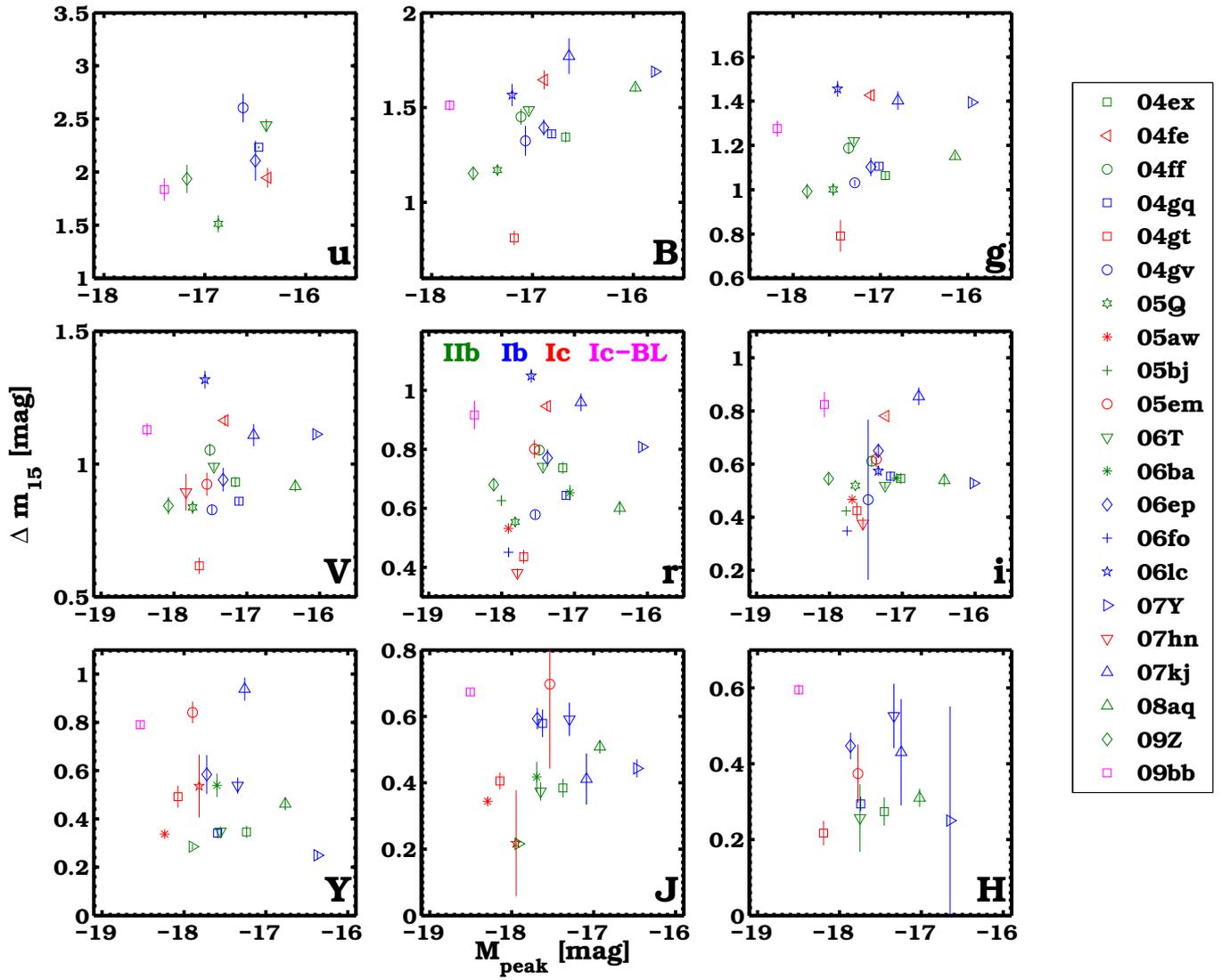}
\caption{\label{philrel}  Peak absolute magnitudes versus $\Delta m_{15}$ for the CSP-I SE~SN sample in nine different  passbands. SNe~IIb, Ib, Ic, and Ic-BL are represented in green, blue, red and magenta, respectively.}
\end{figure}

To end this section, in Fig.~\ref{philrel} we plot the peak absolute magnitudes of our SE~SN sample vs. the light-curve decline rate parameter 
$\Delta m_{15}$ (see Sect.~\ref{sec:dm15}).
Inspection of these parameters reveals mostly scatter plots in the various passbands. 
However, in the $B$ band (and possibly also in the $u$ band)  the SNe~IIb and SNe~Ib exhibit a correlation between the two quantities in the sense that the more luminous objects tend to have broader light curves. A Spearman correlation test between the two quantities in the $B$ band reveals a highly significant correlation with p-value of 0.034. On the contrary, the correlation is not statistically significant in the $u$ band. The correlation in $B$ band  is 
reminiscent of
the well-known luminosity decline-rate relation of thermonuclear SNe~Ia \citep{phillips93}. This trend was not found  in the bolometric light-curve analysis presented by \citet{prentice16} or in the $ugriz$ light curves of the SDSS SNe~Ib/c studied in \citet{taddia15}. 
It is possible  this correlation obtained from the CSP-I sample is  due to the detailed treatment of host reddening \citep[see][]{stritzinger17b}, which has a significant impact on the inferred peak absolute $B$-band magnitude. However, the accuracy of the CSP-I data themselves compared to that found in the literature  may also be a significant contributing factor.

\section{SEDs and UVOIR bolometric light curves}
\label{sec:boloprop}

To capitalize on the extended wavelength range covered by the CSP-I observations, spectral energy distributions (SEDs) are constructed ranging from the $u$ (320~nm) band redward to the $H$ (1800~nm) band. 
Building a complete set of SEDs for each SN first requires the interpolation of each filtered light curve.
Interpolation is accomplished with Gaussian process spline functions \citep[see][]{stritzinger17a}, enabling measurements of both the optical and NIR flux at common epochs. 
Next, the magnitudes are corrected for dust extinction using reddening values computed by \citet{stritzinger17b}. 
The extinction-corrected magnitudes are then converted to specific fluxes at the effective wavelength of each filter. 

Unfortunately it was not possible to obtain complete light-curve coverage for some of the objects in the $u$ and/or NIR passbands.
In the $u$ band this is typically due to a combination of low intrinsic brightness and fast evolution of the light curve, while at NIR wavelengths, gaps in follow-up are largely due to limitations of observational resources. 
To account for gaps in the $u$-band post-maximum follow-up, we resort to extrapolation when necessary.
Specifically, a constant $u-B$ color computed from photometry typically obtained after $+$15d was adopted, and when combined with the $B$-band light curve, provides an accurate extrapolation of the $u$-band flux. 
If $u$-band photometry is completely missing, we make use of bolometric corrections (see below). 
Constructing SEDs that encompass some measure of the flux blue-wards of the atmospheric cutoff, we extrapolate from the wavelengths covered by the $u$ band to zero flux at 2000~\AA. 
This  has  been shown to provide a reasonable approximation of the flux in this wavelength region based on UV observations of a literature-based SE~SN sample \citep{lyman14}. 
For SNe lacking NIR follow-up observations, we resort to extrapolation based on black-body (BB) fits to the optical-band SEDs, and the corresponding Rayleigh-Jeans tail accounts for flux red-ward of $H$ band for the entire sample. 
By the end of this process, each SN has a set of SEDs with conservative corrections accounting for missing observations and flux emitted at the wavelength regions extending beyond those covered by the CSP-I passbands.

\begin{figure}
\centering
\includegraphics[width=18cm]{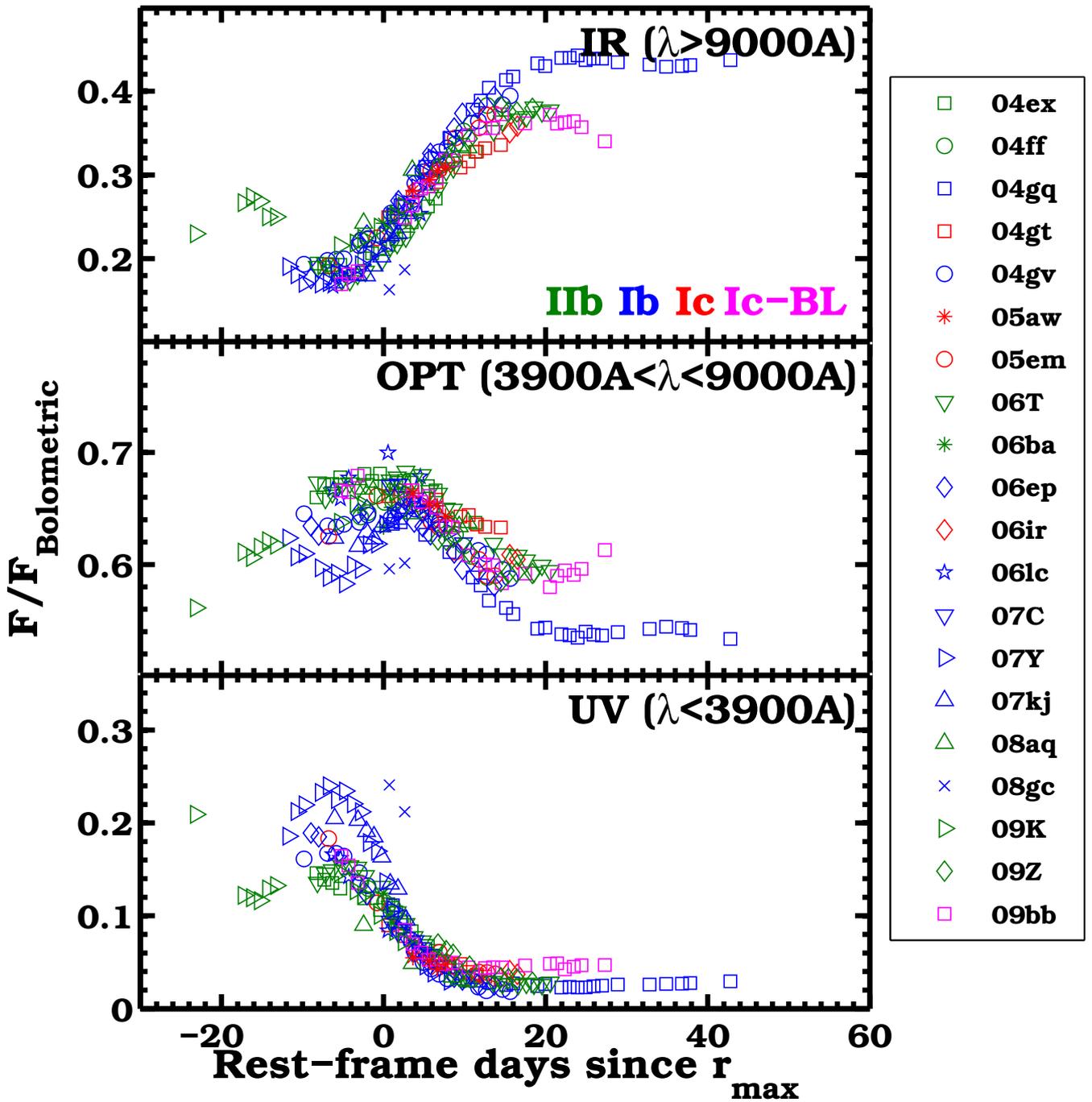}
\caption{\label{contribSED}Contribution of the UV, optical and IR fluxes to the bolometric flux of 20 CSP-I SE~SNe with at least one epoch of observations spanning from $u$ to $H$ band. The optical flux dominates at all epochs, the NIR flux becomes important at late time, whereas the UV fraction is non-negligible only before maximum. SNe~IIb, Ib, Ic, and Ic-BL are represented in green, blue, red and magenta, respectively.}
\end{figure}

For the SNe with complete coverage between $u$ and $H$ band, the contribution to the total UVOIR flux in the UV
($\lambda \leq  3900$ \AA),  optical (OPT; 3900 \AA~$< \lambda <9000$ \AA) and NIR ($\lambda > 9000$ \AA) passbands can be determined as a function of phase. 
Doing so for the best observed objects  
provides the information shown in
Fig.~\ref{contribSED}, which expresses the fraction of flux from these wavelength regions as a function of $t(r)_{\rm max}$ 
 The fraction of flux in the optical  
always dominates, with the UV flux being non-negligible prior to $t(r)_{\rm max}$ and the NIR flux becoming increasingly important after $t(r)_{ max}$. 
 These findings are similar to those shown by \citet{lyman14}, where slightly different wavelength ranges are considered.

The UV corrections obtained from extrapolation to zero flux at 2000~\AA\ consist of $\approx$10\% of the total flux around peak, whereas the mid- and far-IR  corrections consist of only $\approx$3\% of the total flux at similar epochs. 
At $+$20d  after peak the UV correction fraction drops to $\approx$3\%, while  the mid- to far-IR corrections rise to $\approx$5\%.

\begin{figure*}
\centering
\includegraphics[width=18cm]{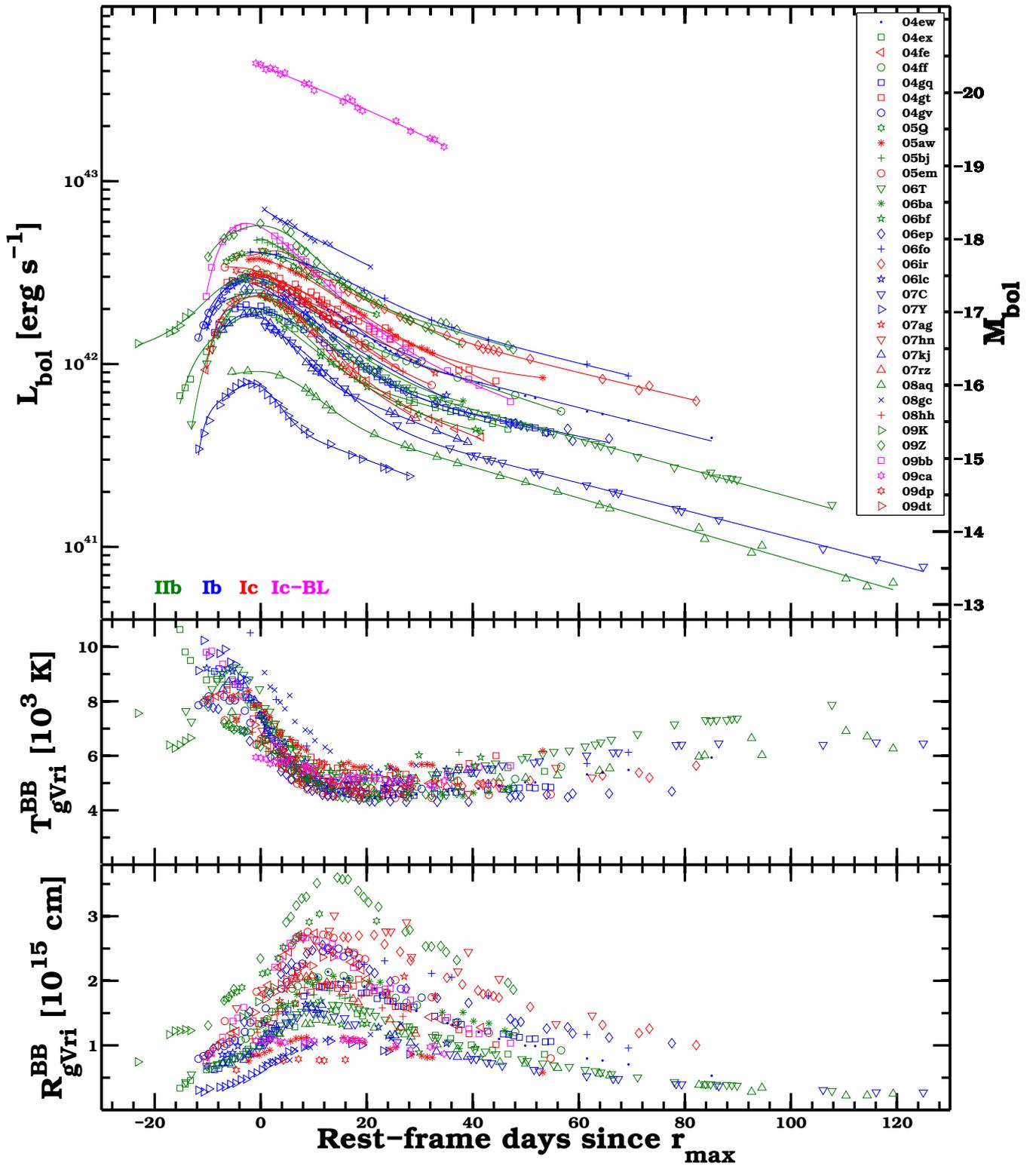}
\caption{\label{LTR} 
Plotted in the top panel are the bolometric light curves of 33 SE~SNe. Each of the light curves was fit with the function presented in Eq.~\ref{eq:1}, and the best fit is shown by solid colored lines. Shown in the middle and bottom panels is the temporal evolution of $T^{BB}_{gVri}$ and radius from the BB fit, respectively. SNe~IIb, Ib, Ic, and Ic-BL are represented in green, blue, red and magenta, respectively.}
\end{figure*}

To produce a UVOIR light curve for a given SN, its time-series of SEDs are integrated over wavelength, and then the resulting total flux is placed on the absolute flux scale through the multiplication of the factor 4$\pi D_L^2$; where $D_L$ is the luminosity distance to the host galaxy. 
In the case of those objects without any $u$-band photometry, we resort to constructing the UVOIR light curve by making use of the $g$-band photometry, the $g-i$ color, and the bolometric corrections presented by \citet{lyman14}.
Through the comparison between the UVOIR light curves produced via the integration of SEDs and
by the use of bolometric corrections, both techniques are found to provide fully consistent results over all epochs, in line with the precision discussed by \citet[][their appendix B]{lyman14}. 
The obtained UVOIR light curves of the CSP-I SE~SN sample are plotted in the top panel of Fig.~\ref{LTR} and made available online on the Pasadena-based CSP-I webpage\footnote{
\href{http://csp.obs.carnegiescience.edu/}{http://csp.obs.carnegiescience.edu/data/}}. 
The associated uncertainties of the UVOIR luminosities are dominated by the error of the distance ($\Delta L/L~\approx~2\Delta D/D$), which are on the order of 7\% (see the errors on the distances in Table~\ref{listsn}). The majority of objects reach peak luminosities ranging between 1--10$\times$10$^{42}$~erg~s$^{-1}$. 
SN~2009ca is an outlier with $L_{\rm max} \approx 4\times10^{43}$~erg~s$^{-1}$. 

 Each UVOIR light curve was fit with Eq.~\ref{eq:1} and the results are over-plotted in Fig.~\ref{LTR} (top panel) as colored solid lines. 
 This provides  parameters characterizing the shape of these light curves, namely the epoch of bolometric peak [$t(bol)_{\rm max}$], the corresponding luminosity  [$L(bol)_{\rm max}$], the decline-rate parameter [$\Delta m_{15}(bol)$], and the slope of the linear decaying phase. 

\begin{figure}
\centering
\includegraphics[width=15cm]{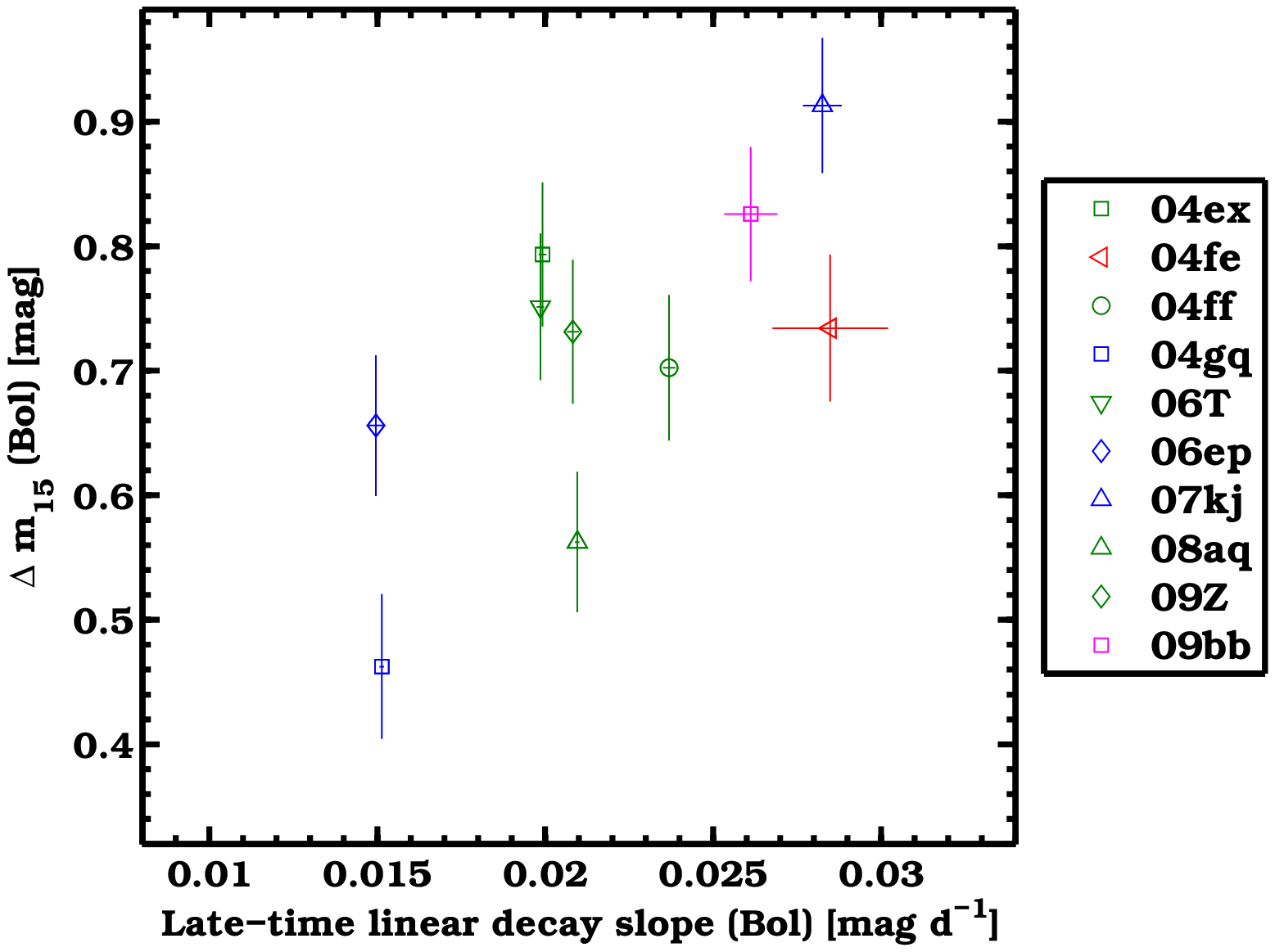}\\
\includegraphics[width=15cm]{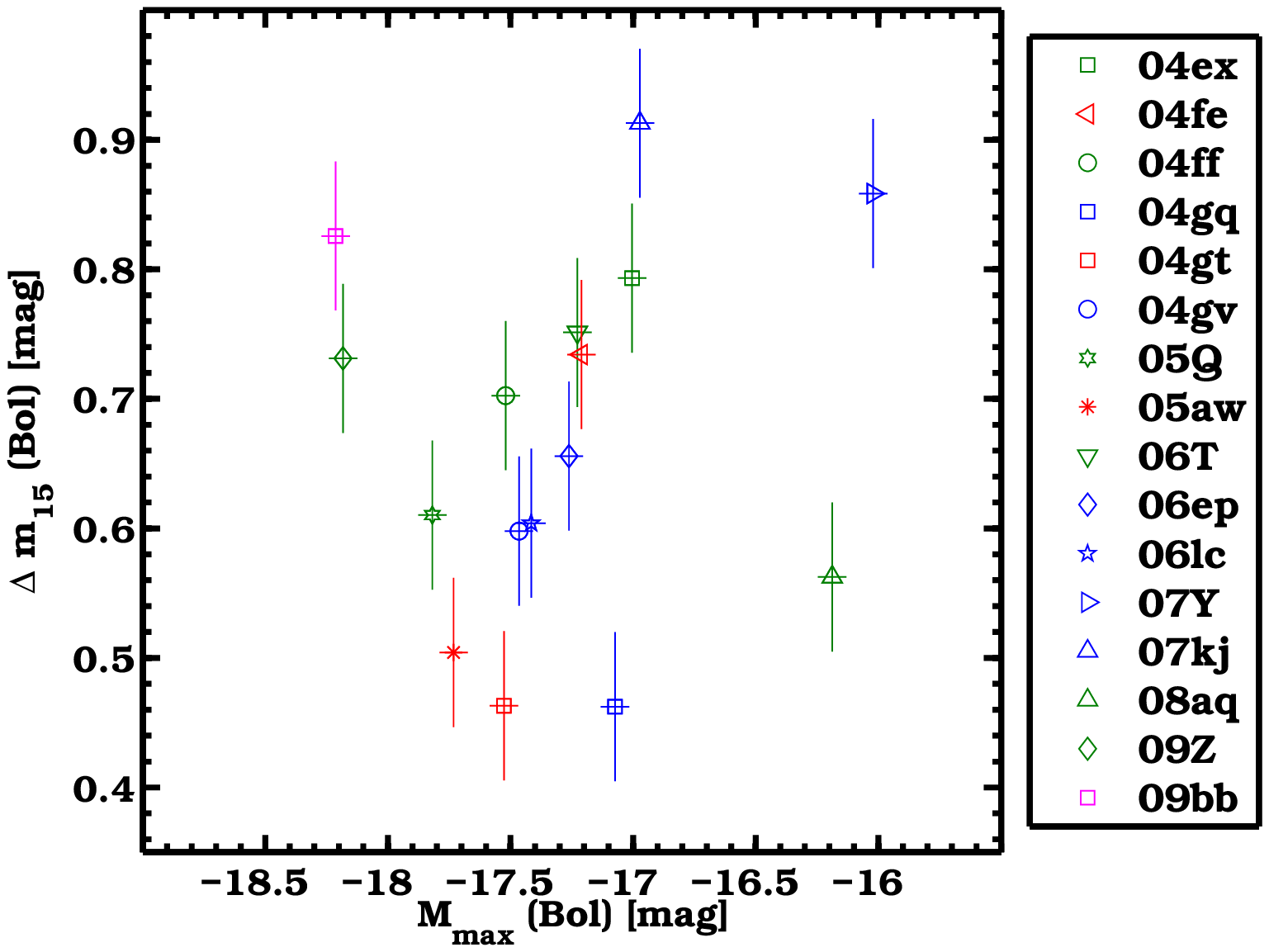}
\caption{\label{dm15vsslope_bolo}$\Delta m_{15}$ of the bolometric light curves versus the late-time linear decay slope (top panel) and the peak bolometric magnitude (bottom panel). A possible correlation is observed in the first case which 
might be explained by a range of values of the $E_{K}/M_{ej}$ ratio (see Sect.~\ref{sec:boloprop}). SNe~IIb, Ib, Ic, and Ic-BL are represented in green, blue, red and magenta, respectively.}
\end{figure}

We find a correlation between $\Delta m_{15}(bol)$ and the late time slope (for the objects with at least one bolometric estimate $+$40d after $t(r)_{\rm max}$. This is consistent with the same trend observed in the $V$ and $r$ bands, and 
it is shown in Fig.~\ref{dm15vsslope_bolo} (top-panel). 
This correlation might be explained in terms of the ratio between energy and ejecta mass.  SE~SNe with larger $E_{K}/M_{ej}$ ratios will be less effective in trapping gamma-rays, and therefore will show steeper slopes at late times (see the parameter $T_0$ in \citealp{wheeler15}). At early epochs, a larger $E_{K}/M_{ej}$ ratio implies a shorter diffusion time and thus a narrower light curve, and therefore a larger $\Delta m_{15}(bol)$.  However, when we check if the objects with broader (narrower) light curves and shallower (steeper) decay rates are also those with lower (higher) $E_{K}/M_{ej}$ ratios (as computed in Sect.~\ref{sec:model}) this is not always the case. 

The correlation between $\Delta m_{15}(bol)$ and $M(bol)_{\rm max}$ is not 
as clear as is found for the $B$ band (see bottom panel of  Fig.~\ref{dm15vsslope_bolo}). 
Finally, excluding the SN~Ic-BL objects, there is no statistically significant difference between the peak luminosities of the various  SE~SN subtypes. The bolometric parameters
discussed in this section are reported in Table~\ref{tab:boloparam}.

\subsection{Black-body fits: Temperature, Photospheric radius, and color-velocity ($V_c$) evolution}
\label{sec:T}

Byproducts of fitting BB functions to the SEDs of the CSP-I SE~SN sample are estimates of the BB temperature  ($T^{BB}_{gVri}$)  and the ``photospheric"' radius ($R_{gVri}^{BB}$) of the emitting region. 
Estimates of these parameters determined from BB fits to the $gVri$-band flux points are plotted in the middle and bottom panel of Fig.~\ref{LTR}.
The evolution of $T^{BB}_{gVri}$ for the sample is remarkably uniform and this holds across subtypes and exhibits a scatter of no more than $\approx$1,000~K beyond $+$5d.  
Prior to maximum the scatter is more pronounced with  $T^{BB}_{gVri}$ found to reach peak values extending from 6,000~K up to 10,000~K. 
By a couple of months past maximum  $T^{BB}_{gVri}$ is found to be
 5,500$\pm$1,000~K, irrespective of the SE~SN subtype.
We emphasize  that the computed $T^{BB}_{gVri}$ values are not sensitively dependent on the exact passbands used in the fit, e.g., if  $g$ band is included or not, and this is a reflection of the photosphere cooling over time.  

Note however that the 
uniformity of the temperatures is 
at least partly a consequence of the assumption on the host-extinction corrections, which were derived assuming intrinsic colors for each SN subtypes \citep{stritzinger17b}. 
This basically means that the extinction corrections to some degree minimizes
the temperature dispersion within each sub-class. 

The R$_{gVri}^{BB}$ is found to increase in all objects, reaching a maximum value around 15 days past $r$-band maximum. After the turnover it follows a slow decline. 
Typical values of the radius at $t_{\rm max}$ are 0.6$-$2.4$\times$10$^{15}$~cm, consistent with results obtained for the SDSS-II SE~SN sample \citep{taddia15}.

\begin{figure}
\centering
\includegraphics[angle=0,width=11cm]{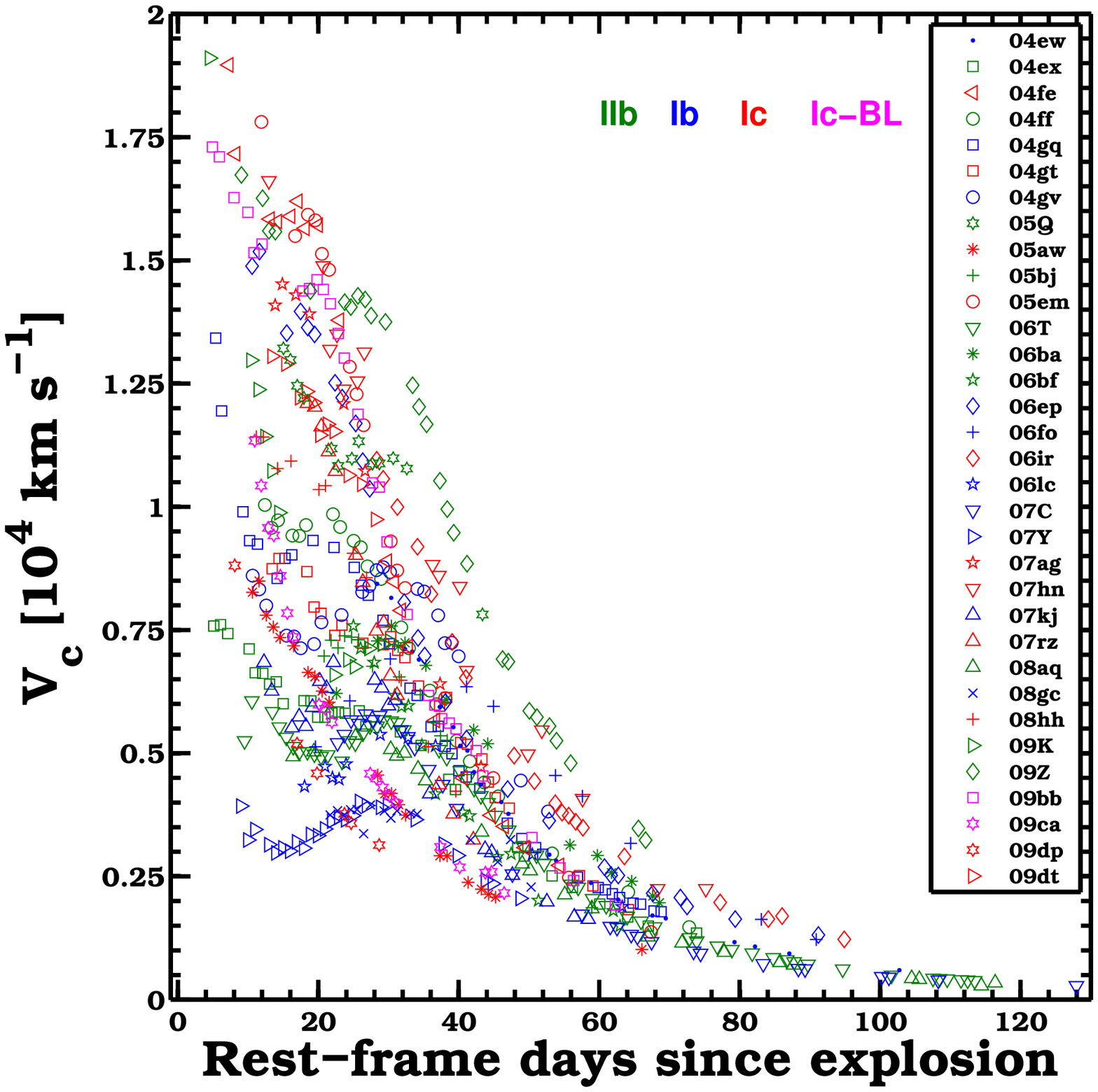}\\
\includegraphics[angle=0,width=11cm]{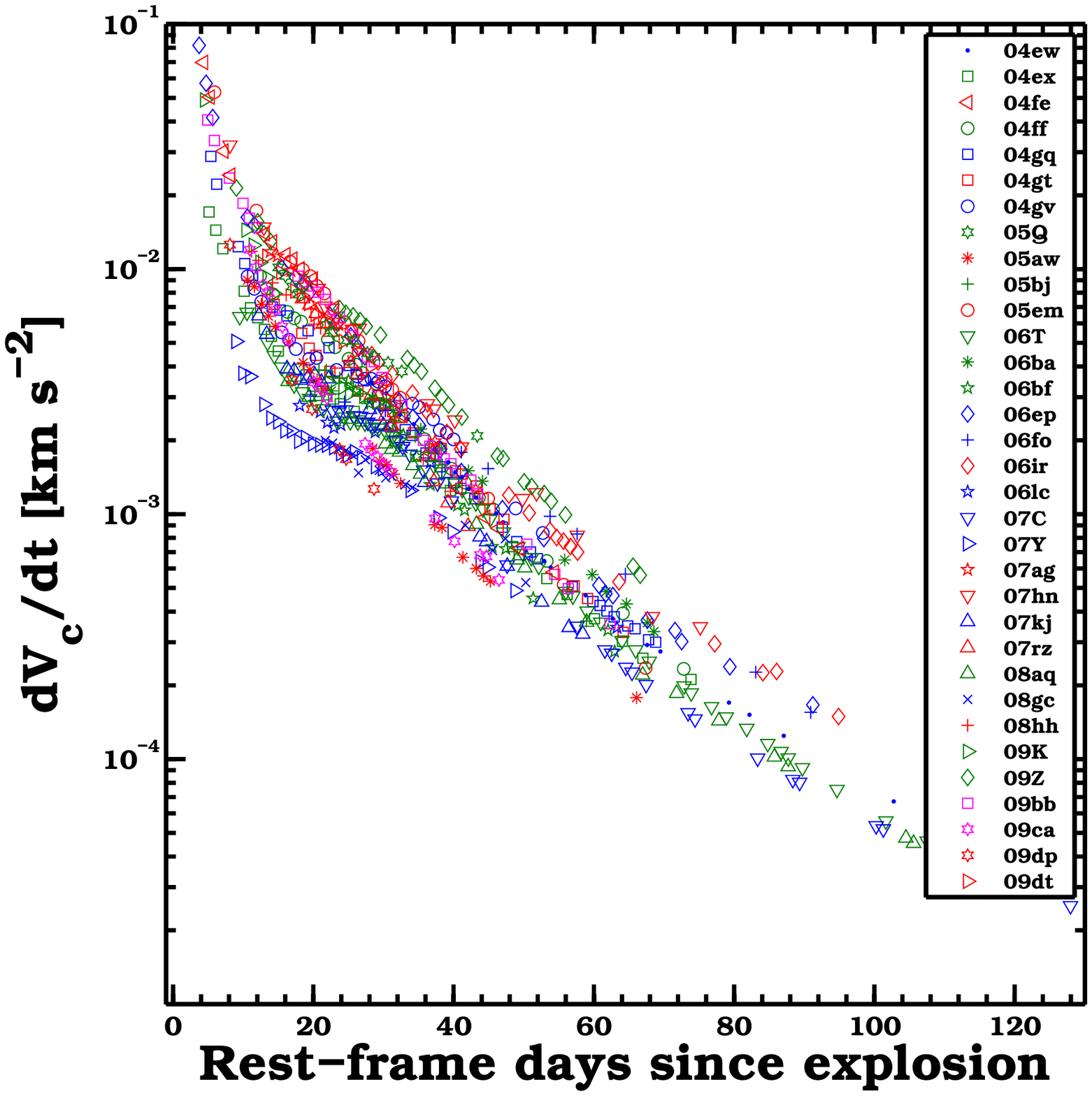}\\
\caption{\label{piroplot} Color velocity ($V_c$) plotted as a function of days past explosion (top-panel), and its gradient (bottom panel).}
\end{figure}

With BB fits to each object's set of SEDs in hand, it is  straightforward to compute the color velocity ($V_c$) parameter and its gradient ($\dot{V_c}$) \citep[][see their Eq.~1]{piro14}. 
The color velocity corresponds to the velocity of the material at $R_{gVri}^{BB}$.  \citeauthor{piro14} argue that high values of $V_c$ and $\dot{V_c}$ are indicative of ejecta material characterized by large density gradients as expected in the outer regions of the expanding ejecta.
Conversely, low values of $V_c$ and $\dot{V_c}$  are indicative of material located in deeper regions of the ejecta that are expanding more slowly than in the outer layers near the surface. 
Plotted in Fig.~\ref{piroplot} (top panel) is $V_c$  vs.  days past explosion (hereafter $t_{exp}$, see Sect.~\ref{sec:exploday}). 
Clearly $V_c$ is highest in the moments following the explosion and subsequently decreases over time. 
We notice that a peak 
in the $V_c$ profiles occurs at $\approx$30d, and this is due to the evolution of $R_{BB}$, which also peaks around that epoch. By  $t_{exp} = +20$d  a little over half of the sample's $V_c$ value drop below $\approx$10,000~km~s$^{-1}$, while by $t_{exp} = +$60d, $V_c$  extends from  as much as $\approx$4,000~km~s$^{-1}$ down to as little as $\approx$1,500~km~s$^{-1}$. 
Each of the SE~SN subtypes are represented at the high end of the  $V_c$  distribution (e.g., SN~Ic~2004fe, SN~Ib~2006ep, and SN~IIb~2009Z), while at the low end only two SNe~Ic (2005aw, 2009dp) are present. 
Indeed, most of the SNe~Ic in the sample appear to exhibit relatively high $V_c$ values at the time of explosion. SNe~Ic also show the highest values of $\dot{V_c}$, again with the exceptions of SN~2005aw and SN~2009dp.  $\dot{V_c}$ is plotted in the bottom panel of Fig.~\ref{piroplot}. 
Examination of the low end of the $V_c$  distribution reveals the presence of several SNe~IIb an Ib with low $V_c$, such as SN~2006T, SN~2006lc, SN~2007Y, SN~2008aq, SN~2007C, SN~2007kj and SN~2008gc. We note that \citet{folatelli14b} recently identified a family of SNe~Ib/IIb that exhibit flat and low ($\approx$4000 and 8000~km~s$^{-1}$) helium velocity evolution extending from before maximum light to past $+$30d. 

\section{Modeling}
\label{sec:model}

We now turn to modeling the CSP-I SE~SN bolometric light curves  in order to  estimate  key explosion parameters including: the explosion energy ($E_{K}$),  the ejecta mass ($M_{ej}$),  the $^{56}$Ni mass, and the degree of $^{56}$Ni mixing in the ejecta.  
In what follows these parameters are computed by both semi-analytical modeling (where $^{56}$Ni mixing is not accounted for) and more sophisticated hydrodynamical modeling. 
To perform this modeling  requires an estimation to the explosion epoch and a measure of the ejecta velocity. 

\subsection{Explosion epochs}
\label{sec:exploday}

\begin{figure*}
\centering
\includegraphics[angle=0,width=18cm]{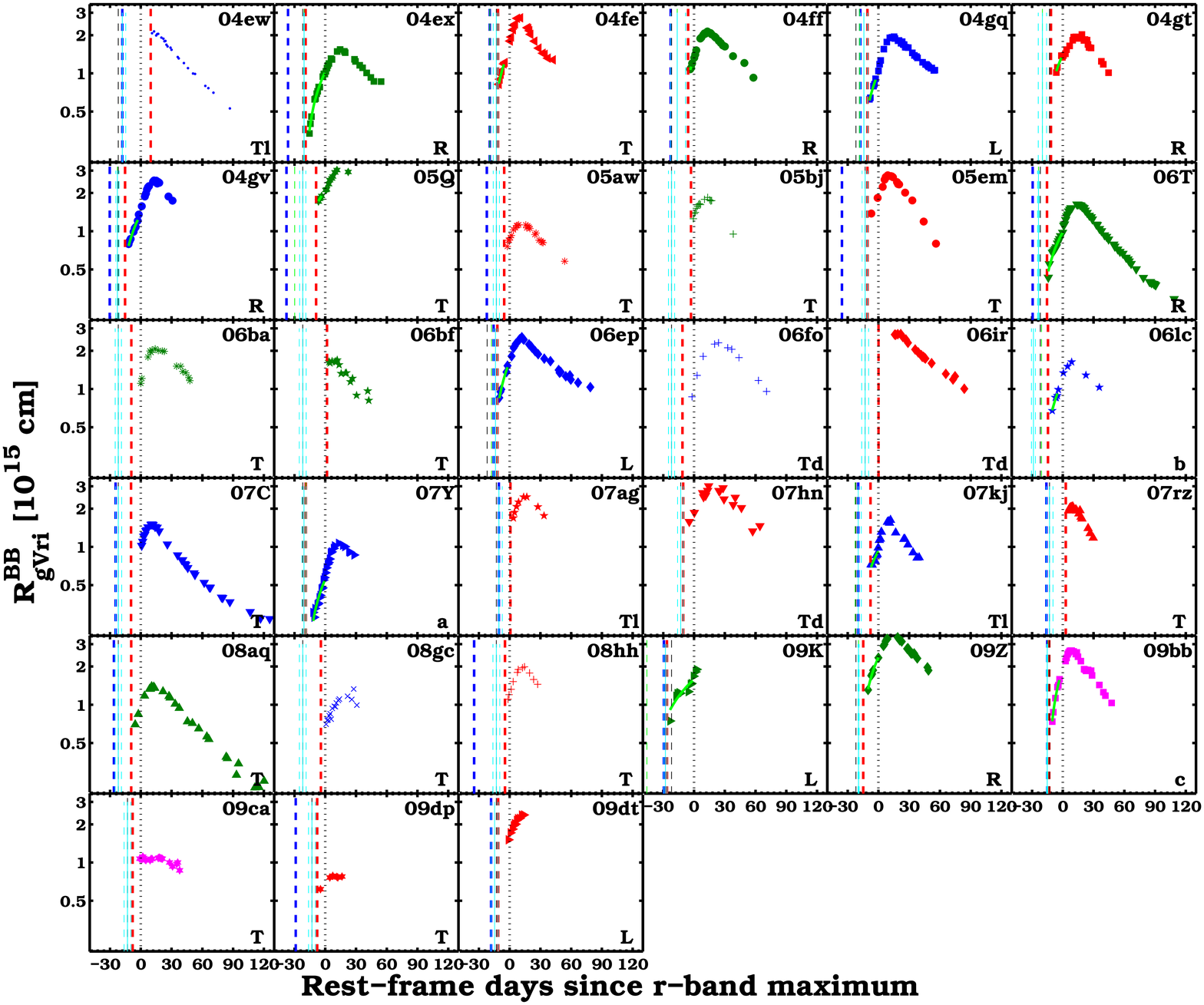}
\caption{\label{findexplo} Illustration of the different techniques used to estimate the explosion epochs for the CSP-I SE~SN sample. The photospheric radius as a function of days since $r$-band maximum is shown for each SN in each sub-panel. The epoch of $r$-band maximum is marked by a black dotted line, the last non-detection epoch is indicated  by a thick, blue dashed line, and the discovery epoch is indicated by a thick, red dashed line. The pre-peak fit to the radius with a PL is indicated by a green solid line and the corresponding explosion epoch estimate is indicated with a green-dashed line. The explosion epochs derived
by assuming a specific rise time for each subtype are marked by a black dashed line. The best explosion epoch estimate obtained for each object is marked by a cyan solid line and its corresponding uncertainties with cyan dashed lines. The method used to obtain the explosion epoch are indicated in each sub-panel. $L$ corresponds to the use of  pre-discovery limits, $R$ corresponds to the use of a PL radius fit, $T$ assumes a rise time; $Tl$ marks the case of assuming the last non-detection epoch as the explosion epoch (as the assumed rise time would have implied a too early explosion as compared to the last non detection), and $Td$ marks when the assumed rise time and discovery epoch are used to determine the uncertainty of the explosion epoch as it occurred close to the inferred explosion, and finally, $a,b,c$ are estimates obtained from the literature (see Table~\ref{dates}). 
Each SN is color-coded so that SNe~IIb, Ib, Ic, and Ic-BL are green, blue, red and magenta, respectively.}
\end{figure*}

To accurately fit the synthetic light curve to the UVOIR light curve of each SN requires an estimate of its explosion epoch. 
 Depending on the discovery details and the subsequent follow-up observations, several techniques  are utilized to estimate the explosion epochs for the SNe in our sample. 
 In cases when the last non-detection and discovery epoch are less than 4 days apart a mean value is adopted. 
 If such limits are not available, the explosion epoch is computed from a power-law (PL) fit to the photospheric radius for all epochs prior to $t(r)_{\rm max}$.
 The adopted PL follows as $r_{ph}(t) \propto(t-t_{expl})^{0.78}$ (see \citealp{piro13}),
 and it is used to predict an explosion epoch constrained to occur between the last non-detection and the discovery epoch.
For objects with poor pre-explosion limits and limited early-time coverage  their explosion epochs are  extrapolated assuming a typical $r$-band rise time  ($t_{r}$).
Here we  adopt $t_{r} = 13\pm3$~days for SNe~Ic and $t_{r} =  22\pm3$~days for SNe~Ib and SNe~IIb \citep[cf.][]{taddia15}. 
Relying on these assumptions enables reasonable explosion epoch estimates for objects with well-constrained values of $t(r)_{\rm max}$ (see Table~\ref{tmax_kcorr}). 
Our best inferred explosion epochs are reported in Table~\ref{dates}, which also provides the method used to estimate them, and details regarding the last non-detection, discovery and confirmation epochs. 
The application of these various methods to fit for the explosion epoch is demonstrated in Fig.~\ref{findexplo}. 

To summarize,  we adopted the average between last non-detection and discovery in four cases (method ``L" in Table~\ref{dates}, where we had good constraints); we used the fit to the black-body radius in seven cases (method ``R" in Table~\ref{dates});  we adopted an average rise time based on the spectroscopic class (method ``T'') for 19 SNe; finally, we adopted explosion epochs from the literature in three cases (see notes a, b and c in in Table~\ref{dates}). We decided to infer the explosion epoch following these methods and to propagate its uncertainty instead of leaving it as a free parameter in the modeling of the bolometric light curves (see Sect.~\ref{sec:arnett}), because the explosion epoch parameter is strongly degenerate with the ratio of energy and ejecta mass, and with the amount of $^{56}$Ni mass intended to be estimated.

\subsection{Photospheric and ejecta velocities}
\label{sec:vel}

\begin{figure}
\centering
\includegraphics[width=10cm]{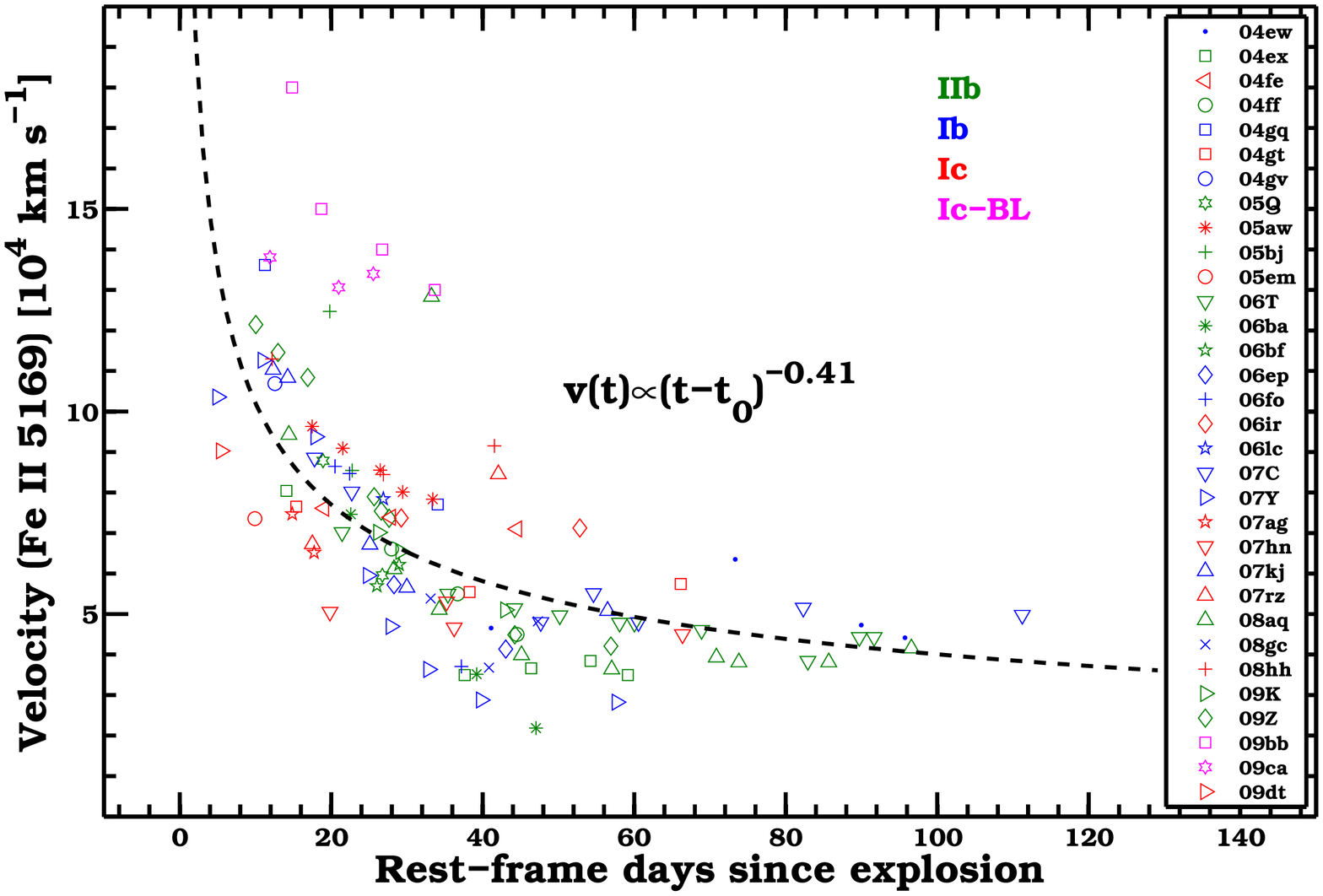}\\
\includegraphics[width=10cm]{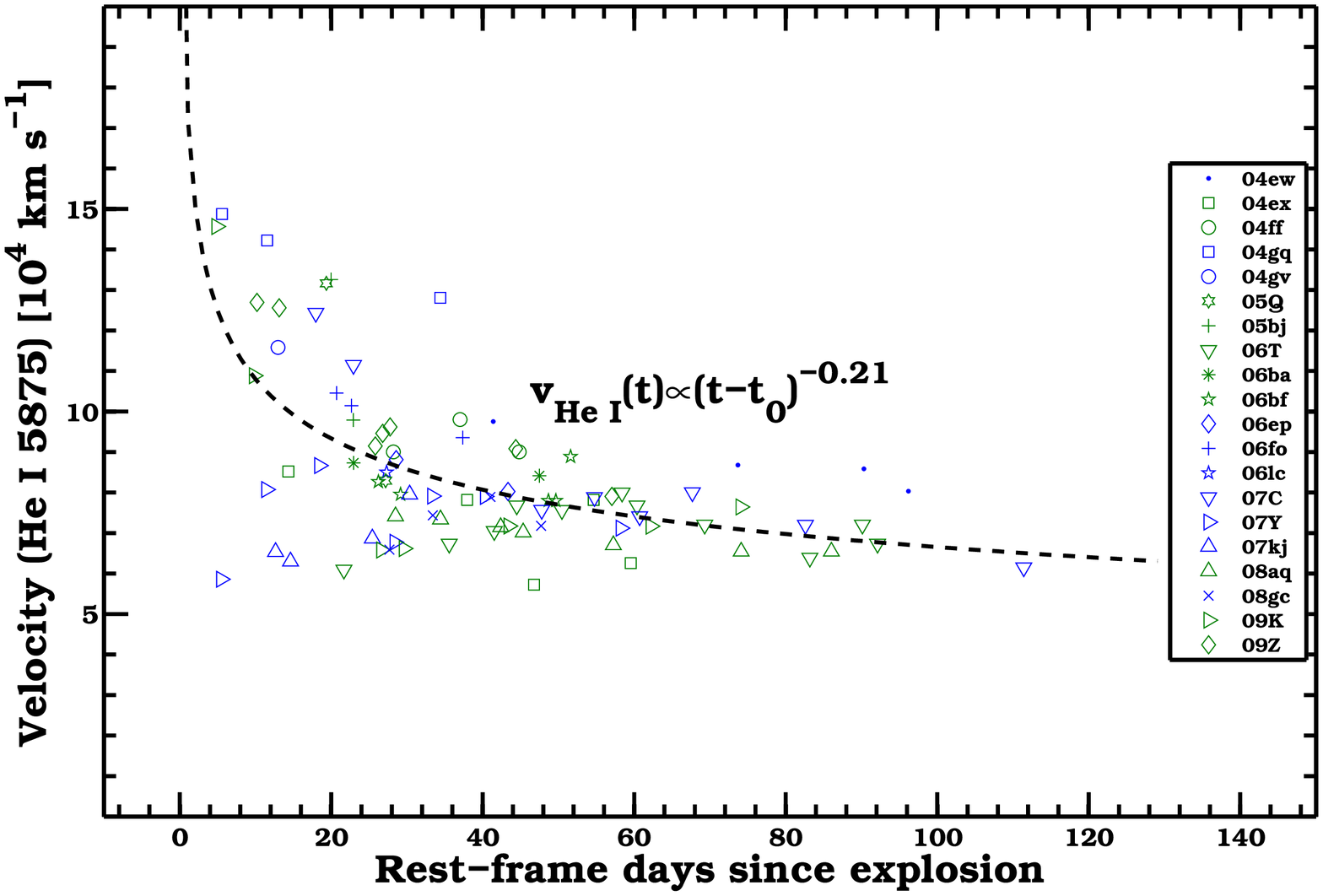}\\
\includegraphics[width=10cm]{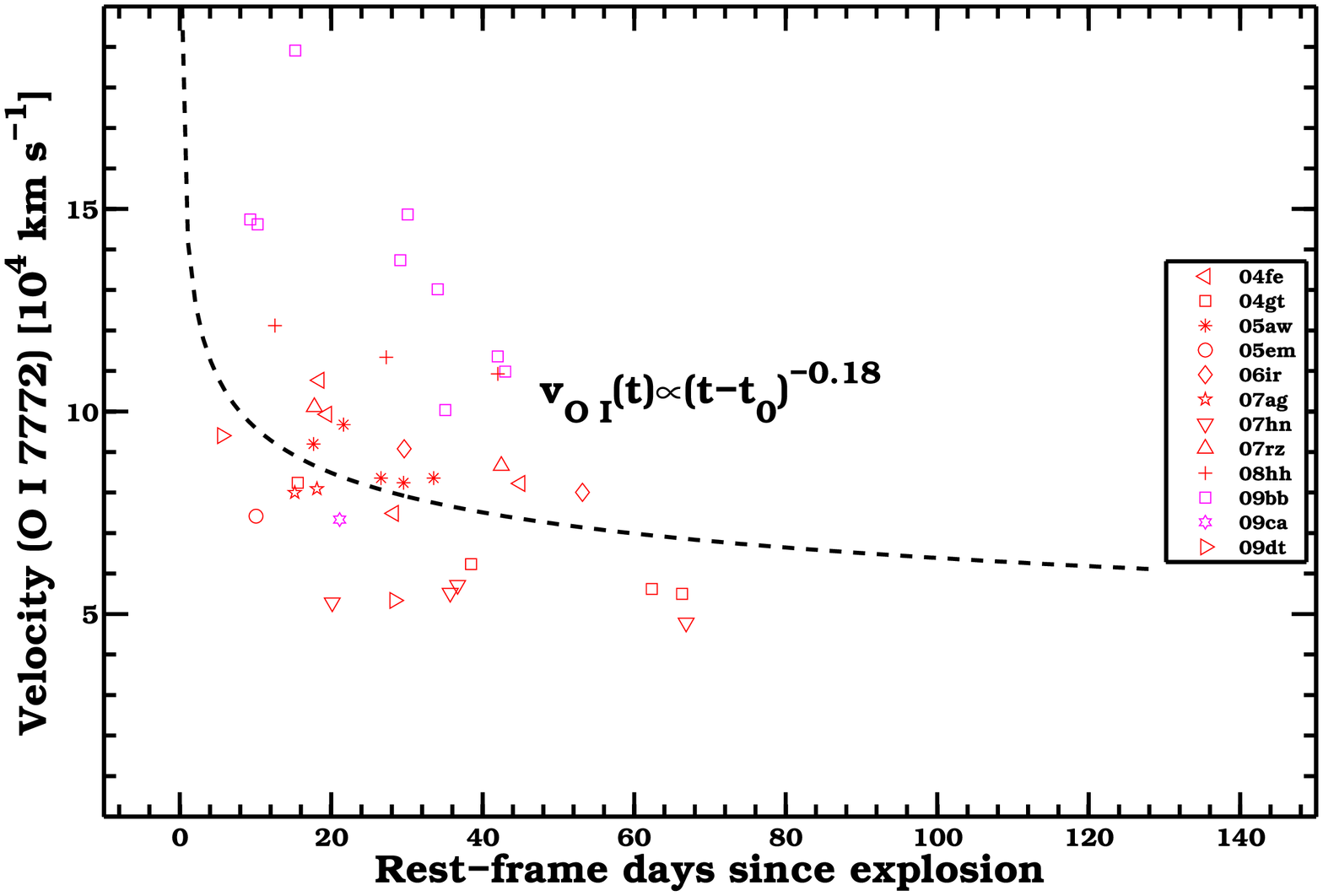}\\
\caption{\label{vel} {\em (Top panel)} Evolution of the Doppler velocity at maximum absorption of the \ion{Fe}{ii}~$\lambda$5169 feature for 32 SE~SNe. 
SNe~IIb, Ib, Ic, and Ic-BL are represented in green, blue, red and magenta, respectively. The SNe~IIb, Ib and Ic follow a similar evolution 
that can be represented by a PL function. 
The SNe~Ic-BL are found to exhibit higher velocities at all epochs and have therefore been excluded from the PL fit (dashed line). {\em (Middle panel)} Velocity evolution of \ion{He}{i}~$\lambda$5876 for SNe~IIb (green) and Ib (blue), fitted by a PL. {\em (Bottom panel)} Velocity evolution of \ion{O}{i}~$\lambda$7772 for SNe~Ic (red) and Ic-BL (magenta), fitted by a PL.}
\end{figure}

Another key input parameter required to fit semi-analytical and hydrodynamical models to the UVOIR light curves is the photospheric velocity ($v_{ph}$). 
Measured as the Doppler velocity at maximum absorption, $v_{ph}$  serves as an important constraint on the ratio between  the $E_K$ and $M_{ej}$. 
In the following,  $v_{ph}$values are adopted from Doppler velocity measurements of the $\ion{Fe}{ii}$ $\lambda$5169 feature \citep[cf.][]{branch02,richardson06}, which are presented in a companion paper by \citet{holmbo17}.
Plotted in Fig.~\ref{vel} are the resulting $v_{ph}$ values vs. days relative to explosion epoch, with the associated uncertainties being on the order of 500~km~s$^{-1}$.
Inspection of  the $v_{ph}$ measurements  reveals  similar values  for each of the SE~SN  subtypes over the same epochs, and the evolution of  $v_{ph}$ is found to be well-represented by a PL function characterized by an index $\alpha=-0.41$ (dashed line in Fig.~\ref{vel}). 
 As expected, the Type~Ic-BL SN~2009bb and SN~2009ca exhibit significantly higher $v_{ph}$ values, several thousand km~s$^{-1}$ higher than
the rest of the sample over the same epochs. 
These two objects are omitted when computing the PL fit.

For the semi-analytic models we use the value of $v_{ph}$ at peak luminosity [$v_{ph}({t_{\rm max})}$] to constrain $E_{K}/M_{ej}$. 
These are computed by fitting a PL to the measured \ion{Fe}{ii}$~\lambda$5169 velocities for each SN and taking the value of the best fit at the peak epoch.
 Assuming the ejecta are spherical and with constant density  the $E_{K}$ to $M_{ej}$ ratio is given by the expression: $E_{K}/M_{ej} = \frac{3}{10}v_{ph}({t_{\rm max})}^2$ \citep{wheeler15}.

Following \citet[][see their Sect. 5.3]{dessart16}, an alternative approach  to constrain the $E_{K}$ to $M_{ej}$  ratio is to determine the quantity $V_{m} = \sqrt{2E_{K}/M_{ej}}$.
In the case of helium rich SNe~IIb and SNe~Ib, the Doppler velocity of 
  the \ion{He}{i}~$\lambda$5875 feature can provide a measure of $V_m$, while for SNe~Ic the \ion{O}{i}~$\lambda$7774 feature is appropriate. 

Doppler velocity measurements of these lines and other spectral features are presented in the companion CSP-I SE~SN spectroscopy paper \citep{holmbo17}. 
The corresponding  \ion{He}{i} and \ion{O}{i} Doppler velocity measurements are plotted in  the central and bottom  panels of Fig.~\ref{vel}, respectively.
The Doppler velocity evolution of these features are well fit by PL functions (dashed lines) characterized by index values of $-$0.21 (\ion{He}{i})  and $-$0.18 (\ion{O}{i}).
When fitting the \ion{Fe}{ii}, \ion{He}{I}, and \ion{O}{i} line velocities, we adopted a unique PL index for all the objects. This is done to more robustly fit the velocity profiles of the events with a low number of spectra. However, we have also tested if this index is well suited for the events with numerous spectra. In particular, in the case of SN~2006T, we found that fitting its velocity profile with the PL index as a free parameter gives a similar index ($-0.35$ instead of $-0.41$) and an interpolated velocity at the maximum epoch which differs by the one derived with fixed index by merely $\approx$290~km~s$^{-1}$, i.e., below the typical velocity uncertainty.
For each SN, the \ion{He}{i} (if Type IIb or Ib) or \ion{O}{i} (if Type Ic) velocities are fit with the proper PL in order to derive the velocity at peak, and this is used to directly estimate $V_m$.

\subsection{Progenitor parameters from Arnett's equations}
\label{sec:arnett}

We first proceed to fit the bolometric light curves with an \citet{arnett82} model, assuming the explosion epochs given in Table~\ref{dates} and $E_{K}/M_{ej} = \frac{3}{10}v_{ph}({t_{\rm max})}^2$. This provides a measure of $E_{K}$, $M_{ej}$ and the $^{56}$Ni mass. 
The specific function to model the UVOIR luminosity is presented by \citet[][see their Eq.~1]{cano13}.  
In the process of computing a light-curve model a constant opacity $\kappa=0.07$~cm$^2$~g$^{-1}$ is adopted, as was done in \citet{cano13} and \citet{taddia15}, and implied by the models of SN~1998bw presented by \citet{chugai00}. 
The fit is done only including luminosity measurements obtained prior to 60 days past the explosion epoch, when the SNe are in their photospheric phase. When computing the best-fit Arnett model 
the gamma-ray escaping fraction was also considered using the method of \citet{wheeler15} and recently  utilized by  \citet{karamehmetoglu17}.

\begin{figure*}
\centering
\includegraphics[angle=0,width=18cm]{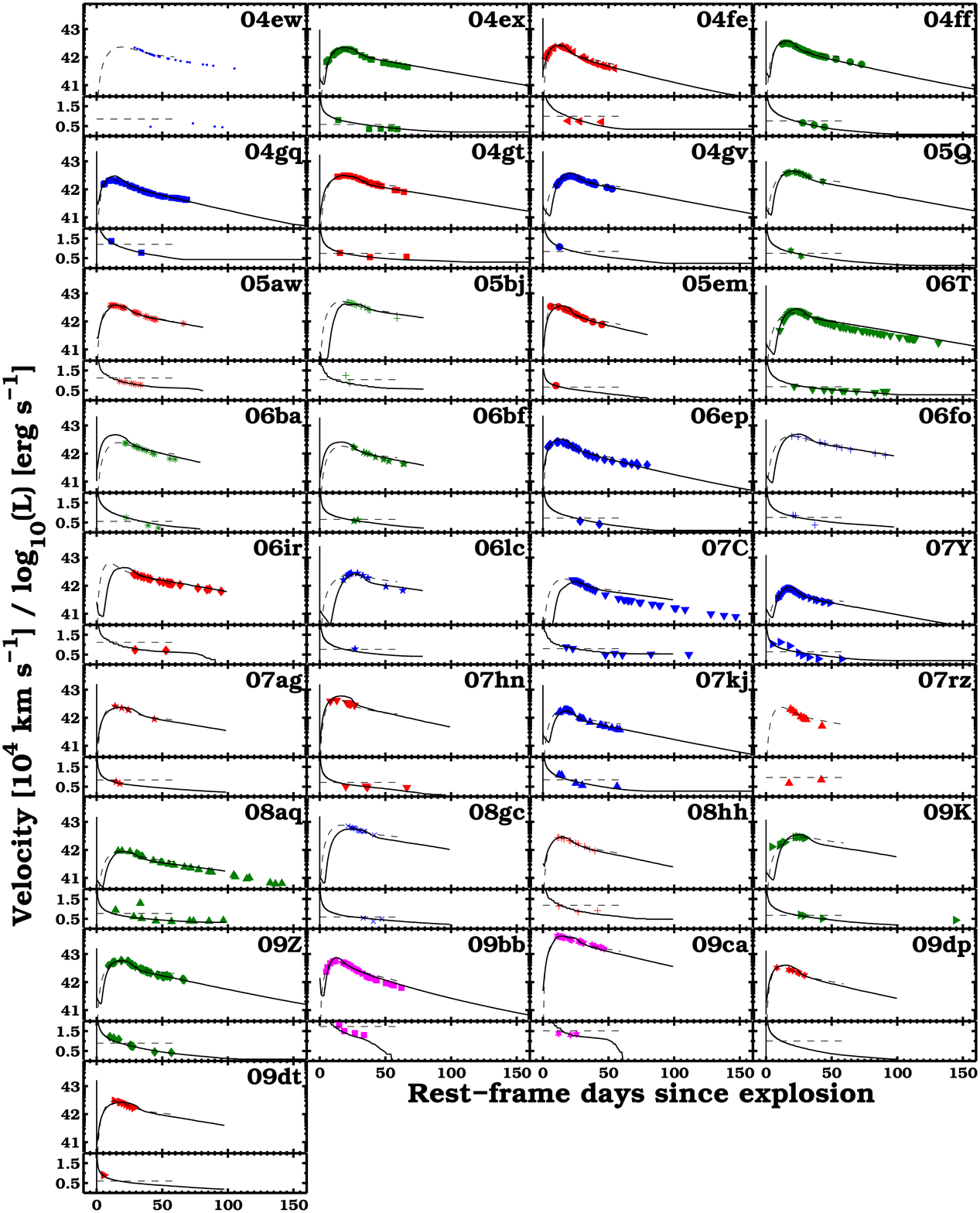}
\caption{\label{model} Semi-analytic and hydrodynamical models of the bolometric light curves and $v_{ph}$ profiles of 33 CSP-I SE~SNe.  Dashed black lines represents to  the best-fit Arnett bolometric model and the $v_{ph}(t_{\rm max})$ velocity derived from \ion{Fe}{ii}~$\lambda$5169 lines. Solid black lines represents   the best-fit hydrodynamical model and corresponding velocity evolution.}
\end{figure*}

Plotted in Fig.~\ref{model} are the UVOIR light curves of the CSP-I SE~SN sample along with the best-fit analytical and hydrodynamical models (see below). 
Also plotted within the panel of each UVOIR light curve is a sub-panel displaying the  
measured $v_{ph}$ values, and the adopted $v_{ph}({t_{\rm max}})$ value at the epoch of $t(r)_{\rm max}$  is also indicated in each sub-panel.
The resulting key explosion parameters obtained from the two methods 
are reported in Table~\ref{param}, along with averaged values for each SE~SN subtype. The error on the $^{56}$Ni mass is dominated by the error on the SN distance, but also includes the error associated with the explosion epoch estimate as well as the fit uncertainty. The errors on $E_{K}$ and $M_{ej}$ are largely dominated by the uncertainty of the explosion epoch.
  SNe~IIb, Ib and Ic show typical ejecta masses of 4.3(2.0)~$M_{\odot}$, 3.8(2.1)~$M_{\odot}$, and 2.1(1.0)~$M_{\odot}$, respectively; kinetic energies are found
 to be 1.3(0.6)$\times$10$^{51}$~erg, 1.4(0.9)$\times$10$^{51}$~erg, and 1.2(0.7)$\times$10$^{51}$~erg, respectively; $^{56}$Ni masses are 0.15(0.07)~$M_{\odot}$, 0.14(0.09)~$M_{\odot}$, and 0.13(0.04)~$M_{\odot}$, respectively. 

\begin{figure}
\centering
\includegraphics[width=18cm]{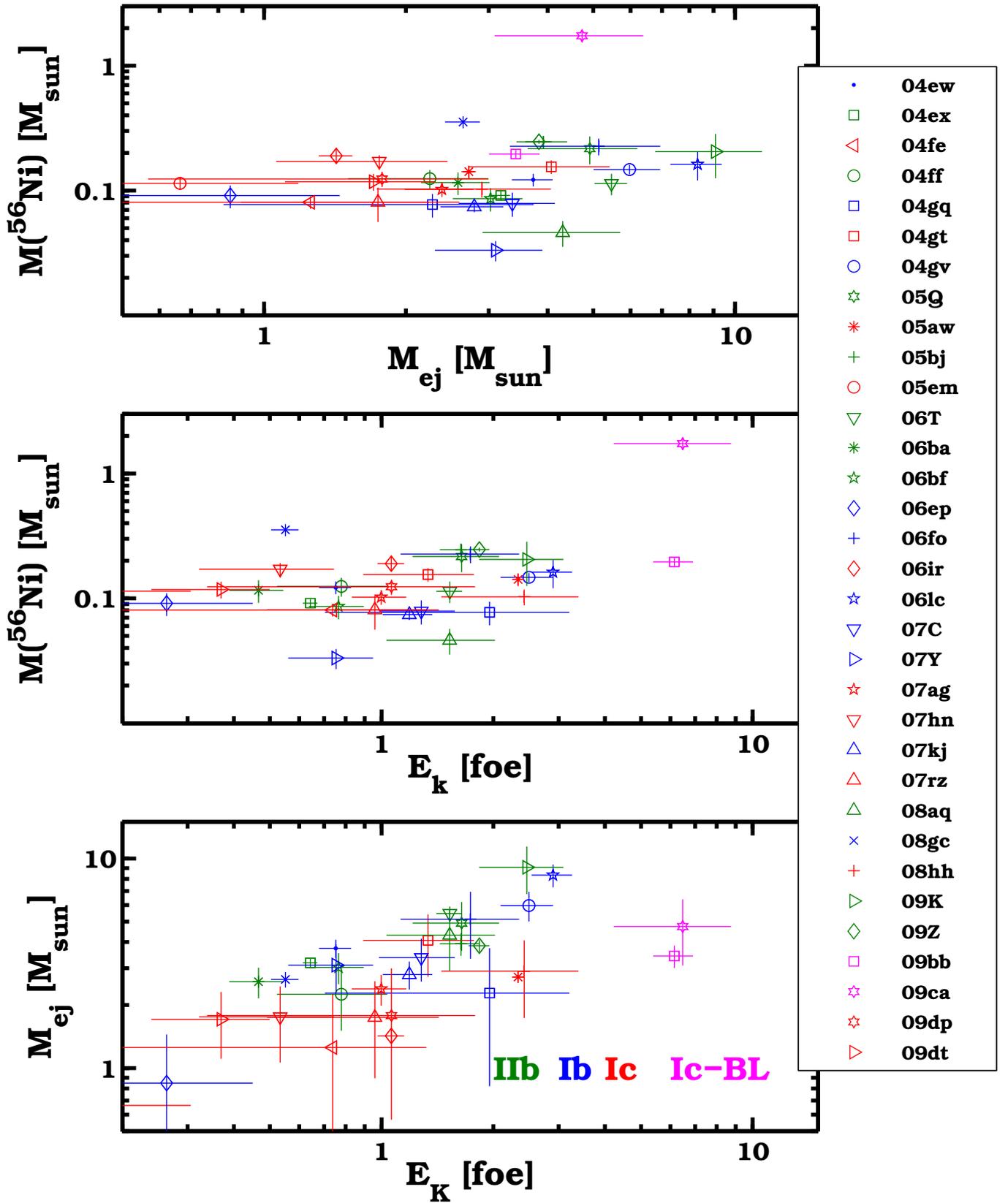}
\caption{\label{param_corr} Correlations between the explosion parameters obtained from the Arnett models of 33 CSP-I SE~SNe. SNe~IIb, Ib, Ic, and Ic-BL are represented in green, blue, red and magenta, respectively.}
\end{figure}

Plotted in  Fig.~\ref{param_corr}  is a clear correlation between $E_{K}$ and $M_{ej}$ as found from the Arnett model, and that there are possible correlations between these two parameters and the $^{56}$Ni mass. Similar results were found by \citet{lyman16}.

\begin{figure*}
\centering
\includegraphics[width=18cm]{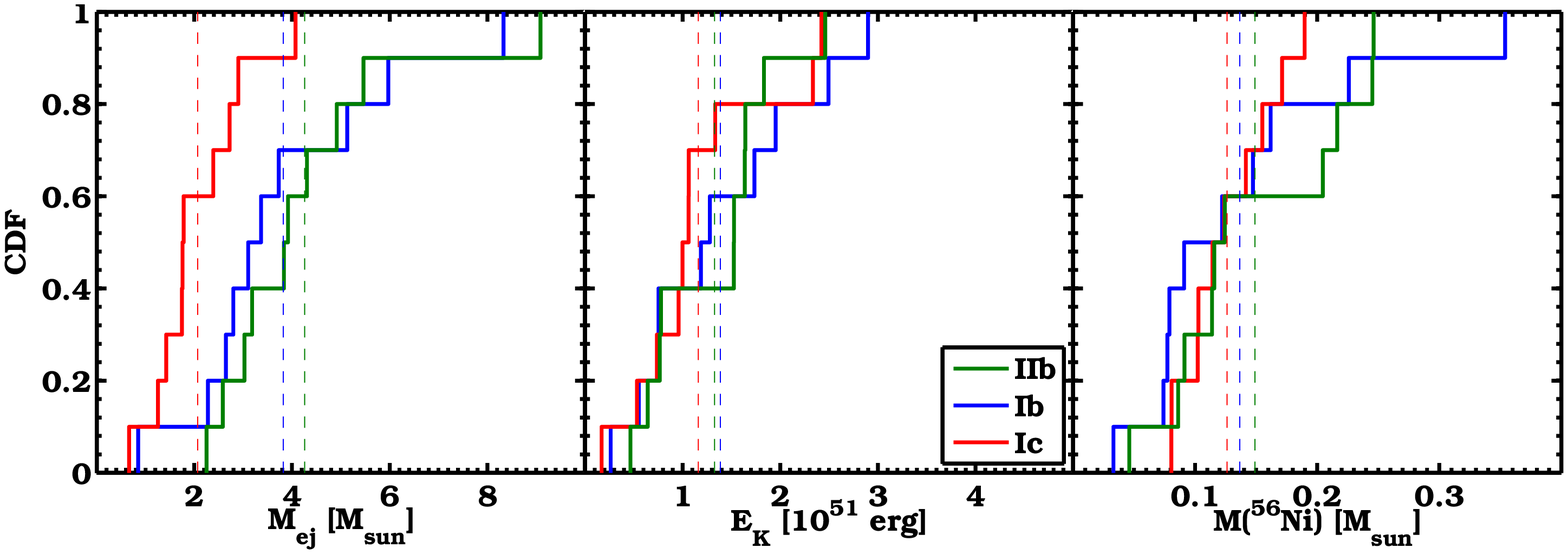}
\caption{\label{param_cdf} Cumulative distributions (solid lines) of the explosion parameters obtained from the Arnett models of 31 CSP-I SE~SNe (the two SNe~Ic-BL are not included). SNe~IIb, Ib, and Ic are represented in green, blue, and red, respectively. The average value of each distribution is marked by a vertical dashed line.}
\end{figure*}

In Fig.~\ref{param_cdf}, the cumulative distributions of the three parameters for the three main classes indicates the only difference among SNe~IIb, Ib and Ic is that SNe~Ic possibly have lower ejecta masses.  
A K--S test reveals the difference is significant (p-value$=$0.007) for the comparison between SNe~Ic and SNe~IIb.

\begin{figure*}
\centering
\includegraphics[width=18cm]{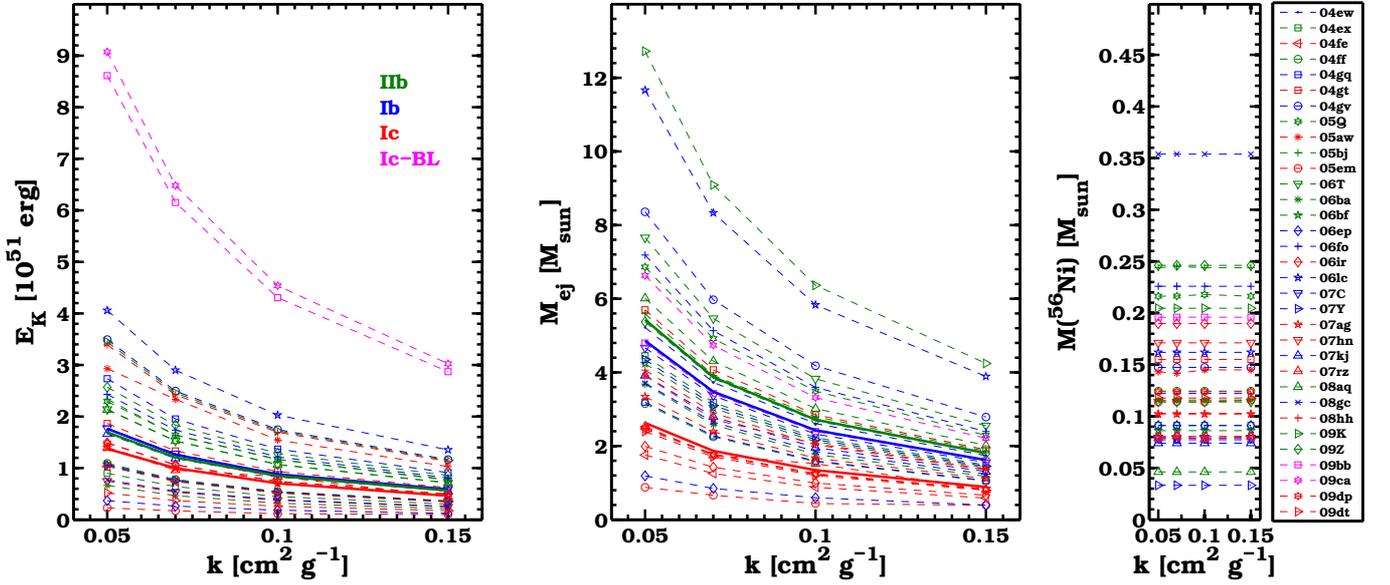}
\caption{\label{testkappa} Best explosion parameters ($E_{K}$,  $M_{ej}$, $M(^{56}Ni)$) from the Arnett models of 33 CSP-I SE~SNe as a function of the opacity. SNe~IIb, Ib, Ic, and Ic-BL are represented in green, blue, red and magenta, respectively. Thick solid lines represent the average for each subtype. Larger opacities imply lower values of $E_{K}$ and $M_{ej}$. For clarity, SN~2009ca is not shown in the $M(^{56}Ni)$ subpanel.}
\end{figure*}

A major limitation in applying semi-analytic modeling techniques to SE~SN UVOIR light curves is the assumption of a constant opacity, denoted $\kappa$. \citet{dessart16} showed how different assumptions on the value of $\kappa$ 
can lead to different results for the best progenitor parameters, and that ultimately, the assumption of constant opacity is quite poor for SE~SNe.
 In the context of the semi-analytic model, we 
explore how our results vary depending on the value adopted for $\kappa$.  Instead of $\kappa=0.07$~cm$^2$~g$^{-1}$, we perform
Arnett fits with $\kappa=0.05$, $0.10$, and $0.15$~cm$^2$~g$^{-1}$. 
In Fig.~\ref{testkappa} the best fit parameters for the four different values of opacity are reported. It is evident how larger opacities can lead to lower values of both $E_{K}$ and $M_{ej}$, without modifying the $^{56}$Ni mass. 
Looking at the average for each subtype, 
a change in opacity from $\kappa=0.05$~cm$^2$~g$^{-1}$ to $\kappa=0.15$~cm$^2$~g$^{-1}$ reduces $E_K$ and the ejecta mass by 67\% for each SE~SN subtype. 

\begin{figure*}
\centering
\includegraphics[width=18cm]{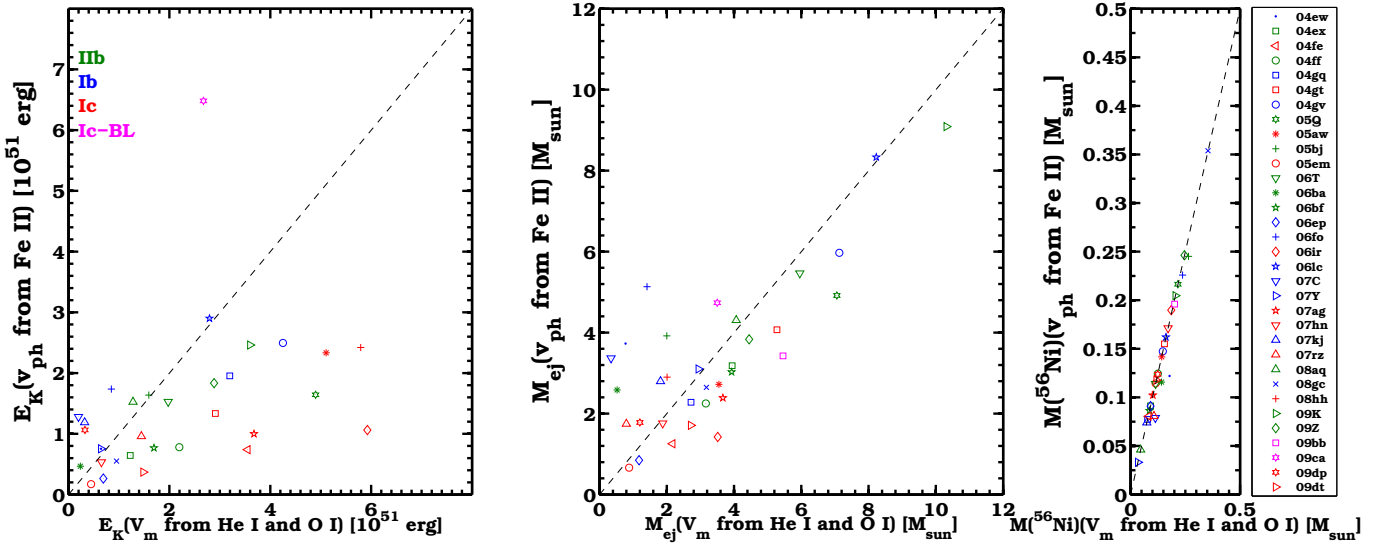}
\caption{\label{param_corr_Vm} Best explosion parameters ($E_{K}$,  $M_{ej}$, $M^{56}$Ni)) from the Arnett models of 33 CSP-I SE~SNe as computed using $v_{ph}$ from \ion{Fe}{ii} versus those computed using $V_m$ from \ion{He}{i} and \ion{O}{i} as explained in \citet{dessart16}. SNe~IIb, Ib, Ic, and Ic-BL are represented in green, blue, red and magenta, respectively. The dashed  lines indicate when the parameters are identical with the two methods. Identical $^{56}$Ni masses are derived, very similar ejecta masses, whereas the energy obtained with the velocities from \ion{He}{i} and \ion{O}{i} are typically larger than  those obtained with the \ion{Fe}{ii} velocity, especially for SNe~Ic. }
\end{figure*}

We also explore how our results are affected by 
using $v_{ph}$ values obtained from the \ion{Fe}{ii} line velocities compared to 
using $V_m$ as derived from the \ion{He}{i} and \ion{O}{i} line velocities at peak.
Assuming a constant opacity (i.e., $\kappa~=~0.07$cm$^2$~g$^{-1}$), this comparison reveals nearly identical 
$^{56}$Ni masses, very similar ejecta masses, while the energies differ, especially for the SNe~Ic. The comparisons between the parameters derived with the two different assumptions on the velocity and using the Arnett model is shown in Fig.~\ref{param_corr_Vm}. 

\subsection{Progenitor parameters from hydrodynamical models}
\label{sec:melina}

Estimates for the explosion parameters are also obtained through
hydrodynamical models compared to the UVOIR light-curve and velocity
evolution of each SN. 
To do so a grid of light-curve models and their associated velocity
evolution is computed using one-dimensional Lagrangian LTE
radiation hydrodynamics calculation \citep{bersten11}, 
based on hydrogen deficient He-core stars \citep[see][for more
details]{bersten12}. 
The grid of models is constructed by exploding a series of relatively
compact  ($R < 3 R_{\sun}$) structures with Helium-core masses of
3.3~$M_{\sun}$ (He3.3), 4~$M_{\sun}$ (He4), 5~$M_{\sun}$ (He5),
6~$M_{\sun}$ (He6), and 8~$M_{\sun}$ (He8). These pre-supernova
models originate from  stellar evolutionary calculations of
single stars with zero-age-main-sequence masses of 12~$M_{\sun}$,
15~$M_{\sun}$, 18~$M_{\sun}$, 20 $M_{\sun}$, and 25~$M_{\sun}$,
respectively \citep{nomoto88}.  

To explode the initial hydrostatic configuration some energy is
artificially injected  near the center of the pre-supernova star, 
yielding the formation of a shock-wave that propagates through and unbinds 
the stars. As it is well known most of the light-curve evolution in
SE~SNe is powered mainly by the energy produced by radioactive
decay because the explosion energy itself is rapidly degraded due to the
compactness of the progenitor. To treat the $\gamma$ photons produced
from radioactive decay we assume gray transfer with a $\gamma$-ray
mean opacity of $\kappa_{\gamma}=  
0.06~Y_{e}$~cm$^{2}$~g$^{-1}$ \citep[see][]{swartz95}, where $Y_{e}$
is the electron to baryon fraction. We allow for any distribution of
$^{56}$Ni inside the ejecta. In this analysis, we have assumed a linear $^{56}$Ni distribution with a maximum value in the central region, and extended inside the configuration out to a specific fraction of the total mass (defined as the mixing parameter; see Table~\ref{param}). 
 Our calculations enable us to
self-consistently determine the propagation of the shock wave through
the star, and follow it through breakout and its subsequent light
curve emission out to late phases. However, we do not calculate the $^{56}$Ni production as a consequence of the explosive nucleosynthesis. We simply assume it as a free parameter of the model to be estimated by fitting the bolometric light curve.

In order to find an optimal model for each object in our sample we
have calculated an extensive grid of models for different values of
the explosion energy, $^{56}$Ni mass and distribution for a given
pre-supernova structure. The grid of hydro models was then compared to our
UVOIR light curves and the photospheric velocity evolution estimated
from \ion{Fe}{ii} $\lambda$5169 (see Section~\ref{sec:vel}). This
allowed us to select models that simultaneously reproduce both
observables thus reducing the known degeneracy between $M_{ej}$ and $E_{\rm exp}$.
We note that the light-curve peak is extremely sensitive to the amount
of $^{56}$Ni produced during the explosion while the width around the
main peak is primarily sensitive to $M_{\rm ej}$ and $E_{\rm
exp}$. If very early observations are available, i.e., before the rise
to the main peak, then it is possible to estimate the size of the 
progenitor via hydrodynamical modeling. However, even with the excellent
coverage of the CSP-I sample, the early cooling phase of the light curves is
missing in most of the objects with the possible exception of the Type~IIb 
SN~2009K. 

\begin{figure}
\centering
\includegraphics[width=18cm]{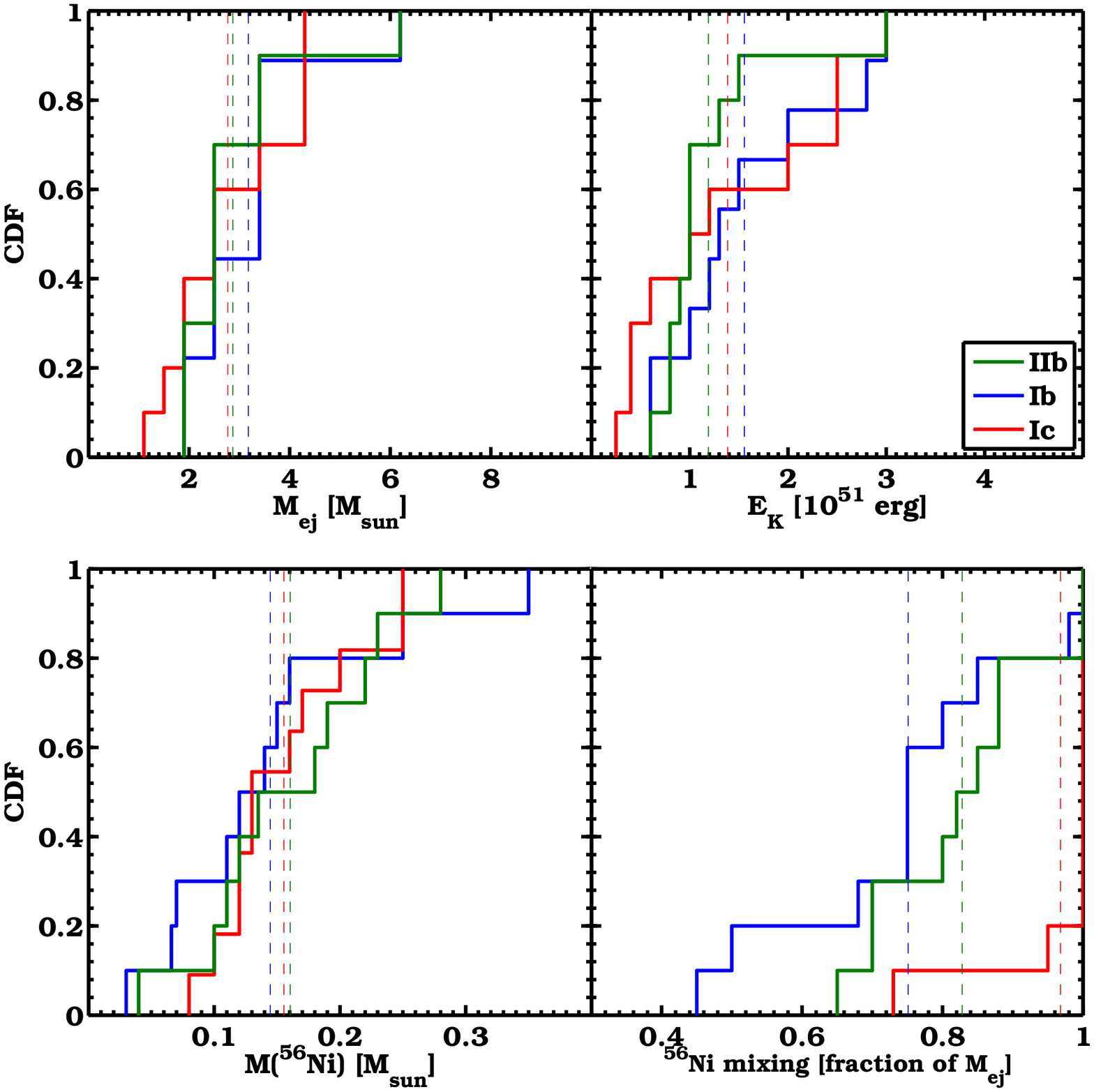}
\caption{\label{cdf_hydro} Cumulative distributions (solid lines) of the explosion parameters obtained from the hydrodynamical models of 31 CSP-I SE~SNe (the two SNe~Ic-BL are not included). SNe~IIb, Ib, and Ic are represented in green, blue, and red, respectively. The average value of each distribution is marked by a vertical dashed line.}
\end{figure}

Best-fit model light curves and velocity profiles are plotted on top
 of the corresponding SN data in Fig.~\ref{model}. The corresponding
 model parameters are listed in Table~\ref{param}. Overall, the
 results are rather similar to those obtained with the Arnett models.
 Figure~\ref{cdf_hydro} shows the cumulative distributions for the
 parameters of the three main subtypes, revealing very similar ejecta
 mass, energy, and $^{56}$Ni mass distributions. SNe~IIb, Ib and Ic
 have average ejecta masses of 2.9(1.3)~$M_{\odot}$, 3.2(1.3)~$M_{\odot}$, and 2.8(1.2)~$M_{\odot}$, respectively; kinetic energies
 are found to be 1.2(0.7)~foe\footnote{A foe is a unit of energy equivalent to $10^{51}$~erg.}, 1.6(0.9)~foe, and 1.4(1.0)~foe,
 respectively; and 
 $^{56}$Ni masses are 0.16(0.07)~$M_{\odot}$, 0.14(0.09)~$M_{\odot}$, 0.16(0.06)~$M_{\odot}$, respectively.   

Interestingly, the average degree of $^{56}$Ni mixing --defined as the fraction of mass enclosed within the maximum radius of the $^{56}$Ni distribution--  is  found to be larger in SNe~Ic compared to SNe~IIb and Ib. Quantitatively, for the CSP-I sample of SNe~Ic the mixing parameter is found to be 1.0 for all the objects except SNe~2006ir and 2005aw (the average is 0.95), as compared to $0.75\pm0.18$  and $0.83\pm0.12$ for the SNe~IIb and Ib, respectively. 
All our SE~SNe are found to have $^{56}$Ni mixed out to $\gtrsim~45\%$ of the ejecta mass.

It is important to note that the mixing parameter is extremely sensitive to
the estimate of the explosion time, which in some cases it is
not tightly constrained. Another factor that can affect our
results, in particular for SN~Ic progenitors, is the initial
progenitor star model. Helium stars were adopted for the initial configurations in our 
calculations, whereas SN~Ic progenitors are thought to be largely stripped of their helium envelopes. 
We adopted helium-rich models since there are currently no helium-free structures available in the literature to use in our hydro calculations for SNe~Ic bolometric light curves.

\section{Discussion}
\label{sec:discussion}

Key explosion parameters for the SN sample were estimated using
 both semi-analytic  and  the hydrodynamical modeling techniques. In our analysis we elected to include all of the objects not observed early enough to directly estimate their peak bolometric light curve. At the end of Table~\ref{param} we report the average $E_{K}$, $M_{ej}$ and $^{56}$Ni mass  obtained from our modeling efforts when excluding these objects. 
We note that if average explosion parameters are estimated using  the entire CSP-I sample  consistent results are obtained  (within the errors) compared to those obtained from just the best-observed subset.  This is an encouraging  finding and suggests our   efforts to estimate the explosion epoch and the peak luminosity  for poorly observed objects does  provide for reasonable estimates on their explosion parameters.

We proceed to compare our semi-analytic results to those obtained from other SE~SN samples in the literature, as well as to compare the parameters derived from the two different methods. Based on the results concerning these parameters, we discuss the implications for the nature of the SE~SN progenitor stars. 

\subsection{Comparison with other samples in the literature}

\begin{figure*}
\centering
\includegraphics[width=18cm]{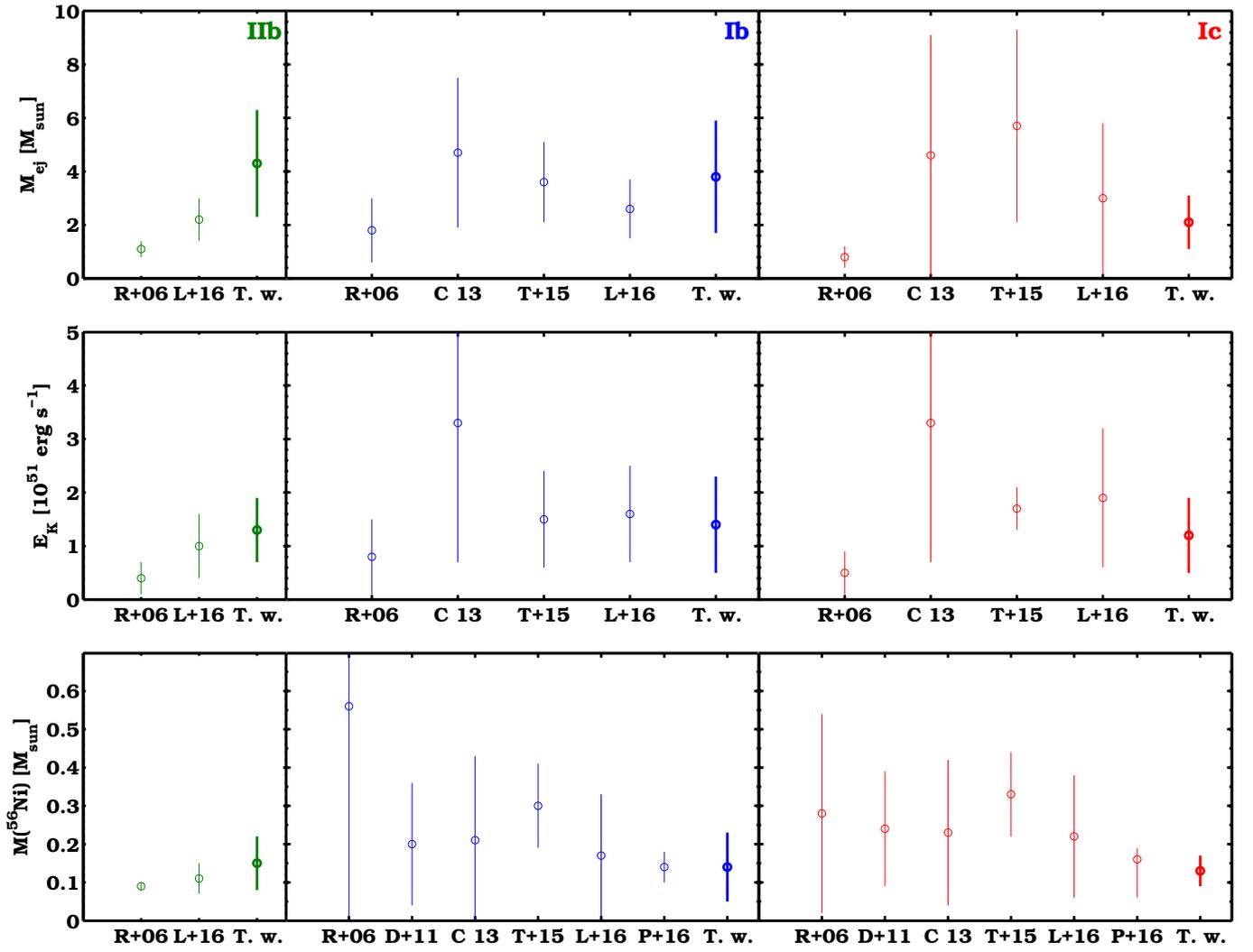}
\caption{\label{compliterature}Explosion and progenitor parameter comparisons among different works in the literature (R+06$=$\citet{richardson06}; D+11$=$\citet{drout11}; C 13$=$\citealp{cano13}; T+15$=$\citealp{taddia15}; L+16$=$\citealp{lyman16}; P+16$=$\citealp{prentice16}; T. w.$=$This work), which include samples of SE~SNe (IIb, Ib, Ic) and make use of semi-analytic models.}
\end{figure*}

In Table~\ref{tab:compliterature} we present the average explosion and progenitor parameters for the different SE~SN subtypes as derived from a number of samples in the literature. We compare these published results with our semi-analytic estimates, and the comparison is  illustrated in Fig.~\ref{compliterature}. 
We stress that we are comparing works where the parameters were
computed with similar models and in particular the ejecta mass and the
explosion energy parameters were derived based on almost identical
assumptions regarding the adopted opacity
(0.06--0.07~cm$^2$~g$^{-1}$). The only exceptions are the $E_K$ and
$M_{ej}$ values derived by \citet{richardson06},  whom adopted
$\kappa \simeq 0.4$~cm$^2$~g$^{-1}$. 

The results of our UVOIR light-curve fits confirms relatively low values ($2.1-4.3~M_{\odot}$) of the ejecta mass for SNe~IIb, Ib and Ic. Among the subtypes, we found SNe~Ib and IIb to exhibit larger average ejecta mass than SNe~Ic. However, within the uncertainty, these averages are still similar, 
as found for example by \citet{lyman16}. 
Turning to the explosion energy, each of the subtypes 
 exhibit values of $E_K = 1.2-1.4\times$10$^{51}$~erg, which is entirely consistent with estimates of previous works. 

The average $^{56}$Ni mass for the three main subtypes ranges between 0.13 and 0.15~$M_{\odot}$. 
These values are somewhat lower than those found from the study of the  
untargeted SDSS-II SN survey, though this discrepancy is due to their sample containing more distant objects  \citep[see][]{taddia15}.

In summary, the comparison between the parameter estimates from the Arnett models indicates no significant differences among the subtypes, with the possible exception of the ejecta mass of SNe~Ic being lower than those of SN~IIb, at least in our work. 

In Table~\ref{tab:compliterature} and in Fig~\ref{compliterature} we do not include a comparison for the SNe~Ic-BL, since we only have two objects in our sample. However, we notice that, in agreement with other works in the literature, our SNe~Ic-BL clearly exhibit higher values of $E_{K}$ and $^{56}$Ni compared to the other subtypes.

\subsection{Comparison between hydrodynamical and semi-analytic models}

One of the properties that we can derive with the hydrodynamical models is the degree of $^{56}$Ni mixing. This is not possible with  Arnett's model, where the radioactive material is assumed to be centrally located in the ejecta. The mixing parameter is important to estimate because a $^{56}$Ni distribution that reaches the outer ejecta can affect the light-curve shape, in particular on the rising part.
In our models we found SNe~Ic are more mixed than SNe~Ib and IIb. All of the SE~SNe are found to be affected by a large degree of $^{56}$Ni mixing. With the exception of one event,  all the SNe~Ic are found to be fully mixed. Studies on the mixing of $^{56}$Ni in core-collapse SN ejecta date back to SN~1987A  \citep[see][and references therein]{shigeyama90,woosley95}. Numerical modeling show the possibility that low $^{56}$Ni mixing can imply the absence of He lines in the spectra despite the presence of He in the ejecta \citep[e.g.,][]{dessart12_OriginIbc}. However, recently we have found evidences for significant mixing in the SNe Ic from studying the  SDSS-II SE~SN sample \citep{taddia15}. Recently, \citet{cano14} has also presented evidence for significant mixing from their analysis of the Type~Ib SN~1999dn.

\begin{figure}
\centering
\includegraphics[width=16cm]{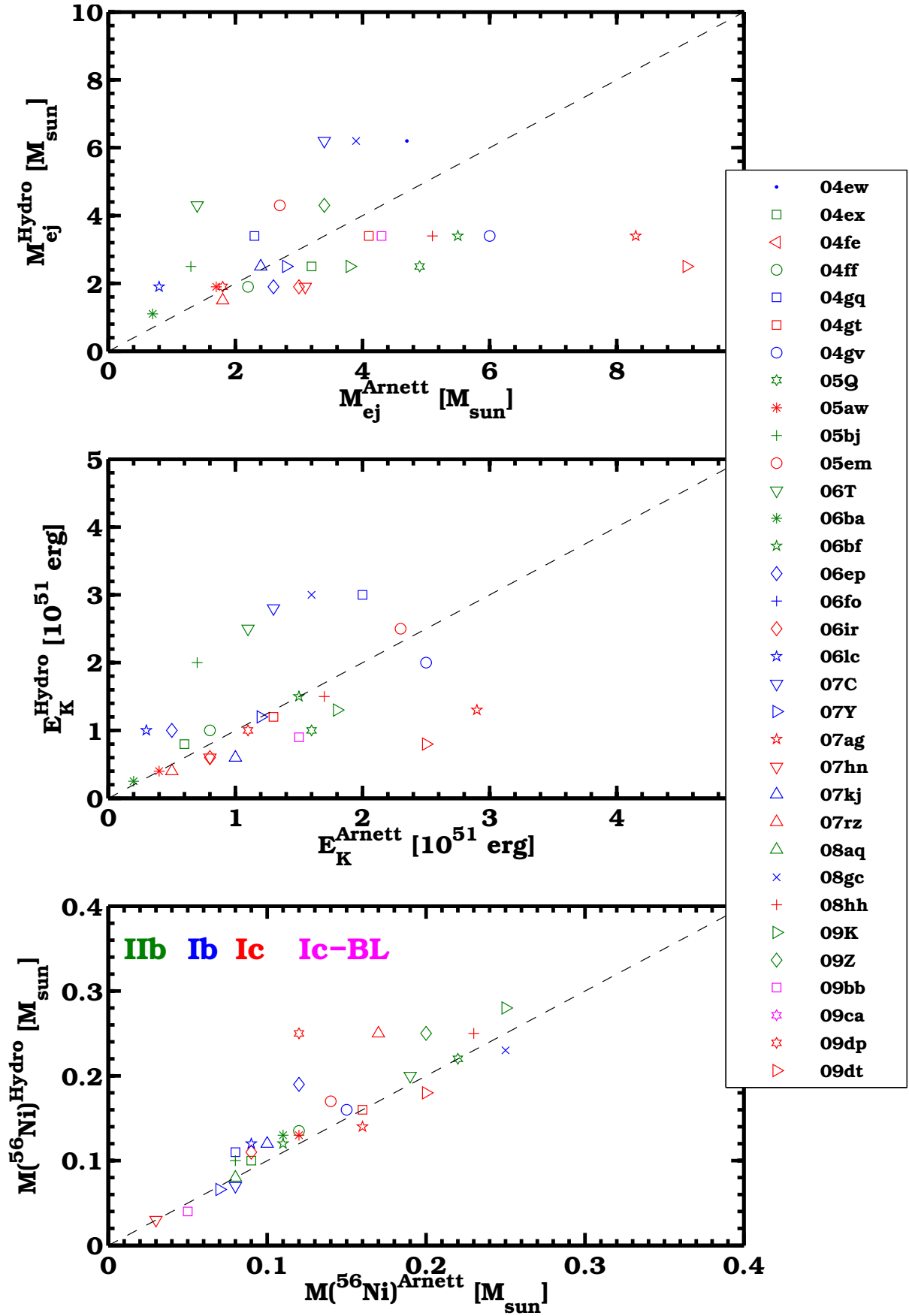}
\caption{\label{hydroarnett}Explosion and progenitor parameter comparisons between the hydrodynamical models and the semi-analytic models.}
\end{figure}

Upon comparison of the other key explosion parameters derived from the Arnett and  hydrodynamical models, as shown in Fig.~\ref{hydroarnett}, 
we find that $^{56}$Ni masses, $M_{ej}$ and $E_{K}$ are in good agreement, with the Arnett models providing slightly larger ejecta masses and kinetic energies for  three objects. 

\citet{dessart16} has recently suggested  Arnett models can give very different light curves compared to those obtained from hydrodynamical models. 
This suggestion is inconsistent with our results and we conclude that adopting  a mean opacity of  $\kappa~=~$0.07~cm$^2$~g$^{-1}$ provides reasonable results to be compared to the more sophisticated hydrodynamical modeling.

\subsection{Implications for the progenitor systems of SE~SNe}
\label{sec:prog}

Stellar-evolution theory shows that single stars below a certain initial mass are unable to strip their outer envelopes, given that their line-driven winds are not strong enough to sustain large mass-loss rates for enough time \citep[see e.g.,][for a recent review on mass loss of massive stars]{smith14}. For relatively low initial-mass stars, to strip the outer hydrogen layers, and in the case of SNe~Ic, also the helium envelopes, mass transfer to a companion star is required \citep[see e.g.,][]{yoon10}. We can consider the ejecta-mass estimates of our SNe and interpret them by using stellar-evolution models in order to infer the nature of their progenitor stars. Following the approach by \citet{lyman16}, we consider that single  progenitor star models with initial masses above 20~$M_{\odot}$, computed by the binary population and spectral synthesis (BPASS) code
(\citealp{eldridge08}; \citealp{eldridge09}), cannot produce ejecta masses below a value of about five $M_{\odot}$.  However, binary models of less massive stars calculated with the same code can easily leave behind lower ($<$~4~$M_{\odot}$) H-poor ejecta masses when they explode.

\begin{figure}
\centering
\includegraphics[width=15cm]{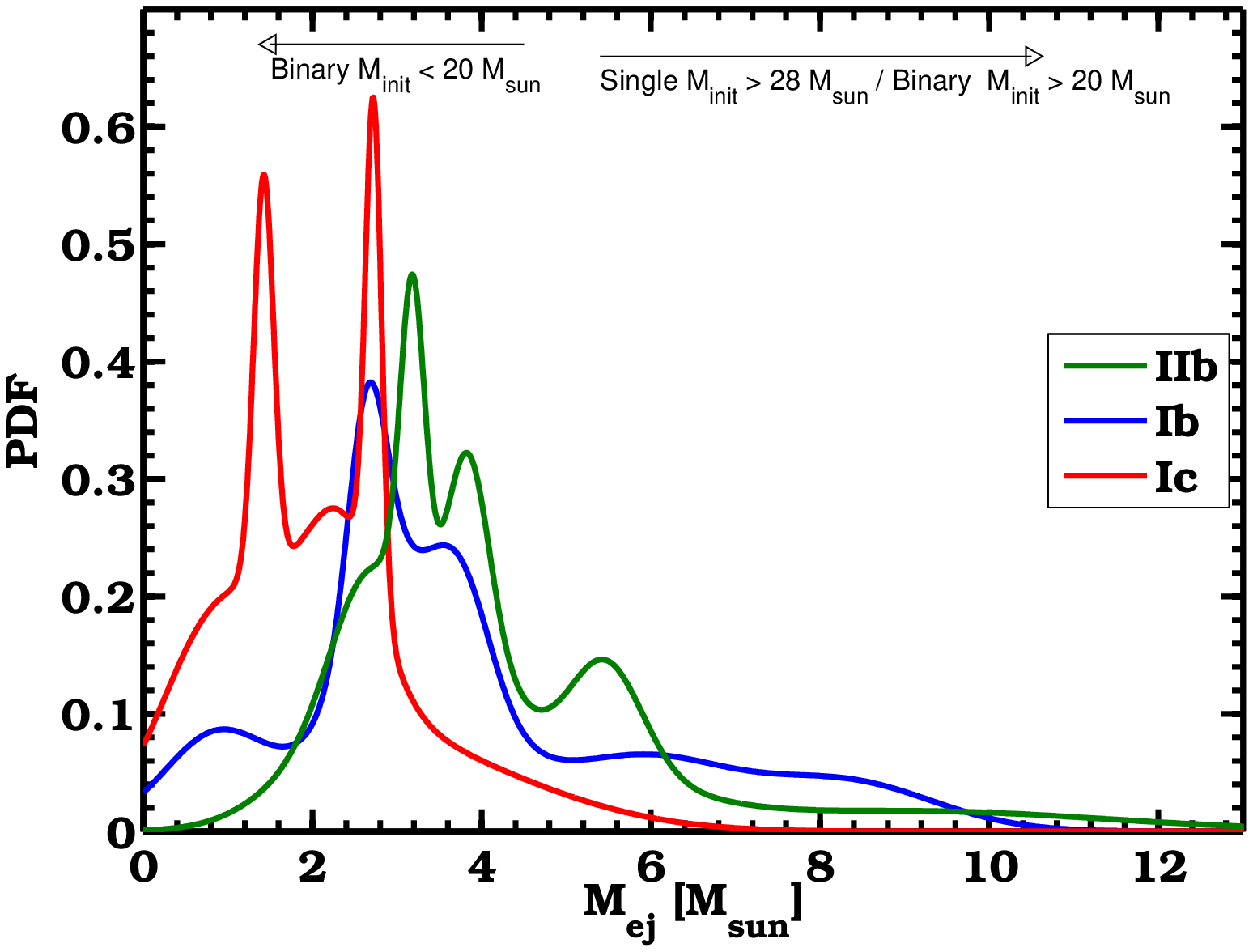}\\
\includegraphics[width=15cm]{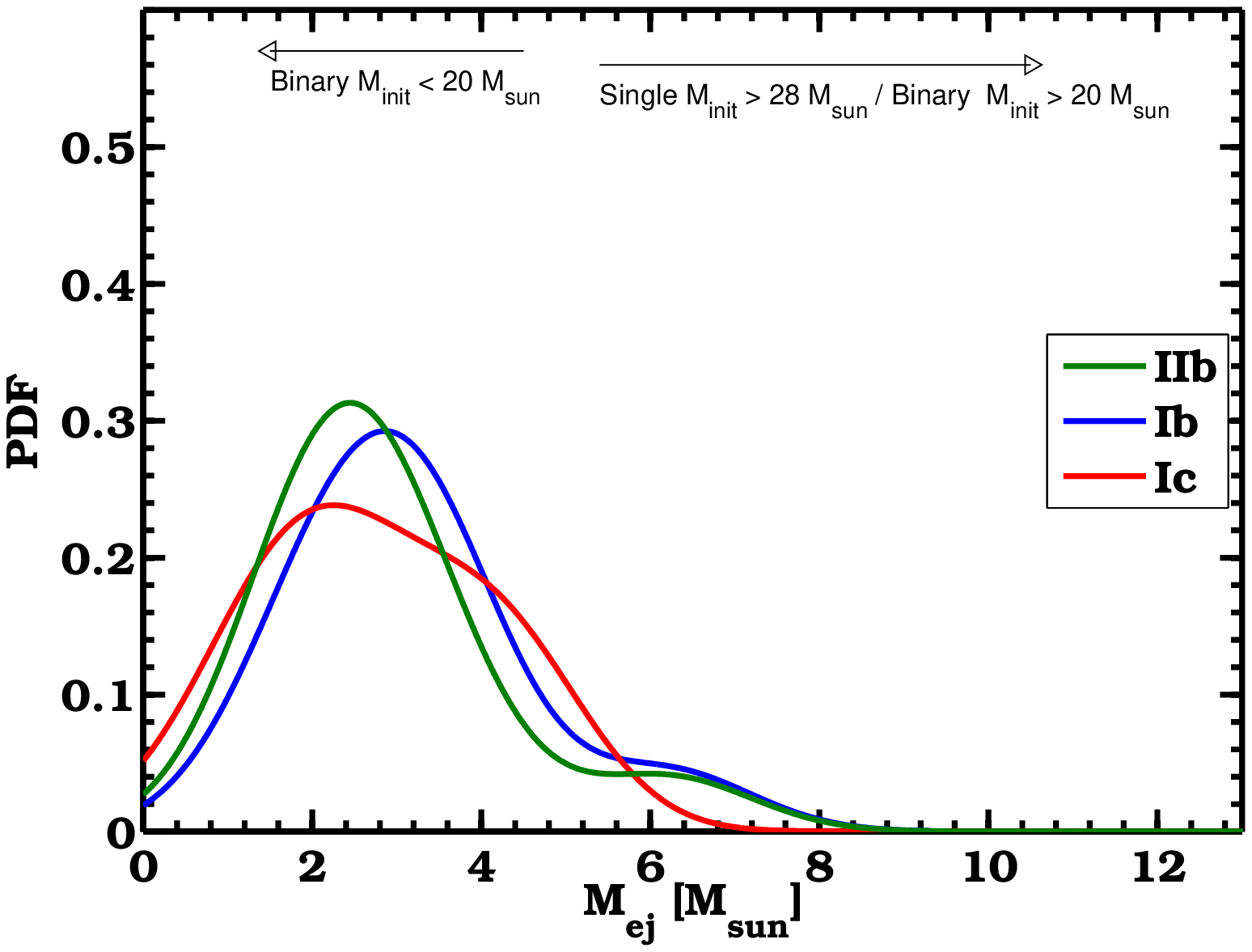}\\
\caption{\label{pdfMej} (Top panel) Probability distribution function of the ejecta masses obtained from the Arnett models of 31 CSP-I SE~SNe. SNe~IIb, Ib, and Ic are represented in green, blue, and red, respectively. (Bottom panel) Probability distribution function of the ejecta masses obtained from the hydrodynamical models of 29 CSP-I SE~SNe.}
\end{figure}

The left panel of Fig.~\ref{param_cdf} and the probability distribution functions (PDFs) in the top panel of Fig.~\ref{pdfMej} show that the Arnett models suggest  
our SE~SNe have ejecta masses $\lesssim$6~$M_{\odot}$. For SNe~Ic the limit is even lower, with only a small probability of events having ejecta masses above $\approx$5~$M_{\odot}$.   

In order to build the probability distribution of the SN ejecta masses, we consider each ejecta mass estimate $M_{ej}$ and its associated error $\sigma$, and construct a
Gaussian distribution centered around $M_{ej}$, with standard deviation equal to $\sigma$, and a normalization equal to one divided by the number of event of each subclass. 
Finally, to obtain the final probability distribution all of the  Gaussian  distributions were summed.

In \citet{lyman16}, the ejecta-mass distribution for SNe~IIb and SNe~Ib peaks at lower values compared to our distributions, favoring the binary scenario for the progenitors of these SNe. In our study, 
 a scenario with a significant majority (92\%) of helium-poor SNe coming from low-mass stars in binary systems still holds. To compute this 92\%, we assumed $M_{ej}~=$~4.5~$M_{\odot}$ from \citeauthor{lyman16} as the upper limit to have low-mass binary progenitors with initial masses $<~20~M_{\odot}$. 
If we instead assume $M_{ej}~=$~5.5~$M_{\odot}$ \citeauthor{lyman16} as the threshold to have single massive progenitors (with initial masses $>~28~M_{\odot}$), only 1.6\% of the SNe~Ic possibly come from massive single stars, or from massive binary progenitors with initial masses $>~20~M_{\odot}$. In the case of  helium-rich SNe, a non-negligible fraction (SNe~IIb$\approx$19\% and SNe~Ib$\approx$21\%) might come from massive binaries or massive single stars. These values are higher than what was found by \citeauthor{lyman16}. 

However, we have to keep in mind that our ejecta mass estimates from Arnett's models are strongly dependent on the assumption regarding the opacity. As clearly shown in the central panel of Fig.~\ref{testkappa}, the ejecta masses of the most massive SE~SNe would drop from $\approx$6~$M_{\odot}$ to $\approx$5~$M_{\odot}$ and from $\approx$5~$M_{\odot}$ to $\approx$4~$M_{\odot}$ if instead of $\kappa=0.07$~cm$^2$~g$^{-1}$ we adopt $\kappa=0.10$~cm$^2$~g$^{-1}$. On the other hand, a lower opacity ($\kappa=0.05$~cm$^2$~g$^{-1}$) would increase the number of SE~SNe with ejecta masses above $\approx$5~$M_{\odot}$, and thus they could possibly arise from single stars. 
Overall, the ejecta masses derived from the Arnett models for our SE~SNe still favor the binary scenario for the majority of their progenitors, though we do not exclude the existence of a small fraction of single massive progenitors. We also note that the discussion above does not consider models with fast rotating SN progenitors, which can make a 40~$M_{\odot}$ inital-mass star producing only $\approx$5.2~$M_{\odot}$ of ejecta \citet[see][]{dessart17}.

In order to overcome the degeneracy inherent to the assumption of constant opacity in the ejecta-mass estimates, we turn to guidance from our hydrodynamical models. 
 To do so, assuming an uncertainty of $\pm$~1$M_{\odot}$ for our ejecta mass estimate from the the hydrodynamical models,  we can draw the probability density functions for the ejecta masses that are shown in the bottom panel of Fig.~\ref{pdfMej}. In this case the SNe~Ic are more similar to SNe~IIb and Ib, which in turn have lower masses than those obtained with application of Arnett's model. The probability of having single massive stars as the progenitors of these SE~SNe is rather low for all the subtypes, in good agreement with the PDF obtained by \citet{lyman16}. 

The large degree of mixing found for SNe~Ic suggests \citep[see also]
[]{taddia15} that the lack of helium is real for this SE~SN subtype, and it is not due to the  helium not being ionized by radioactive material \citep[see][]{dessart12_OriginIbc}.

\section{Conclusions}
\label{sec:conclusion}
We presented the analysis of a sample of 34 SE~SNe from the CSP-I. Our main findings concerning their light-curve shapes are:

\begin{itemize}
\item{SE~SNe show similar light-curve properties among the three main subtypes, in particular similar $\Delta m_{15}$ and peak absolute magnitudes (typically $-18~<~M_{\rm max}(r)~<~-17$~mag). }

\item{$\Delta m_{15}$ is found to correlate with the slope of the light curve during its linear decay. This can be explained in terms of large spread of explosion energy over ejecta mass ($E_K/M_{ej}$), with larger values corresponding to larger $\Delta m_{15}$ and steeper late-time slopes.}

\item{A possible correlation between $\Delta m_{15}(B)$ and peak absolute $B$-band  magnitude is found, reminiscent of a well-known trend followed by thermonuclear SNe~Ia. }
\end{itemize}

 Our main findings  concerning the progenitor properties based on the bolometric modeling are:

  \begin{itemize}
 
  \item{From our hydrodynamical models, typical ejecta masses for SE~SNe are found to be relatively small (1.1--6.2~$M_{\odot}$) and thus incompatible with the majority of events arising from massive single stars.}
  
\item{This result on the mass is similar when we consider the ejecta masses from the semi-analytic models, even though these are known to be affected by the assumptions regarding the constant value of the opacity. We found that assuming $\kappa$~$=$~0.07~cm$^{2}$~g$^{-1}$ provides a good agreement between the results of the hydrodynamical models and those of the semi-analytical models. We also found that inferring the expansion velocity directly from \ion{Fe}{ii} or via \ion{He}{i} and \ion{O}{i} for He-rich and He-poor SNe does not significantly alter the results on the ejecta mass.}
  
   \item{SNe~Ic tend to exhibit a larger degree of mixing among the various SE~SN subtypes, suggesting that the lack of helium in their spectra corresponds to an actual lack of this element in the progenitors.}
  
\end{itemize}

\begin{acknowledgements}
We thank P. Hoeflich, P. A. Mazzali, and A. Piro for useful discussions.
F. Taddia and J. Sollerman gratefully acknowledge support from the Knut and Alice Wallenberg Foundation.  
M.~D. Stritzinger, C. Contreras, and E. Y. Hsiao acknowledge generous support provided by the Danish Agency for Science and Technology and Innovation realized through a Sapere Aude Level 2 grant and the  Instrument-center for Danish Astrophysics (IDA).
M.~D. Stritzinger is supported by a research grant (13261) from  VILLUM FONDEN.
This material is based upon work supported by the US National Science
Foundation (NSF) under grants AST--0306969, AST--0607438,  and AST--1008343, AST--1613426, AST-1613455 and AST-1613472. 
A portion of the work presented here was done at the Aspen Center for Physics, which is supported by NSF grant PHY-1066293. 
This research has made use of the NASA/IPAC Extragalactic Database (NED) which is operated by the Jet Propulsion Laboratory, California Institute of Technology, under contract with the National Aeronautics and Space Administration.
\end{acknowledgements}

\onecolumn

\clearpage

\clearpage

\clearpage

\clearpage

\clearpage

\clearpage
 
 \clearpage
 
 \clearpage

\clearpage

\clearpage

\clearpage

\clearpage

\clearpage

\clearpage

\clearpage

\clearpage

\clearpage

\clearpage

\clearpage

\clearpage

\clearpage

\clearpage

\clearpage

\clearpage

\onecolumn

\clearpage
\begin{deluxetable}{lllllllllllllllll}
\rotate
\tabletypesize{\scriptsize}
\tablewidth{0pt}
\tablecolumns{15}
\tablecaption{CSP-I SE~SN sample.\label{listsn}}
\tablehead{
\colhead{SN} &
\colhead{SN RA} &
\colhead{SN DEC} &
\colhead{SN Type} &
\colhead{Galaxy} &
\colhead{Galaxy RA} &
\colhead{Galaxy DEC} &
\colhead{$A_V^{\rm MW}$} &
\colhead{$R_V^{\rm host}$} &
\colhead{$A_V^{\rm host}$} &
\colhead{Redshift}  &
\colhead{Distance} &
\colhead{2a} &
\colhead{2b} &
\colhead{t-type} &
\colhead{PA} &
\colhead{$d_{SN}$}\\
\colhead{} &
\colhead{(hh:mm:ss)} &
\colhead{(dd:mm:ss)} &
\colhead{} &
\colhead{} &
\colhead{(hh:mm:ss)} &
\colhead{(dd:mm:ss)} &
\colhead{(mag)} &
\colhead{} &
\colhead{(mag)} &
\colhead{} &
\colhead{(Mpc)} &
\colhead{($\arcmin$)}&
\colhead{($\arcmin$)}&
\colhead{}&
\colhead{($^\circ$)}&
\colhead{}}
\startdata
2004ex & 00:38:10.19 & $+$02:43:17.2& IIb     &  NGC 0182                    &00:38:12.38 & +02:43:42.8& 0.061 & 1.3    &     0.26(0.02) &   0.017549  &  70.6$\pm$4.9                       &      2.0    &    1.7  & 1.3     &  75     &  0.72   \\
2004ff & 04:58:46.19 & $-$21:34:12.0& IIb     &  ESO 552- G 040              &04:58:47.06 & -21:34:09.9& 0.090 & 1.3    &     0.30(0.03) &   0.022649  &  92.7$\pm$6.4                       &      1.2    &    0.7  & 2.0     &  50     &  0.44   \\
2004gq & 05:12:04.81 & $-$15:40:54.2& Ib      &  NGC 1832                    &05:12:03.33 & -15:41:16.1& 0.200 & 2.4    &     0.26(0.07) &   0.006468  &  25.1$\pm$4.6\tablenotemark{*}      &      2.6    &    1.7  & 4.0     &  10     &  0.48   \\
2004gt & 12:01:50.37 & $-$18:52:12.7& Ic      &  NGC 4038                    &12:01:53.01 & -18:52:03.4& 0.127 & 2.5    &     1.07(0.15) &   0.005477  &  23.2$\pm$1.6                       &      5.2    &    3.1  & 8.8     &  80     &  0.25   \\
2004gv & 02:13:37.42 & $-$00:43:05.8& Ib      &  NGC 0856                    &02:13:38.36 & -00:43:02.2& 0.090 & 2.4    &     0.08(0.03) &   0.019927  &  79.6$\pm$5.5                       &     1.24    &   0.96  & 0.3     &  20     &  0.49   \\
2005aw & 19:15:17.44 & $-$54:08:24.9& Ic      &  IC 4837A                    &19:15:16.17 & -54:07:57.0& 0.168 & 1.4    &     0.65(0.06) &   0.009498  &  41.5$\pm$2.9                       &      4.1    &    0.6  & 2.5     & 165     &  0.21   \\
2005em & 03:13:47.71 & $-$00:14:37.0& Ic      &  IC 0307                     &03:13:45.21 & -00:14:29.2& 0.263 & \ldots &     0          &   0.025981  & 105.0$\pm$7.2                       &     1.78    &   0.62  & 1.0     &  73     &  2.91   \\
2006T  & 09:54:30.21 & $-$25:42:29.3& IIb     &  NGC 3054                    &09:54:28.60 & -25:42:12.4& 0.205 & 1.3    &     0.42(0.05) &   0.008091  &  31.6$\pm$5.9\tablenotemark{*}      &      3.8    &    2.3  & 3.0     & 118     &  0.25   \\
\enddata
\tablenotetext{*}{Redshift independent measurement.} 
\tablenotetext{**}{Distance computed via NED cosmological calculator.} 
\tablenotetext{***}{PA from SIMBAD.} 
\tablenotetext{+}{Extinction from EW(\ion{Na}{i}~D).} 
\tablecomments{Coordinates, Galactic extinctions ($A_V^{\rm MW}$), redshifts, distances, major (2a) and minor (2b) axes of the galaxies are from NED, t-types and position angles (PA) are from the Asiago Supernova Catalogue (ASC). A single horizontal line separates the objects observed in both optical and NIR from those observed only in the optical.}
\end{deluxetable}

\clearpage
\begin{deluxetable}{llllllllll}
\tablewidth{0pt}
\tabletypesize{\scriptsize}
\tablecaption{Optical- and NIR-band epochs of peak maximum  (JD$-$2450000).\label{tmax_kcorr}}
\tablehead{
\colhead{SN}       &
\colhead{t$_{\rm max}(u)$ }&
\colhead{t$_{\rm max}(B)$ }&
\colhead{t$_{\rm max}(g)$ }&
\colhead{t$_{\rm max}(V)$ }&
\colhead{t$_{\rm max}(r)$ }&
\colhead{t$_{\rm max}(i)$ }&
\colhead{t$_{\rm max}(Y)$ }&
\colhead{t$_{\rm max}(J)$ }&
\colhead{t$_{\rm max}(H)$ }}
\startdata
 2004ex  & 3306.08(0.07)  & 3306.69(0.03)  & 3307.06(0.01)  & 3308.22(0.02)  & 3309.10(0.03)  & 3310.55(0.02)  & 3314.87(0.03)  & 3310.22(0.01)  & 3313.01(0.11)    \\
 2004ff  & 3310.95(1.37)* & 3312.60(0.03)  & 3312.64(0.05)  & 3314.32(0.03)  & 3314.67(0.01)  & 3316.34(0.04)  & 3318.82(1.37)* & 3319.87(1.37)* & 3320.02(1.40)*   \\
 2004gq  & 3356.53(0.11)  & 3357.90(0.02)  & 3358.50(0.05)  & 3359.86(0.04)  & 3361.56(0.05)  & 3363.92(0.02)  & 3367.56(0.02)  & 3370.35(0.37)  & 3366.92(0.02)    \\
 2004gt  & 3359.51(1.37)* & 3360.37(0.08)  & 3361.97(0.44)  & 3362.51(0.11)  & 3363.23(0.04)  & 3365.94(0.20)  & 3368.14(0.07)  & 3368.58(0.01)  & 3368.99(0.02)    \\
 2004gv  & 3364.62(0.01)  & 3365.27(0.31)  & 3365.51(0.04)  & 3367.09(0.01)  & 3368.60(0.02)  & 3370.36(0.24)  & 3371.80(0.01)  & 3371.70(0.12)  & 3377.18(0.02)    \\
 2005aw  & 3455.45(1.37)* & 3456.48(1.37)* & 3457.02(1.36)* & 3458.06(1.36)* & 3459.17(0.07)  & 3460.02(0.01)  & 3464.60(0.01)  & 3464.31(0.02)  & 3464.52(1.40)*   \\
 2005em  & 3647.94(1.37)* & 3648.97(1.36)* & 3649.51(1.36)* & 3648.95(0.03)  & 3651.66(0.02)  & 3652.45(0.10)  & 3660.15(0.13)  & 3657.43(3.09)  & 3657.13(0.86)    \\
 2006T   & 3779.10(0.03)  & 3779.61(0.01)  & 3780.08(0.01)  & 3781.00(0.02)  & 3781.93(0.02)  & 3783.43(0.01)  & 3784.06(0.08)  & 3784.98(0.10)  & 3787.21(1.65)    \\
\enddata                                                        
\tablecomments{A horizontal line separates the objects observed in both optical and NIR from those observed only in the optical.}
\tablenotetext{*}{Obtained from the relation shown in Fig.~\ref{maxvslambda}.}
\tablenotetext{**}{Obtained from the fit to the templates shown in Fig.~\ref{template}.}
\end{deluxetable}

\clearpage
\begin{deluxetable}{cccccccccc}
\tablewidth{0pt}
\tabletypesize{\tiny}
\tablecaption{Optical and NIR band $\Delta m_{15}$ values.\label{dm15_kcorr}.}
\tablehead{
\colhead{SN}       &
\colhead{$\Delta m_{15}(u)$ }&
\colhead{$\Delta m_{15}(B)$ }&
\colhead{$\Delta m_{15}(g)$ }&
\colhead{$\Delta m{15}(V)$ }&
\colhead{$\Delta m_{15}(r)$ }&
\colhead{$\Delta m_{15}(i)$ }&
\colhead{$\Delta m_{15}(Y)$ }&
\colhead{$\Delta m_{15}(J)$ }&
\colhead{$\Delta m_{15}(H)$ } \\
\colhead{}       &
\colhead{({\rm mag}) }&
\colhead{({\rm mag}) }&
\colhead{({\rm mag}) }&
\colhead{({\rm mag}) }&
\colhead{({\rm mag}) }&
\colhead{({\rm mag}) }&
\colhead{({\rm mag}) }&
\colhead{({\rm mag}) }&
\colhead{({\rm mag}) }}
\startdata
    2004ex  &     0.00(0.06) & 1.34(0.03) & 1.06(0.02) & 0.93(0.02) & 0.74(0.02) & 0.55(0.02) & 0.35(0.02) & 0.38(0.03) & 0.27(0.04)    \\
    2004ff  &     \ldots     & 1.45(0.04) & 1.19(0.03) & 1.05(0.02) & 0.80(0.01) & 0.61(0.02) & \ldots     & \ldots     & \ldots        \\
    2004gq  &     2.23(0.01) & 1.36(0.02) & 1.11(0.01) & 0.86(0.01) & 0.64(0.01) & 0.55(0.02) & 0.34(0.02) & 0.58(0.04) & 0.29(0.02)    \\
    2004gt  &     \ldots     & 0.81(0.04) & 0.79(0.07) & 0.62(0.03) & 0.44(0.02) & 0.42(0.03) & 0.49(0.04) & 0.41(0.02) & 0.22(0.03)    \\
    2004gv  &     2.60(0.13) & 1.32(0.08) & 1.03(0.01) & 0.83(0.01) & 0.58(0.01) & 0.47(3.13) & 0.00(0.02) & 0.00(0.02) & 0.00(0.02)    \\
    2005aw  &     \ldots     & \ldots     & \ldots     & \ldots     & 0.53(0.01) & 0.47(0.00) & 0.34(0.01) & 0.34(0.01) & \ldots        \\
    2005em  &     \ldots     & \ldots     & \ldots     & 0.92(0.04) & 0.80(0.03) & 0.62(0.03) & 0.84(0.04) & 0.70(0.25) & 0.37(0.08)    \\
    2006T   &     2.44(0.05) & 1.49(0.02) & 1.22(0.01) & 0.99(0.01) & 0.74(0.01) & 0.52(0.01) & 0.35(0.02) & 0.37(0.03) & 0.26(0.09)    \\
\enddata
\tablecomments{ $\Delta m_{15}$ values for the CSP SE~SNe observed at early epochs. A single horizontal line separates the objects observed in both optical and NIR from those observed only in the optical.}
\end{deluxetable}

\clearpage
\begin{deluxetable}{cccccccccc}
\tablewidth{0pt}
\tabletypesize{\scriptsize}
\tablecaption{Peak optical and NIR apparent magnitudes.\label{tab:apppeak_kcorr}}
\tablehead{
\colhead{SN}       &
\colhead{$u_{\rm max}$ }&
\colhead{$B_{\rm max}$ }&
\colhead{$g_{\rm max}$ }&
\colhead{$V_{\rm max}$ }&
\colhead{$r_{\rm max}$ }&
\colhead{$i_{\rm max}$ }&
\colhead{$Y_{\rm max}$ }&
\colhead{$J_{\rm max}$ }&
\colhead{$H_{\rm max}$ }\\
\colhead{}       &
\colhead{({\rm mag}) }&
\colhead{({\rm mag}) }&
\colhead{({\rm mag}) }&
\colhead{({\rm mag}) }&
\colhead{({\rm mag}) }&
\colhead{({\rm mag}) }&
\colhead{({\rm mag}) }&
\colhead{({\rm mag}) }&
\colhead{({\rm mag}) }}
\startdata
    2004ex  &     19.05(0.06) & 18.11(0.02) & 17.78(0.01) & 17.42(0.01) & 17.30(0.01) & 17.36(0.01) & 17.11(0.02) & 16.95(0.02) & 16.85(0.03)      \\
    2004ff  &     \ldots      & 18.37(0.02) & 18.04(0.01) & 17.73(0.01) & 17.62(0.01) & 17.59(0.01) & \ldots      & \ldots      & \ldots           \\
    2004gq  &     16.33(0.01) & 15.83(0.01) & 15.56(0.01) & 15.36(0.01) & 15.26(0.01) & 15.12(0.01) & 14.58(0.02) & 14.50(0.03) & 14.34(0.01)      \\
    2004gt  &     \ldots      & 16.31(0.03) & 15.90(0.02) & 15.40(0.02) & 15.10(0.02) & 14.90(0.03) & 14.18(0.04) & 14.00(0.02) & 13.85(0.03)      \\
    2004gv  &     18.18(0.03) & 17.67(0.02) & 17.43(0.01) & 17.20(0.01) & 17.11(0.01) & 17.14(0.30) & 16.80(0.02) & 16.68(0.02) & 16.48(0.02)      \\
    2005aw  &     \ldots      & \ldots      & \ldots      & \ldots      & 15.75(0.01) & 15.78(0.01) & 15.12(0.01) & 15.02(0.01) & \ldots           \\
    2005em  &     \ldots      & \ldots      & \ldots      & 17.83(0.02) & 17.79(0.03) & 17.92(0.02) & 17.31(0.03) & 17.65(0.18) & 17.38(0.08)      \\
    2006T   &     17.55(0.03) & 16.47(0.02) & 16.09(0.01) & 15.70(0.01) & 15.51(0.01) & 15.55(0.01) & 15.15(0.02) & 15.02(0.02) & 14.87(0.03)      \\
\enddata                                                        
\tablecomments{Peak optical and NIR magnitudes for the CSP SE~SNe observed at early epochs. A horizontal line separates the objects observed in both optical and NIR from those observed only in the optical.}
\end{deluxetable}

\clearpage
\begin{deluxetable}{cccccccccc}
\tablewidth{0pt}
\tabletypesize{\scriptsize}
\tablecaption{Peak absolute magnitudes in the optical and NIR bands\label{tab:abspeak_kcorr}}
\tablehead{
\colhead{SN}       &
\colhead{$u_{\rm max}$}&
\colhead{$B_{\rm max}$}&
\colhead{$g_{\rm max}$}&
\colhead{$V_{\rm max}$}&
\colhead{$r_{\rm max}$}&
\colhead{$i_{\rm max}$}&
\colhead{$Y_{\rm max}$}&
\colhead{$J_{\rm max}$}&
\colhead{$H_{\rm max}$}\\
\colhead{}       &
\colhead{({\rm mag}) }&
\colhead{({\rm mag}) }&
\colhead{({\rm mag}) }&
\colhead{({\rm mag}) }&
\colhead{({\rm mag}) }&
\colhead{({\rm mag}) }&
\colhead{({\rm mag}) }&
\colhead{({\rm mag}) }&
\colhead{({\rm mag}) } }
\startdata

2004ex   &  $-$15.98(0.17) & $-$16.67(0.16) & $-$16.94(0.15) & $-$17.16(0.15) & $-$17.16(0.15) & $-$17.02(0.15) & $-$17.24(0.15) & $-$17.38(0.15) & $-$17.45(0.15)  \\
2004ff   &  \ldots         & $-$17.12(0.16) & $-$17.36(0.16) & $-$17.51(0.15) & $-$17.48(0.15) & $-$17.42(0.15) & \ldots         & \ldots         & \ldots          \\
2004gq   &  $-$16.47(0.42) & $-$16.81(0.41) & $-$17.02(0.41) & $-$17.11(0.41) & $-$17.12(0.41) & $-$17.16(0.40) & $-$17.59(0.40) & $-$17.62(0.40) & $-$17.74(0.40)  \\
2004gt   &  \ldots         & $-$17.18(0.26) & $-$17.46(0.24) & $-$17.65(0.22) & $-$17.70(0.19) & $-$17.62(0.17) & $-$18.07(0.16) & $-$18.15(0.16) & $-$18.19(0.15)  \\
2004gv   &  $-$16.62(0.16) & $-$17.07(0.16) & $-$17.29(0.16) & $-$17.48(0.15) & $-$17.54(0.15) & $-$17.47(1.57) & $-$17.77(0.15) & $-$17.87(0.15) & $-$18.05(0.15)  \\
2005aw   &  \ldots         & \ldots         & \ldots         & \ldots         & $-$17.91(0.16) & $-$17.69(0.15) & $-$18.24(0.15) & $-$18.29(0.15) & \ldots          \\
2005em   &  \ldots         & \ldots         & \ldots         & $-$17.55(0.15) & $-$17.55(0.15) & $-$17.36(0.15) & $-$17.90(0.15) & $-$17.53(0.23) & $-$17.77(0.17)  \\
2006T    &  $-$16.39(0.43) & $-$17.04(0.42) & $-$17.30(0.42) & $-$17.45(0.41) & $-$17.43(0.41) & $-$17.24(0.41) & $-$17.55(0.41) & $-$17.65(0.41) & $-$17.75(0.41)  \\
\enddata                                                        
\tablecomments{Peak absolute magnitudes in the different bands for the CSP SE~SN sample observed at early epochs. A single horizontal line separates the objects observed in both optical and NIR from those observed only in the optical. The associated uncertainty is the sum in quadrature of the uncertainty in the distance modulus, the uncertainty in the extinction, and the uncertainty in the apparent peak magnitude value.}
\end{deluxetable}

\clearpage
\begin{deluxetable}{ccccc}
\tablewidth{0pt}
\tabletypesize{\scriptsize}
\tablecaption{Bolometric light-curve parameters\label{tab:boloparam}}
\tablehead{
\colhead{SN}       &
\colhead{t$_{\rm max}$(bol)}&
\colhead{M$_{\rm max}$(bol)}&
\colhead{$\Delta m_{15}$(bol)}&
\colhead{Late-time slope (bol)}\\
\colhead{}       &
\colhead{(JD)}&
\colhead{(mag)}&
\colhead{(mag)}&
\colhead{(mag~days$^{-1}$)}}
\startdata

  2004ex    &   2453309.54(0.01) & -17.00(0.06)& 0.79(0.06)& 0.0199(0.0001)               \\
  2004ff    &   2453313.52(0.22) & -17.52(0.06)& 0.70(0.06)& 0.0237(0.0002)               \\
  2004gq    &   2453358.08(0.01) & -17.07(0.06)& 0.46(0.06)& 0.0151(0.0001)               \\
  2004gt    &   2453362.32(0.45) & -17.53(0.06)& 0.46(0.06)& 0.0353(0.0001)               \\
  2004gv    &   2453367.46(0.27) & -17.47(0.05)& 0.60(0.05)& \ldots                       \\
  2005aw    &   2453457.97(0.03) & -17.73(0.06)& 0.50(0.06)& \ldots   \\
  2006T     &   2453781.54(0.15) & -17.23(0.06)& 0.75(0.06)& 0.0199(0.0001)               \\
  2006bf    &   \ldots           & \ldots          & \ldots          & 0.0092(0.0000)     \\
\enddata                                                        
\tablecomments{Bolometric light-curve parameters for the CSP SE~SNe observed before bolometric peak, as obtained by fitting the bolometric light curves with Eq.~\ref{eq:1}. A horizontal line separates the objects observed in both optical and NIR from those observed only in the optical.}
\end{deluxetable}

\clearpage
\begin{deluxetable}{llccccccl}
\tablewidth{0pt}
\tabletypesize{\scriptsize}
\tablecaption{\label{dates}JD and magnitude of last-non detection, discovery, and confirmation epochs and estimated explosion epoch for 33 CSP-I SE~SNe.}
\tablehead{
\colhead{SN}       &
\colhead{Discovery}&
\colhead{Last non-detection}&
\colhead{Discovery}&
\colhead{Confirmation}&
\colhead{Last non-detection}&
\colhead{Discovery }&
\colhead{Confirmation}&
\colhead{Explosion date}\\
\colhead{}&
\colhead{telegram }&
\colhead{(JD--2450000)}&
\colhead{(JD--2450000)}&
\colhead{(JD--2450000)}&
\colhead{(mag) }&
\colhead{(mag) }&
\colhead{(mag) }&
\colhead{(JD--2450000)}}
\startdata   
  2004ex & IAUC 8418 & 3272.77  &3289.84   &3291.83  &$>$19.0  &  17.7    &  17.7     &  3288.40$^{0.33}_{-0.33}$\tablenotemark{R}\\        
  2004ff & IAUC 8425 & 3291.91  &3308.90   &3309.91  &$>$19.0  &  18.0    &  18.0     &  3298.16$^{8.22}_{-6.25}$\tablenotemark{R}\\        
  2004gq & IAUC 8452 & 3343.88  &3350.86   &3351.43  &$>$19.5  &  15.5    &  15.9     &  3347.37$^{3.49}_{-3.49}$\tablenotemark{L}\\        
  2004gt & IAUC 8454 & 3136.75  &3351.58   &3355.51  &$>$15.7  &  14.9    &  14.6     &  3343.33$^{4.08}_{-4.08}$\tablenotemark{R}\\        
  2004gv & IAUC 8454 & 3338.24  &3353.17   &3354.07  &$>$18.6  &  17.6    &  17.4     &  3345.77$^{1.54}_{-1.54}$\tablenotemark{R}\\        
  2005aw & CBET 127  & 3436.82  &3453.77   &3454.75  &$>$17.9  &  15.3    &  15.3     &  3446.17$^{3.00}_{-3.00}$\tablenotemark{R}\\        
  2005em & IAUC 8604 & 3615.93  &3640.94   &3641.88  &$>$19.5  &  18.1    &  18.0     &  3638.66$^{2.28}_{-3.00}$\tablenotemark{T}\\        
  2006T  & CBET 385  & 3752.45  &3766.49   &3767.35  &$>$18.0  &  17.2    &  17.4     &  3758.14$^{0.92}_{-0.92}$\tablenotemark{R}\\
\enddata                                                        
\tablecomments{A horizontal line separates the objects observed in both optical and NIR from those observed only in the optical.}
\tablenotetext{a}{From \citet{stritzinger09}.}
\tablenotetext{b}{From \citet{taddia15}.}
\tablenotetext{c}{From \citet{pignata11}.}
\tablenotetext{L}{From good pre-explosion limits.}
\tablenotetext{R}{From the fit of the photospheric radius before $r_{\rm max}$.}
\tablenotetext{T}{From the rise time.}
\end{deluxetable}

\begin{deluxetable}{lc|ccc|cccc}
\tablewidth{0pt}
\tabletypesize{\scriptsize}
\tablecaption{\label{param}Explosion parameters for 33 CSP-I SE~SNe from the semi-analytic and hydrodynamical modeling of their bolometric light curves.}
\tablehead{
\colhead{SN}       &
\colhead{Type}       &
\colhead{$M_{ej}$}&
\colhead{$E_{K}$}&
\colhead{$\rm M(^{56}Ni)$}&
\colhead{$M_{ej}$}&
\colhead{$E_{K}$}&
\colhead{$\rm M(^{56}Ni)$}&
\colhead{$\rm ^{56}Ni$ mixing}\\
\colhead{}&
\colhead{}&
\colhead{($M_\odot$)}&
\colhead{($10^{51}$ erg)}&
\colhead{($M_\odot$)}&
\colhead{($M_\odot$)}&
\colhead{($10^{51}$ erg) }&
\colhead{($M_\odot$)}&
\colhead{}}
\startdata 
  \multicolumn{7}{c}{         \hspace{2.5cm} Arnett model       \hspace{3.5cm}  Hydrodynamical model} \\
  2004ex    &    IIb      &   3.2(0.1) & 0.6(0.1) & 0.09(0.01)     &     2.5    &    0.8  &  0.10   &   0.88   \\
  2004ff    &    IIb      &   2.2(0.7) & 0.8(0.3) & 0.12(0.02)     &     1.9    &    1.0  &  0.135  &   0.88   \\
  2004gq    &    Ib       &   2.3(1.5) & 2.0(1.3) & 0.08(0.02)     &     3.4    &    3.0  &  0.11   &   0.98   \\
  2004gt    &    Ic       &   4.1(1.3) & 1.3(0.4) & 0.16(0.02)     &     3.4    &    1.2  &  0.16   &   1.00   \\
  2004gv    &    Ib       &   6.0(1.0) & 2.5(0.4) & 0.15(0.01)     &     3.4    &    2.0  &  0.16   &   0.68   \\
  2005aw    &    Ic       &   2.7(0.1) & 2.3(0.1) & 0.14(0.02)     &     4.3    &    2.5   &  0.17   &   0.95   \\
  2005em    &    Ic       &   0.7(0.5) & 0.2(0.1) & 0.11(0.01)     &     1.1    &    0.25  &  0.13   &   1.00   \\
\enddata
\tablecomments{A single  horizontal line separates the objects observed in both optical and NIR from those observed only in the optical. Uncertainties on the average of each sub-sample are the standard deviations.}
\end{deluxetable}

\begin{deluxetable}{l|lccc|}
\tablewidth{0pt}
\tabletypesize{\scriptsize}
\tablecaption{\label{tab:compliterature}Comparison of the explosion parameters for different SE~SN samples in the literature, from semi-analytic models.}
\tablehead{
\colhead{Sample}       &
\colhead{Type}       &
\colhead{$M_{ej}$}&
\colhead{$E_{K}$}&
\colhead{$M{\rm(^{56}Ni)}$}\\
\colhead{}&
\colhead{}&
\colhead{($M_\odot$)}&
\colhead{($10^{51}$ erg)}&
\colhead{($M_\odot$)}}
\startdata 
 
 {\bf \citet{richardson06}}&IIb (2)    & 1.1$\pm$0.3 & 0.4$\pm$0.3 & 0.09$\pm$0.01 \\  
&Ib    (6)                       & 1.8$\pm$1.2 & 0.8$\pm$0.7 & 0.56$\pm$0.55 \\  
&Ic   (7)                        & 0.8$\pm$0.4 & 0.5$\pm$0.4 & 0.28$\pm$0.26 \\  
\hline
{\bf \citet{drout11}}&Ib (11)    & $-$ & $-$ & 0.20$\pm$ 0.16\\  
&Ic    (10)                       & $-$ & $-$ & 0.24$\pm$0.15 \\  
\hline
{\bf \citet{cano13}}&Ib  (19)      &  4.7$\pm$2.8   & 3.3$\pm$2.6       & 0.21$\pm$0.22      \\
                    &Ic  (13)       &  4.6$\pm$4.5   & 3.3$\pm$2.6       & 0.23$\pm$0.19      \\
\hline		 
		 {\bf \citet{taddia15}}&Ib (6)  &     3.6$\pm$1.5       &  1.5$\pm$0.9   & 0.30$\pm$0.11        \\
\enddata

\tablecomments{Uncertainties on the average of each sub-sample are the standard deviations.}
\end{deluxetable}

\end{document}